\documentclass[a4paper,11pt,headings=big]{scrartcl}
\pdfoutput=1
\usepackage[utf8]{inputenc}
\usepackage{dsfont, amsmath,amssymb, amsfonts,slashed}
\usepackage{geometry}
\usepackage{graphicx}
\usepackage{cite}
\usepackage{feyn}
\usepackage{bbm}
\usepackage{xcolor}
\definecolor{blue}{rgb}{0,0,0.5}
\usepackage[colorlinks,linkcolor=blue,citecolor=blue,urlcolor=blue]{hyperref}
\usepackage{multirow}
\usepackage{rotating}
\allowdisplaybreaks

\newcommand{\tr}[1]{\mathrm{tr} \left[ #1 \right]}
\newcommand{\T}[1]{\mathsf{T}^{#1}}

\newcommand{\axial}{\mathfrak{a}}
\newcommand{\im}{{i}}

\newcommand{\elem}[1]{{#1^{0}}}
\newcommand{\elemcoupl}[1]{{#1_{0}}}

\addtokomafont{disposition}{\rmfamily\boldmath}
\addtokomafont{caption}{\small}
\addtokomafont{captionlabel}{\small\bfseries}

\title{Direct and indirect signals of natural composite Higgs models}
\author{\large
Christoph Niehoff,
Peter Stangl,
and David M. Straub%
\footnote{E-Mail:
\texttt{\href{mailto:christoph.niehoff@tum.de}{christoph.niehoff@tum.de},
\href{mailto:peter.stangl@tum.de}{peter.stangl@tum.de},
\href{mailto:david.straub@tum.de}{david.straub@tum.de}}}
}
\date{\normalsize\itshape
Excellence Cluster Universe, TUM, Boltzmannstr.~2, 85748~Garching, 
Germany}

\begin{document}

\maketitle

\bigskip

\begin{abstract}\noindent
We present a comprehensive numerical analysis of a four-dimensional model with 
the
Higgs as a composite pseudo-Nambu-Goldstone boson that features a calculable
Higgs potential and protective custodial and flavour symmetries to 
reduce electroweak fine-tuning. We employ a novel numerical technique 
that allows us for the
first time to study constraints from radiative electroweak symmetry breaking,
Higgs physics, electroweak precision tests, flavour physics, and direct
LHC bounds on fermion and vector boson resonances in a single framework.
We consider four different flavour symmetries in the composite sector, one of 
which we show to not be viable anymore in view of strong precision constraints. 
In the other cases, all constraints can be passed with a sub-percent 
electroweak fine-tuning.
The models can explain the excesses recently observed in $WW$, 
$WZ$, $Wh$ and $\ell^+\ell^-$ resonance searches by ATLAS and CMS and the 
anomalies in angular observables and branching ratios of rare semi-leptonic $B$ 
decays observed by LHCb. Solving the $B$ physics anomalies predicts the 
presence of a dijet or $t\bar t$ resonance around 1~TeV just below the 
sensitivity of LHC run 1.
We discuss the prospects to probe the models at run 2 of the LHC.
As a side product, we identify several gaps in the searches for vector-like 
quarks at hadron colliders,
that could be closed by reanalyzing existing LHC data.
\end{abstract}

\newpage
\setcounter{tocdepth}{2}
\tableofcontents

\section{Introduction}

The Standard Model Higgs boson faces a severe naturalness problem since the 
presence of heavy states associated to a more fundamental theory would lead to 
enormous corrections to its mass, requiring an extreme fine-tuning to explain 
the observed value. This conundrum can be solved if the Higgs boson is not an 
elementary scalar but a bound state of some new strong interaction. The 
lightness of the composite Higgs with respect to the -- as yet 
unobserved -- composite resonances finds a natural explanation if the Higgs is 
a 
pseudo-Nambu-Goldstone boson (pNGB) of an approximate global symmetry of the 
strong sector \cite{Kaplan:1983fs,Dugan:1984hq}.
To avoid the flavour problems of technicolour theories, the mechanism of 
partial 
compositeness can be invoked to generate the masses of the SM particles 
\cite{Kaplan:1991dc}.
This mechanism is closely related to the ``geometric'' generation of fermion
mass hierarchies from wave function overlaps in models with warped extra 
dimensions \cite{Grossman:1999ra,Gherghetta:2000qt,Huber:2000ie,Csaki:2003sh},
and in fact much of the progress in composite Higgs models in the last decades has 
been made using holographic models 
\cite{Contino:2003ve,Agashe:2004rs,Panico:2007qd}.
In these models, the Higgs potential becomes 
calculable, leading to additional  predictivity. Recently, also 
four-dimensional 
models have been constructed where the Higgs potential is calculable at 
one-loop 
level \cite{DeCurtis:2011yx,Panico:2011pw,Marzocca:2012zn}\footnote{See 
\cite{Bellazzini:2014yua} for an overview of the model landscape and 
\cite{Panico:2015jxa} for a comprehensive review of 4D pNGB Higgs models.}.
These models have the advantage that they are simpler in structure than 
the 5D theories, but more general, as they need not necessarily be the 
low-energy limit of a 5D holographic theory. Our aim in this paper is to study
one particular implementation of the 4D pNGB Higgs, taking into account
all relevant experimental constraints.

Direct and indirect constraints on composite pNGB Higgs models have already 
been 
discussed in the literature
(for recent analyses see e.g.\
\cite{Grojean:2013qca,Contino:2015mha}
for electroweak precision tests, 
\cite{Barbieri:2012tu,Konig:2014iqa,Azatov:2014lha}
for flavour physics,
\cite{Csaki:2008zd,Low:2010mr,Gillioz:2012se,Montull:2013mla,Carena:2014ria}
for Higgs physics, 
\cite{DeSimone:2012fs,Gillioz:2012se,Redi:2013eaa,
Li:2013xba,Azatov:2013hya,Delaunay:2013pwa,Reuter:2014iya,
Gripaios:2014pqa,Matsedonskyi:2014mna, Backovic:2015bca} 
for quark partner searches, and
\cite{Agashe:2009bb,Redi:2013eaa,Pappadopulo:2014qza,Lane:2014vca,Thamm:2015zwa,
Kaminska:2015ora}
for vector resonance searches). 
However, a complete and simultaneous numerical 
analysis of all relevant constraints on a particular model is still lacking. 
This is due in part to the fact that a mere parameter scan, as is typically 
done 
in supersymmetric extensions of the SM, is not feasible since the parameter 
space does not ``factorize'' into Standard Model (SM) and new physics (NP) 
parameters; due to partial 
compositeness, the masses of SM particles and the angles and phase of the
Cabibbo-Kobayashi-Maskawa (CKM) matrix are non-trivial 
functions of many model parameters. As a consequence, finding viable parameter 
points from a random set of model parameters becomes untractable. This problem 
becomes even more severe once the Higgs potential is taken into account, as the 
Higgs mass and VEV often arise from an interplay between gauge and fermion 
loops 
which again depends on many parameters. For these reasons, numerical analyses 
of 
composite Higgs models often have to rely on simplifying assumptions, 
e.g.\ only considering third generation fermions and their partners --
which does not allow including flavour constraints, for instance. Full 
numerical studies of indirect constraints have been performed in warped extra 
dimensional models
(without a pNGB Higgs)
by making use of approximate analytical expressions for the SM parameters
\cite{Casagrande:2008hr,Blanke:2008zb,Blanke:2008yr,Albrecht:2009xr,
Bauer:2009cf,Casagrande:2010si}, but this only works for particular 
representations
of the additional fermions\footnote{Namely if the left-handed elementary quark 
doublet
mixes with a single composite $SU(2)_L$ doublet, unlike in the model to be 
studied below.}.
We have overcome these problems by generalising a numerical method first 
proposed 
in \cite{Straub:2013zca} and with the help of a high-performance computing 
cluster.
This allows us for 
the first time to scrutinize one specific model taking into account all 
relevant 
experimental constraints and to identify novel correlations.

In selecting a model to analyze in detail, our focus has been to maximize
naturalness and predictivity, but to be as economic as possible concerning both
particle content and number of parameters. The model should 
thus fulfill the following requirements.
\begin{itemize}
\item The symmetry breaking coset should contain custodial symmetry to avoid 
excessive 
contributions to the $T$ parameter, but no extra Higgs states. This singles out 
$SO(5)/SO(4)$.
\item The $Zb_L\bar b_L$ coupling should be custodially protected from 
tree-level corrections.
This leaves two possible 
choices of quark partner representations under the $SO(4)$ symmetry
\cite{Agashe:2006at}. We choose the 
one where all quark partners can be embedded in two fundamental 
representations, 
as in the MCHM5 \cite{Contino:2006qr}.
\item The Higgs potential should be calculable. This can be achieved by imposing 
the 
Weinberg sum rules \cite{Marzocca:2012zn,Pomarol:2012qf}.
These are automatically fulfilled in deconstructed models 
like the 4DCHM \cite{DeCurtis:2011yx} or the DCHM \cite{Panico:2011pw}.
We choose the 4DCHM, because it features a 
finite one-loop effective potential already for two sites.\footnote{%
In the two-site DCHM, a logarithmic divergence spoils the predictivity for the 
Higgs VEV, but the Higgs mass can still be computed \cite{Matsedonskyi:2012ym}.}
\item The contribution to $\Delta F=2$ observables, i.e.\ meson-antimeson 
mixing,
should be suppressed compared to the naive 
anarchic expectation to avoid the $\epsilon_K$ problem 
\cite{Csaki:2008zd,Blanke:2008zb,Bauer:2009cf}.
Several mechanisms have been proposed to address this problem (apart from 
invoking
accidental cancellations).
We focus on the assumption that the composite sector is exactly invariant under 
a 
large flavour symmetry which is only broken minimally (i.e. by the amount 
required to reproduce CKM mixing) by the composite-elementary mixings.%
\footnote{For alternative mechanisms, see
\cite{Fitzpatrick:2007sa,Csaki:2008eh,Santiago:2008vq,Csaki:2009wc,
Delaunay:2010dw,Redi:2012uj,Matsedonskyi:2014iha,Cacciapaglia:2015dsa}%
.} This 
arguably corresponds to one of the strongest assumptions one can make on the 
flavour structure of partial compositeness, which is why we view it as a 
natural 
starting point in the search for a model that passes precision tests, but is 
natural in the electroweak sense.
\end{itemize}

In summary, we focus on the two-site 4DCHM with quark partners in two 
fundamentals of $SO(5)$, which we will call M4DCHM5 in the remainder of the 
paper. 
For the flavour structure of this model, we will consider the four different 
possibilities studied qualitatively already in \cite{Barbieri:2012tu}:
an effective $U(3)^3$ \cite{Cacciapaglia:2007fw,Redi:2011zi,Barbieri:2012uh}
or $U(2)^3$ \cite{Barbieri:2012uh,Barbieri:2012tu}
flavour symmetry with flavour-invariant
composite-elementary mixings either for left- or right-handed quarks in both 
cases,
dubbed left- or right-compositeness, respectively.

We stress that, while we aim to include as many experimental 
constraints as possible, our analysis is on a conceptually different level 
compared to analyses of weakly-coupled renormalizable extensions of the SM, e.g. 
the MSSM. 
This is because the models we are studying are non-renormalizable with a cutoff 
in the few-TeV region and contain a sector with strong couplings. Consequently, 
the models are not only less ambitious, but also less predictive since 
contributions from cutoff-scale operators or strong interaction effects could 
potentially spoil the picture obtained from naive computations in the two-site 
picture. Nevertheless, for the observables we are considering, the calculable 
effects often already lead to stringent experimental constraints and we find it 
unlikely 
that cutoff-scale physics comes to the rescue by cancelling these effects. It 
should however be kept in mind that many of the predictions are afflicted with
considerable theoretical uncertainties.

\section{Model setup}\label{sec:model}

In this section, we briefly review the M4DCHM and its Lagrangian. For details, 
the reader is referred to the original publication~\cite{DeCurtis:2011yx}. 
The relation to similar models is discussed in \cite{Marzocca:2012zn}.

\subsection{Bosonic part}
The M4DCHM can be understood as a deconstructed description of an 
extra-dimensional Gauge-Higgs-Unification model with a bulk gauge group $SO(5)$ 
that is broken down by boundary conditions on the branes to $SO(4)$ and the SM 
gauge group. To make the model phenomenologically viable, the symmetries are 
enlarged to include a bulk colour sector and an additional $U(1)_X$ to match 
the 
hypercharge assignments of the SM. So, from a 4D point of view, there is a 
strongly interacting composite sector subject to a global symmetry breaking 
pattern\footnote{The M4DCHM also contains an additional 
symmetry breaking $(SO(5)_L \times SO(5)_R) / SO(5)_{L+R}$ to account for the 
presence of heavy resonances.} $(SU(3)_c \times SO(5) \times U(1)_X) / (SU(3)_c 
\times SO(4) \times 
U(1)_X)$ and an elementary SM-like sector with gauge group 
$\elem{SU(3)} \times \elem{SU(2)} \times \elem{U(1)} $. 

In the two-site model one considers only one level of heavy resonances, thus 
the 
spectrum contains resonances $\rho_\mu^A$  for the $SO(5)$ as well as heavy 
gluons and a heavy $\rho_{X\mu}$. These resonances mix with their elementary 
counterparts such that the diagonal group becomes the remaining SM gauge group 
and hypercharge is given as\footnote{See appendix \ref{sec:SO5} for our 
convention for the $SO(5)$ generators.} 
\begin{equation}
 Y = \T{3_R} + X.
\end{equation}

The bosonic sector of the theory contains the gauge part as well as the sigma 
model describing the global symmetry breaking,
\begin{equation}
 \mathcal{L}_\mathrm{bosonic} = \mathcal{L}_\mathrm{gauge} + \mathcal{L}_\sigma.
\end{equation}
The gauge Lagrangian,
\begin{align} 	
 \mathcal{L}_\mathrm{gauge} &= -\frac{1}{4} \tr{\elem{G}_{\mu\nu} 
\elem{G}^{\mu\nu}} -\frac{1}{4} \tr{\elem{W}_{\mu\nu} \elem{W}^{\mu\nu}} 
-\frac{1}{4} \elem{B}_{\mu\nu} \elem{B}^{\mu\nu}  && \text{(elementary)}
\nonumber\\
                            &\quad\, - \frac{1}{4} \tr{{\rho_G}_{\mu \nu} 
{\rho_G}^{\mu\nu} } - \frac{1}{4} \tr{\rho_{\mu \nu} \rho^{\mu\nu} }   
-\frac{1}{4} \, \rho_{X\mu \nu} {\rho_X}^{\mu \nu} && \text{(composite)}
\nonumber\\
                            &\quad\, + \frac{f_G^2}{4} \, \left( 
\elemcoupl{g}_3 
\, \elem{G}_{\mu} - g_G \, {\rho_G}_\mu \right)^2 +  \frac{f_X^2}{4} 
\left( \elemcoupl{g}' \, \elem{B}_\mu - g_X \rho_{X\mu} \right)^2, && 
\text{(mixing)}
\label{eq:Lgauge}
\end{align}
contains the usual kinetic terms for the elementary $\elem{SU(3)} 
\times \elem{SU(2)} \times \elem{U(1)} $ gauge fields, where 
\begin{equation}
 \elem{W}_\mu = \elem{W}^{a_L}_\mu \, \T{a_L},
\end{equation}
 as well as kinetic terms for the gluon-, $SO(5)$- and 
$U(1)_X$-resonances. For the $\rho_\mu$ resonances it will be useful 
to 
group them into $SU(2)_L$, $SU(2)_R$ and coset components 
(following \cite{Marzocca:2012zn}, the 
latter we will call ``axial resonances'' in the following),
 \begin{equation}
  \rho_\mu = \rho^A_\mu \, \T{A} =  {\rho_L}^{a_L}_\mu \, \T{a_L} + 
{\rho_R}^{a_R}_\mu \, \T{a_R} + \axial_\mu^{\hat{a}}\, \T{\hat{a}}.
 \end{equation}
We also introduce explicit mixing terms between the $SU(3)$- and 
$U(1)$-resonances with their elementary counterparts, which are 
characterized by the scales $f_G$ and $f_X$.

The sigma model Lagrangian
\begin{equation}
  \mathcal{L}_\sigma = \frac{f_1^2}{4} \tr{\left( \mathcal{D}_\mu \Omega_1 
\right)^\dagger \left( \mathcal{D}^\mu \Omega_1 \right)}+ \frac{f_2^2}{2}\left[ 
\left( \mathcal{D}_\mu \Omega_2 \right)^\mathrm{t} \left( \mathcal{D}^\mu 
\Omega_2 \right)\right]_{55}
 \end{equation}
contains covariant derivatives acting on the sigma model fields $\Omega_1$ and 
$\Omega_2$, which are given as
\begin{align}
 \mathcal{D}_\mu \Omega_1 &= \partial_\mu \Omega_1 - \im \left( \elemcoupl{g} 
\, 
\elem{W}_{\mu} + \elemcoupl{g}' \elem{B}_\mu \, \T{3_R}\right) \Omega_1 + \im 
\,g_\rho \, \Omega_1 \, \rho_\mu, \\
 \mathcal{D}_\mu \Omega_2 &= \partial_\mu \Omega_2  - \im \, g_\rho \, \rho_\mu 
\, \Omega_2.
\end{align}
Note that the sigma model fields are uncharged under the global 
$U(1)_X$ symmetry (and, of course, they do not carry colour charges).

We adopt the so-called holographic gauge for the 
sigma model fields, which is inspired by a convenient gauge chosen in the 
corresponding 5D gauge theory,
 \begin{equation}
  \Omega_1(x) = \mathcal{U} = \exp\left[ \im \frac{\sqrt{2}}{f_1} \sigma_{\hat 
a}(x) \, \T{\hat a} \right], \qquad \Omega_2(x) = \mathds{1}_5.
 \end{equation}

 There is also the SM gauge freedom that has to be fixed. Here we adopt the SM 
unitary gauge, such that $\sigma_{\hat{a}}(x) = (0,0,0,h(x))$. In this gauge 
the 
Goldstone matrix takes the form
\begin{equation}
\mathcal{U} :=  \exp\left[ \im \frac{\sqrt{2}}{f_1} \sigma_{\hat a}(x) \, 
\T{\hat a} \right] = \left( \begin{array}{ccccc}

                    1&&&&\\

                    &1&&&\\

                    &&1&&\\

                    &&&\cos\left( \frac{h(x)}{f_1} \right)
&\sin\left( \frac{h(x)}{f_1} \right)\\
&&&-\sin\left( \frac{h(x)}{f_1} \right)
&\cos\left( \frac{h(x)}{f_1} \right)\\
\end{array}
 \right).
\end{equation}

Writing the Lagrangian as above in holographic gauge leads to a mixing term of 
the form
\begin{equation}
 \frac{1}{\sqrt{2}} \, g_\rho \, f_1 \,\, \axial_4^\mu \, \partial_\mu h.
\end{equation}
One can get rid of this term by a field redefinition,
\begin{align}
 \axial_4^\mu & \rightarrow  \axial_4^\mu - \frac{\sqrt{2}}{g_\rho} 
\frac{f}{f_2^2} \partial^\mu h \,, &
 h & \rightarrow  \frac{f_1}{f} h \,,
\end{align}
where $f$ is given by $f^{-2} := f_1^{-2} + f_2^{-2}$. By this transformation 
the mixing term vanishes and the composite Higgs kinetic term is canonically 
normalized. As a result, all dependencies on the 
Higgs field are given via
\begin{equation}
 s_h = \sin \left( \frac{h}{f} \right).
\end{equation}

\subsection{Fermionic part}
For the fermionic part of the model, we distinguish between the quark and the 
lepton part,
\begin{equation}
 \mathcal{L}_\mathrm{fermionic} = \mathcal{L}_\mathrm{quark} + 
\mathcal{L}_\mathrm{lepton}. 
\end{equation}
As in the boson sector, the quark Lagrangian contains elementary, composite and 
mixing parts,
\begin{align}
 \mathcal{L}_\mathrm{quark} &= \im \bar{\elem{q}}_L \slashed{\mathcal{D}} 
\elem{q}_L + \im \overline{\elem{u}}_R \slashed{\mathcal{D}} \elem{u}_R + \im 
\overline{\elem{d}}_R \slashed{\mathcal{D}} \elem{d}_R  
 & \text{(elementary)} \nonumber \\
 &\quad\, + \im \overline{\Psi}_\mathrm{comp} \slashed{\mathcal{D}} 
\Psi_\mathrm{comp} + \im \overline{\widetilde{\Psi}}_\mathrm{comp} 
\slashed{\mathcal{D}} \widetilde{\Psi}_\mathrm{comp}  & \text{(composite)}  
\nonumber\\
 &\quad\, - m_U \left( \overline{Q}_u Q_u + \overline{S}_u S_u  \right) - 
m_{\widetilde{U}} \left( \overline{\widetilde Q}_u \widetilde{Q}_u + 
\overline{\widetilde S}_u \widetilde{S}_u \right)  \nonumber \\
 &\quad\, - \left( m_{Y_U} + Y_U \right) \overline{S}_{u L} \widetilde{S}_{u R} 
- m_{Y_U} \overline{Q}_{u L} \widetilde{Q}_{u R} \,\, + \,\, \mathrm{h.c.}  
\nonumber \\
 &\quad\, +\Delta_{u_L} \, \overline{\xi}_{u L} \mathcal{U} \left( Q_{u R}+S_{u 
R} \right) + \Delta_{u_R} \, \overline{\xi}_{u R} \mathcal{U} \left( 
\widetilde{Q}_{u L} + \widetilde{S}_{u L} \right) \,\, + \,\, \mathrm{h.c.}  & 
\text{(mixing)} \nonumber \\
 &\quad\, +(u \leftrightarrow d) 
\label{eq:Lquark}
 \end{align}
Here we have two bidoublets, $Q$ and $\widetilde{Q}$, and two singlets, $S$ and 
$\widetilde{S}$, for every flavour. 
For the kinetic terms we used an $SO(5)$ notation where we combined the 
singlets and bidoublets into $SO(5)$ fundamentals: $\Psi_\mathrm{comp} 
= (Q, S)_{u,d}$, $\widetilde{\Psi}_\mathrm{comp} = (\widetilde{Q}, 
\widetilde{S})_{u,d}$. For these the covariant derivatives are then defined as
\begin{equation}
 \mathcal{D}_\mu \Psi_\mathrm{comp} = \left( \partial_\mu - \im g_G \, 
\rho_{G \mu} - \im g_\rho \, \rho_\mu - \im q_X \, g_X \, \rho_{X\mu} 
\right) 
\Psi_\mathrm{comp}
\end{equation}
and the same for $\widetilde{\Psi}_\mathrm{comp}$. The $U(1)_X$ charges are 
assigned to match the hypercharge of the SM. Thus, the fundamentals 
$\Psi_\mathrm{comp}^{(u)}$ and $\Psi_\mathrm{comp}^{(d)}$ have $q_X^{(u)} = 
\frac{2}{3}$ and  $q_X^{(d)} =-\frac{1}{3}$.

The elementary fields are embedded into (incomplete) $SO(5)$ fundamentals via
\begin{eqnarray}
 \xi_{u L} = \frac{1}{\sqrt{2}} \left( \begin{array}{c}
                                     d_L \\ -\im d_L \\ u_L \\ \im u_L \\ 0
                                    \end{array} \right) &,& 
 \xi_{u R} = \left( \begin{array}{c}
                                       0 \\ 0 \\ 0 \\ 0 \\ u_R
                                      \end{array}
 \right), \\
  \xi_{d L} = \frac{1}{\sqrt{2}} \left( \begin{array}{c}
                                     u_L \\ \im u_L \\ -d_L \\ \im d_L \\ 0
                                    \end{array} \right) &,& 
 \xi_{d R} = \left( \begin{array}{c}
                                       0 \\ 0 \\ 0 \\ 0 \\ d_R
                                      \end{array}
 \right)
\end{eqnarray}

Since we are mainly interested in the interplay between quark 
flavour measurements and the Higgs potential, we do not consider effects of 
partial lepton compositeness in this work. Indeed, if the compositeness of the 
left- and right-handed lepton chiralities are comparable, they are required to 
be small due to the leptons' lightness and their impact on the observables to 
be considered below is expected to be small. Moreover, flavour-changing 
interactions are strongly constrained by negative searches for charged lepton 
flavour violating processes. In practice, we simply consider elementary leptons 
with direct bilinear couplings to the Higgs field,
\begin{equation}
 \mathcal{L}_\mathrm{lepton} = \im \overline{l}_L \slashed{D} l_L + \im 
\overline{\ell}_R \slashed{D} \ell_R - \frac{m_\mathrm{SM}}{v} \, 
\overline{l}_L 
\cdot \left( \begin{array}{c} 0 \\ h \end{array} \right) \ell_R \,\, + \,\, 
\mathrm{h.c.},
\end{equation}
where, just as for the elementary quarks, the covariant derivatives are 
understood as 
couplings to the elementary $\elem{SU(3)} \times \elem{SU(2)} 
\times \elem{U(1)} $ gauge fields.
We note however that a significant degree of compositeness for some of the 
leptons could be motivated experimentally, e.g.\ to reconcile radiative 
electroweak symmetry breaking (EWSB) with naturalness in the absence of light 
top partners \cite{Carmona:2014iwa} or to explain the hints for violation of 
lepton flavour non-universality in $B$ decays \cite{Niehoff:2015bfa}. These 
effects are beyond the scope of our present analysis.

Let us note here that many of the above model parameters are 
correlated if they originate from an extradimensional gauge theory, e.g.\
coupling constants are generated by overlap integrals of Kaluza-Klein mode 
functions.
In our numerical analysis, we will not impose such relations but instead try to 
be as general as possible to explore the viability of purely 4D pNGB Higgs 
models, regardless of whether a dual 5D description exists.

\subsection{Flavour structure}

As noted in the introduction, we assume the strong sector to be invariant 
under a flavour symmetry, only to be broken by the composite-elementary 
mixings. We consider four possibilities (see \cite{Barbieri:2012tu} for a 
thorough comparison),
\begin{itemize}
\item In $U(3)^3_\text{LC}$ (LC for left-handed compositeness), the strong 
sector is invariant under a $U(3)$ symmetry\footnote{The cube in 
$U(3)^3$ refers to the fact that after the breaking, the SM quark sector is 
approximately invariant under a $U(3)_q\times U(3)_u \times U(3)_d$.} that is 
broken by the {\em right-handed} composite-elementary mixings.
\item In $U(3)^3_\text{RC}$ (RC for right-handed compositeness), the strong 
sector is invariant under a $U(3)\times U(3)$ which is broken by the 
{\em left-handed} composite-elementary mixings.
\item The $U(2)^3_\text{LC}$ and  $U(2)^3_\text{RC}$ models are analogous, but 
restricted to a smaller symmetry only acting on the first two generations of 
composite quarks.
\end{itemize}
In the $U(3)^3$ models, the matrices in the composite part of the fermion 
Lagrangian (the 3rd and 4th lines in \eqref{eq:Lquark}) are proportional to 
the identity, while in the $U(2)^3$ models, they are of the form 
$\text{diag}(a,a,b)$. The main difference is the form of the 
composite-elementary mixings. We give their explicit forms in 
appendix~\ref{sec:deltas}.

\subsection{Higgs potential}\label{sec:veff}

The explicit breaking of the global symmetries by the
mixings of the composite resonances with the elementary sector
generates an effective potential for the NGB field (to be identified with the 
SM Higgs) at the quantum level, so that it acquires a mass and a VEV, breaking 
electroweak symmetry.
In general, the effective potential is UV-sensitive and not necessarily 
calculable.
In the M4DCHM it is finite at one loop, making the model predictive under the 
assumption that higher loop contributions are subleading with respect to the 
(calculable) one-loop contribution. We will rely on this assumption in the 
following.

At one loop, the effective potential is given in terms of all the $n$-point 
correlation functions of the Higgs and therefore contains a gauge as well as a 
fermion contribution. It can be calculated by the Coleman-Weinberg 
formula \cite{Coleman:1973jx}
\begin{align}
 V_\mathrm{eff}(h) &= \sum \frac{c_i}{64 \pi^2} \left( 2 \, \tr{M_i^2(h)} \, 
\Lambda^2 - \tr{\left(M_i^2 (h)\right)^2} \log \left[ \Lambda^2 \right] \right. 
\nonumber \\
 & \qquad \qquad \qquad \qquad \left. + \,\, 
\tr{\left(M_i^2 (h)\right)^2 \log \left[ M_i^2(h) } \right] 
\right),
\end{align}
where $M_i^2(h)$ denote the Higgs-dependent 
mass-(mixing)-matrices\footnote{For fermions these are given by $M^2(h) = 
M(h)^\dagger M(h)$.} which we give in appendix \ref{app:massmatriecs} and
\begin{equation}
 c_i = \left\{ \begin{array}{rl} 3 & \text{for neutral gauge 
bosons} \\ 6 & \text{for charged gauge 
bosons} \\ -12 & \text{for (coloured) Dirac fermions} \\ \end{array} \right.  .
\nonumber
\end{equation}
Here we explicitly showed the dependence on the cutoff of the theory. For a 
non-renormalizable effective theory these UV dependent terms spoil the 
predictivity, such that one has to demand the following relations for 
ensuring the calculability of the Higgs potential:
\begin{align}
  &\tr{M_i^2(h)}- \tr{M_i^2(h=0)}=0, \\
  &\tr{(M_i^2 (h))^2} - \tr{(M_i^2 (h=0))^2} =0,
\end{align}
where it was taken into account that the constant term of the potential is not 
physical. These relations are just a reformulation of the Weinberg sum rules of 
the fermion and gauge sector that are usually imposed to guarantee a finite 
potential \cite{Marzocca:2012zn,Pomarol:2012qf}. For the quark sector, these 
relations represent a 
generalisation of the Weinberg sum rules to the three family case. 

In deconstructed models the Higgs potential is usually protected by the higher 
dimensional gauge symmetry, such that the Weinberg sum rules are automatically 
satisfied. This is the case for the M4DCHM as well
\cite{DeCurtis:2011yx,Marzocca:2012zn}. Note that in this case also the scale 
dependence cancels 
from the effective potential. Then the expression for the potential simplifies 
to
\begin{equation} \label{eq:Veff}
 V_\mathrm{eff}(h) = \sum \limits_{\text{all particles}} \frac{c_i}{64 \pi^2} 
m_i^4 (h) \, \log( m_i^2 (h) ),
\end{equation}
where $m_i (h)$ denote the masses in the mass basis, i.e. the singular values 
of the mass matrices.

\begin{figure}[tbp]
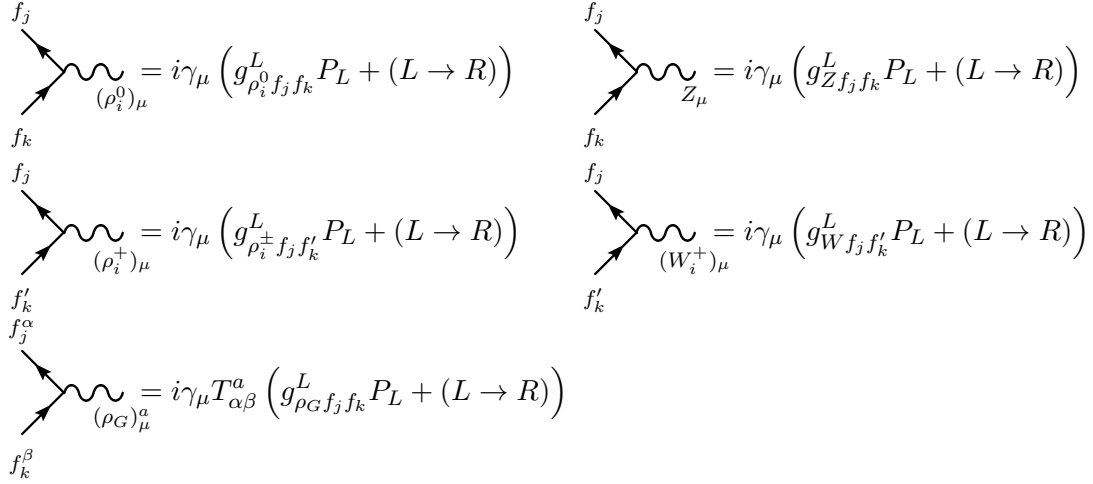

\begin{align*}
\Diagram{\vertexlabel^{f_j} \\
fdV \\
& g\vertexlabel_{(\rho^0_i)_\mu} \\
\vertexlabel_{f_k} fuA \\}~
&=i\gamma_\mu \left(g_{\rho^0_if_jf_k}^L P_L + (L\to R) \right)
&
\Diagram{\vertexlabel^{f_j} \\
fdV \\
& g\vertexlabel_{Z_\mu} \\
\vertexlabel_{f_k} fuA \\}~
&=i\gamma_\mu \left(g_{Zf_jf_k}^L P_L + (L\to R) \right)
\\[0.7cm]
\Diagram{\vertexlabel^{f_j} \\
fdV \\
& g\vertexlabel_{(\rho^+_i)_\mu} \\
\vertexlabel_{f'_k} fuA \\}~
&=i\gamma_\mu \left(g_{\rho^\pm_if_jf'_k}^L P_L + (L\to R) \right)
&
\Diagram{\vertexlabel^{f_j} \\
fdV \\
& g\vertexlabel_{(W^+_i)_\mu} \\
\vertexlabel_{f'_k} fuA \\}~
&=i\gamma_\mu \left(g_{Wf_jf'_k}^L P_L + (L\to R) \right)
\\[0.7cm]
\Diagram{\vertexlabel^{f_j^\alpha} \\
fdV \\
& g\vertexlabel_{(\rho_G)_\mu^a} \\
\vertexlabel_{f_k^\beta} fuA \\}~
&=i\gamma_\mu T^a_{\alpha\beta}\left(g_{\rho_Gf_jf_k}^L P_L + (L\to R) \right)
&
\end{align*}
\caption{Notation for Feynman rules used in this work.}
\label{fig:feyn}
\end{figure}

\section{Experimental constraints}\label{sec:exp}

In this section, we discuss all the experimental constraints that we impose in 
our analysis.
Since approximate analytical expressions for most of the observables have 
already been
provided elsewhere
(see in particular 
\cite{Barbieri:2012tu,Straub:2013zca,Konig:2014iqa,Matsedonskyi:2014iha}),
we focus on discussing the numerical computation and on
specifying our treatment of theoretical and experimental uncertainties.
Section~\ref{sec:sm} specifies how we compute the masses and couplings of the 
SM states, including among others the Higgs mass and VEV as well as the CKM 
matrix, section~\ref{sec:indirect} discusses indirect constraints, including 
electroweak
precision observables and flavour physics, while section~\ref{sec:direct} deals
with the direct bounds on fermion and vector resonances.

Since the numerical computation of the observables involves masses and 
couplings in the mass eigenstate basis (mb), we fix our notation for the 
couplings 
by specifying the Feynman rules in figure~\ref{fig:feyn}. In the gauge basis 
(gb) 
the Lagrangian contains non-diagonal mass matrices for vector 
bosons and fermions as well as interaction terms connecting both kinds of 
fields, which are (schematically) given as follows
\begin{equation}
 \mathcal{L} \supset \left[ M_\mathrm{g} \right]_{i j} A_{\mu \, i} A_j^\mu  - 
\left[ M_\psi \right]_{i j} \overline{\Psi}_i \Psi_j + \left[ g_\mathrm{gb} 
\right]_{i j k} \, \overline{\Psi}_i \gamma_\mu \Psi_j \, A_k^\mu
.
\end{equation}
After EWSB one can go to the mass basis by unitary transformations,
\begin{equation}
 \Psi_{L,R \,\, i}^\mathrm{(gb)} = \left[ V_\psi^{(L,R)} \right]_{i j} \, 
\Psi_{L,R \,\, i}^\mathrm{(mb)}, \qquad \qquad A_{\mu \, i}^\mathrm{(gb)} = 
\left[ V_g \right]_{i j} \, A_{\mu \, i}^\mathrm{(mb)},
\end{equation}
such that the couplings as defined in figure~\ref{fig:feyn} can be calculated as
\begin{equation}
  \left[ g^\mathrm{(mb)} \right]_{a b c} = [ V_\psi^\dagger ]_{a i} 
\, \left[ V_\psi \right]_{j b} \, \left[ V_g \right]_{k c} \, \left[ 
g^\mathrm{(gb)} \right]_{i j k}.
\end{equation}

\subsection{Standard model masses and couplings}\label{sec:sm}

\subsubsection{Higgs VEV and masses of SM states}

The tree-level masses for fermions and gauge bosons are obtained by 
diagonalizing the mass matrices given in appendix \ref{app:massmatriecs} after 
EWSB. Since the interaction terms of the Goldstone bosons generate mixing terms 
between elementary fields and the composite resonances, all masses of fields 
with SM-like quantum numbers will depend on the VEV taken by the pNGB, i.e. 
they depend on $s_h^* = \sin( \left< h \right> / f )$. In the model used in 
this work this quantity is not a free parameter, but it is calculable as the 
minimum of the loop-generated effective potential.

In practice, we calculate the $s_h$-dependent masses of all particles and use 
(\ref{eq:Veff}) to calculate the effective Higgs potential. Then, the correct 
value of $s_h^*$ is obtained by numerically minimizing the potential. We also 
explicitly demand that the found minimum is non-trivial, otherwise we discard 
the parameter point.

The value of $s_h^*$ is fixed in our numerical analysis by imposing the 
tree-level value of the Fermi constant in muon decay as a constraint,
\begin{equation}
G_\mu^\text{tree} = \frac{1}{\sqrt{2} v_\text{SM}^2}  =
\frac{1}{\sqrt{2} (s_h^*)^2 f^2 }
\,,
\end{equation}
which is valid up to negligible vector resonance exchange contributions.
Since we only include the tree-level $G_\mu$, we add a relative theoretical 
uncertainty of $1\%$. Loop corrections to $G_\mu$ are effectively included in 
terms of the $T$ parameter, see section~\ref{sec:ST}.

Once $s_h^*$ is known, the Higgs mass can be calculated as the curvature of 
the effective potential at its minimum,
\begin{equation}
 m_h^2 = \left. \partial_h^2 V_\mathrm{eff}(h) \right|_{h = \left< h \right>}  
=  \frac{1-s_h^2}{f^2} \left. \partial_{s_h}^2 V_\mathrm{eff}(s_h) \right|_{s_h 
= s_h^*}. 
\end{equation}

For the $W$, $Z$, and top masses, we directly interpret the masses obtained 
from diagonalizing the mass matrices as $\overline{\text{MS}}$ running masses 
at the scale $m_t$. We add a relative theory uncertainty of $5\%$ to account 
for this crude assumption. In principle, we could compute the one-loop 
matching corrections to the masses to get a more reliable estimate. In 
practice, this is not feasible because the composite-elementary mixing means 
that the numerical computation of a large self-energy matrix, e.g.\ 
$27\times27$ in the case of the top quark, would be necessary which quickly 
leads to excessive computing times.

For the light quark masses, we also interpret the tree-level masses as
$\overline{\text{MS}}$ running masses at $m_t$; then we use 
\texttt{RunDec} \cite{Chetyrkin:2000yt} to run them to the relevant scales 
where they can be compared to the PDG averages \cite{Agashe:2014kda}.

\subsubsection{CKM matrix}\label{sec:ckm}

In the SM, CKM elements are determined from a global fit to weak decays 
mediated by tree-level $W$ exchange as well as loop-induced meson-antimeson 
mixing observables. In the presence of NP, the latter are susceptible to NP 
contributions as will be discussed below in section~\ref{sec:df2}. But even the 
tree-level processes receive corrections in a composite Higgs framework that 
lead to relevant constraints. The reason is that the $3\times3$ quark mixing 
matrix is no longer unitary in the presence of composite-elementary mixing, but 
becomes part of a larger ($27\times27$) mixing matrix among quarks and quark 
partners. Deviations from CKM unitarity, predicted by the SM, can thus be used 
to constrain quark compositeness.

To compare to the absolute values of CKM elements measured in experiments,
one can define effective CKM elements from ratios of $W$ couplings,
\begin{equation}
|V_{ij}| = \frac{|g_{Wu_id_j}^L|}{|g_{W\ell\nu}^L|} \,.
\end{equation}
The $|V_{ij}|$ obtained in this way can be directly compared to the elements
extracted in experiments assuming the SM as long as right-handed $W$ couplings 
and contributions from tree-level heavy resonance exchange can be neglected. We 
do take these two effects into account in our numerics, although they turn out 
to be negligible.

In our numerical analysis, we include five CKM elements that are directly 
measured in tree-level processes.
\begin{itemize}
\item $|V_{ud}|$ from superallowed nuclear beta decays,
\item $|V_{us}|$ from $K\to \pi\ell\nu$ decays,
\item $|V_{ub}|$ from inclusive $B\to X_u\ell\nu$ and exclusive $B\to 
\pi\ell\nu$ decays,
\item $|V_{cb}|$ from inclusive $B\to X_c\ell\nu$ and exclusive $B\to 
D^*\ell\nu$ decays,
\item $|V_{tb}|$ from the cross-section of $t$-channel single top production at 
LHC.
\end{itemize}
The measured values and references are given in table~\ref{tab:exp}. In the 
case of $|V_{ub}|$ and $|V_{cb}|$, there are long standing discrepancies 
between the determinations from inclusive vs.\ exclusive $B$ decays. Since 
these tensions cannot be resolved in our model, we use the PDG prescription 
\cite{Agashe:2014kda} to 
rescale the discrepant measurements. We multiply the uncertainties of 
$|V_{ub}|$ by a factor of $1.9$ compared to the ones given in 
table~\ref{tab:exp}, and a factor of $2.9$ in the case of $|V_{cb}|$.

$|V_{ud}|$ and $|V_{us}|$ are important because in the SM, they are constrained 
by the unitarity condition on the first row of the CKM matrix, 
\begin{equation}
1 = |V_{ud}|^2+|V_{us}|^2+|V_{ub}|^2\approx |V_{ud}|^2+|V_{us}|^2 \,,
\end{equation}
where $|V_{ub}|$ is numerically negligible. The 
smallness of $|V_{ub}|$ and $|V_{cb}|$ is also why, in the SM, 
$|V_{tb}|\approx1$ holds up to a permille level correction. Partial 
compositeness can lead to a deviation from both relations (see e.g.\ 
\cite{Redi:2011zi,Barbieri:2012tu,Grojean:2013qca}).

Finally, we also include the CKM angle $\gamma$ that is measured via the 
interference of $b\to c\bar u s$ and $b\to u\bar c s$ amplitudes in $B\to DK$ 
decays. Again in the case where right-handed $W$ couplings and direct vector 
resonance contributions can be neglected, $\gamma$ can be computed from the 
tree-level $W$ couplings as
\begin{equation}
\gamma = \arg\left(
-\frac{g_{Wud}^L\,g_{Wub}^{L*}}{g_{Wcd}^L\,g_{Wcb}^{L*}}
\right)
 \,.
\end{equation}
This expression is independent of phase conventions. For the experimental 
value in tab.~\ref{tab:exp}, we symmetrize the value obtained by the CKMfitter 
collaboration from a fit to all experiments.

\subsection{Indirect constraints}\label{sec:indirect}

\begin{table}[tbp]
\centering
\renewcommand{\arraystretch}{1.15}
\begin{tabular}{llllll}
\hline
$R_b$ & $0.21629(66)$ & \cite{ALEPH:2005ab}
& $|V_{ud}|$ & $0.97417(21)$ & \cite{Hardy:2014qxa} \\
$R_c$ & $0.1721(30)$ & \cite{ALEPH:2005ab} 
& $|V_{us}|$ & $0.2249(8)$ & \cite{Aoki:2013ldr} \\
$R_h$ & $20.804(50)$ & \cite{ALEPH:2005ab}
& $|V_{ub}|_\text{ex}$ & $(3.72 \pm 0.16)\times10^{-3}$ & 
\cite{Lattice:2015tia} \\
$\mu^{gg}_{WW}$ & $0.86\pm0.17$ & \cite{Khachatryan:2014jba,ATLAS:2015bea}
& $|V_{ub}|_\text{in}$ & $(4.33 \pm 0.28)\times10^{-3}$ & \cite{Lees:2011fv} \\
$\mu^{gg}_{ZZ}$ & $1.18\pm0.39$ & \cite{Khachatryan:2014jba,ATLAS:2015bea}
& $|V_{cb}|_\text{ex}$ & $(3.904 \pm 0.075)\times10^{-2}$ & 
\cite{Bailey:2014tva} \\
$\mu^{gg}_{gg}$ & $1.12\pm0.22$ & \cite{Khachatryan:2014jba,ATLAS:2015bea}
& $|V_{cb}|_\text{in}$ & $(4.221\pm 0.078)\times10^{-2}$ & 
\cite{Alberti:2014yda} \\
$\mu^{gg}_{\tau^+\tau^-}$ & $0.97\pm0.39$ & 
\cite{Khachatryan:2014jba,ATLAS:2015bea}
& $|V_{tb}|$ & $0.998 \pm 0.041$ & \cite{Khachatryan:2014iya} \\
$\Delta M_K$ & $3.483(6)\times10^{-15}\,\text{GeV}$ & 
\cite{Agashe:2014kda} &
$|\epsilon_K|$ & $2.228(11)\times10^{-3}$ & \cite{Agashe:2014kda} \\
$\Delta M_d$ & $0.510(3)\,\text{ps}^{-1}$ & \cite{Amhis:2014hma} &
$S_{\psi K_S}$ & $0.682(19)$ & \cite{Amhis:2014hma} \\
$\Delta M_s$ & $17.761(22)\,\text{ps}^{-1}$ & \cite{Amhis:2014hma} &
$\phi_s$ & $-0.010\pm0.039$ & \cite{Aaij:2014zsa} \\
$S$ & $0.05\pm0.11$ & \cite{Baak:2014ora} &
$\gamma$ & $(72.9\pm6.7)^\circ$ & \cite{Charles:2015gya} \\
$T$ & $0.09\pm0.13$ & \cite{Baak:2014ora} &
\\
\hline
\end{tabular}
\caption{Values of the experimental constraints used in the numerical analysis. 
For details and the treatment of theoretical uncertainties, see main text.}
\label{tab:exp}
\end{table}

\subsubsection{\texorpdfstring{$S$}{S} and \texorpdfstring{$T$}{T} parameters}
\label{sec:ST}

By construction, the $T$ parameter does not receive a contribution at tree 
level in pNGB models based on the $SO(5)/SO(4)$ coset. At one loop, the 
dominant 
contribution typically comes from fermion loops involving, in particular, the 
top partners. In addition, the modification of the gauge 
boson couplings to the 
Higgs and the electroweak would-be Goldstone bosons leads to an 
``infrared-log'' 
contribution \cite{Barbieri:2007bh,Orgogozo:2012ct}.
Finally, also loops involving the heavy spin-1 resonances can contribute
(see \cite{Contino:2015mha} for a recent discussion).
For simplicity, in our analysis we restrict ourselves to the fermion 
contribution, which is finite and gauge-independent. It can be computed 
numerically as
\begin{equation}
\alpha_\text{em} T  = \frac{\Pi^T_{WW}}{m_W^2}-\frac{\Pi^T_{ZZ}}{m_Z^2}
\end{equation}
where the masses are tree-level masses, and
\begin{multline}
-16\pi^2\Pi^T_{VV} = \sum_{f_i,f_j} H\!\left(m_{f_i}^2,m_{f_j}^2\right)
\left(|g_{Vf_if_j}^L|^2+|g_{Vf_if_j}^R|^2\right)
\\
+4 m_{f_i}m_{f_j} B_0\!\left(m_{f_i}^2,m_{f_j}^2\right)
\text{Re}\!\left(g_{Vf_if_j}^{L*}g_{Vf_if_j}^R\right)
\end{multline}
is the fermion contribution to the transverse part of the vacuum polarization. 
The sum runs over all SM fermions and quark resonances. The Passarino-Veltman 
function is defined as in \cite{Hahn:1998yk} and the function $H$ can be found 
e.g.\ in \cite{Pierce:1996zz}.

In contrast to $T$, the $S$ parameter arises already at tree level, effectively 
leading 
to a lower bound on the mass scale of the spin-1 resonances. In models where 
$T=T_\text{SM}$ at tree level, the NP contribution to $S$ can be obtained 
numerically as
\begin{equation}
\left. \alpha_\text{em} S \,\right|_{T=0} = \frac{1}{4} 
\left(s_W^2-\sin^2\theta_\text{eff}\right) \,,
\end{equation}
where $s_W^2 = 1-m_W^2/m_Z^2$ and the effective weak mixing angle is defined 
via 
the leptonic forward-backward asymmetry,
\begin{align}
x &= \frac{g_{Zee}^R+g_{Zee}^L}{g_{Zee}^R-g_{Zee}^L}
\,,
&
\sin^2\theta_\text{eff} &= \frac{1+x}{4}
\,.
\end{align}

Experimentally, a recent global fit of electroweak precision data finds 
\cite{Baak:2014ora}
\begin{align}
S&=0.05\pm0.11
\,,
&
T&=0.09\pm0.13
\,,
\end{align}
with a correlation coefficient of $+0.9$. Since we neglect gauge contributions 
to $T$ and all loop contributions to $S$, in our numerical analysis we further 
assume uncorrelated theory uncertainties of $0.05$ for $S$ and $0.10$ for $T$, 
which we combine with the correlated experimental uncertainties. The size of 
these theory uncertainties is chosen to encompass the typical size of the 
neglected ``IR-log'' contributions to $S$ and $T$.

\subsubsection{\texorpdfstring{$Z$}{Z} decays}\label{sec:z}

Due to the large degree of compositeness required for the left-handed top quark 
(and thus also $b$ quark), the partial width of the $Z$ into $b$ quarks 
measured 
at LEP provides a powerful constraint on models with partial compositeness. 
While our model features a custodial protection of this coupling, the 
observable 
is still important to constrain the subleading composite-elementary mixing of 
the $b_L$. In the flavour-symmetric models, also the partial widths into 
lighter 
quarks lead to constraints. We include the following observables in our 
analysis,
\begin{align}
 R_b &= \frac{\Gamma(Z\to b\bar b)}{\Gamma(Z\to q\bar q)}
 \,,&
 R_c &= \frac{\Gamma(Z\to c\bar c)}{\Gamma(Z\to q\bar q)}
 \,,&
 R_h &= \frac{\Gamma(Z\to q\bar q)}{\Gamma(Z\to \ell\bar \ell)}
 \,,&
\label{eq:RZ}
\end{align}
where $\Gamma(Z\to q\bar q)$ implies a sum over all quarks but the top.
We compute only the tree-level corrections at zero momentum to these 
observables (see \cite{Grojean:2013qca} for a discussion of effects beyond 
this limit). We add the higher-order SM contributions (see 
\cite{Freitas:2014hra}) numerically to reproduce 
the correct SM predictions in the absence of NP contributions. The experimental
measurements are listed in table~\ref{tab:exp}.

A comment is in order on the
loop corrections to $Z\to\bar b_Lb_L$, which we have not taken into account.
Although corrections are already generated at tree level in the M4DCHM5, these
are suppressed by the custodial protection mechanism, which however is not 
active at loop level.
In \cite{Carena:2007ua,Barbieri:2007bh,Anastasiou:2009rv} it was
shown that in similar models as the ones we are studying,
there is a correlation between fermionic loop corrections to the $T$ parameter
and the loop correction to $Z\to\bar b_Lb_L$. For a heavy new physics scale,
this can be understood as being due to renormalization-group mixing of 
dimension-6 operators invariant under the SM gauge symmetries. Considering the
operators (in the notation of \cite{Grzadkowski:2010es} and in the basis where the down-type quark mass
matrix is diagonal)
\begin{align}
Q_{\phi q}^{(1)} &= (\phi^\dagger i \overleftrightarrow{D}_\mu \phi) (\bar q_3 \gamma^\mu q_3) \,,
&
Q_{\phi q}^{(3)} &= (\phi^\dagger i \overleftrightarrow{D}_\mu^a \phi) (\bar q_3 \tau^a\gamma^\mu q_3) \,,
\end{align}
the correction to the left-handed $Z$ coupling to bottom quarks $\delta g_{Zbb}^L$ satisfies
\begin{equation}
 \delta g_{Zbb}^L \propto \left( C_{\phi q}^{(3)} + C_{\phi q}^{(1)} \right).
\end{equation}
Custodial protection implies $C_{\phi q}^{(1)}\approx-C_{\phi q}^{(3)}$
up to subleading mixing effects, implying a vanishing
correction $\delta g_{Zbb}^L$ at the matching scale.
Since the two Wilson coefficients run differently
\cite{Jenkins:2013wua,Alonso:2013hga,Brod:2014hsa}, a non-zero correction is induced at the
electroweak scale
which is proportional to the matching scale value of $C_{\phi q}^{(3)}$.\footnote{
The Wilson coefficient $C_{\phi u}$ associated with the Operator
$Q_{\phi u} = (\phi^\dagger i \overleftrightarrow{D}_\mu \phi) (\bar u_3 \gamma^\mu u_3)$
that also enters the RGE induced contributions to $\delta g_{Zbb}^L$ vanishes due to custodial protection.}
The quantum corrections leading to this running induce at the same time
a non-zero $T$ parameter which is also proportional to the matching scale value of $C_{\phi q}^{(3)}$
and thus correlated to $\delta g_{Zbb}^L$.
For a positive contribution to the $T$ parameter, the sign of this correlation 
leads to a negative contribution to $R_b$ that is disfavoured by 
experiment \cite{Carena:2007ua,Barbieri:2007bh,Anastasiou:2009rv}.
Thus, parameter points with a large positive contribution to the $T$ parameter
might be excluded by taking into account the one-loop corrections to $\delta 
g_{Zbb}^L$.
A challenge of taking this loop contribution into account is that it involves
Passarino-Veltman functions at non-zero external momentum with three 
propagators. Due to
the large number of states in the M4DCHM5, this would significantly increase 
computing time, so
we are not able to take this effect into account.
It should thus be kept in mind that our results might be optimistic in the 
sense 
that we might keep points that are possibly excluded.
A dedicated analysis of the impact of higher order corrections to $\delta 
g_{Zbb}^L$ would be worthwhile.
We also note that the tension between the constraints on the $T$ parameter and 
$\delta g_{Zbb}^L$ 
might be relaxed by including an additional level of resonances 
\cite{Anastasiou:2009rv}.

\subsubsection{Higgs production and decay}\label{sec:higgs}

We compute the modification of the Higgs partial widths $r_X=\Gamma(h\to 
X)/\Gamma(h\to X)_\text{SM}$ at tree level for $X=WW$, $ZZ$, $b\bar b$, and 
$\tau^+\tau^-$, and at one-loop level for $X=gg$ and $\gamma\gamma$. We take 
into account the loop contributions from all SM and heavy fermions and vector 
bosons. The signal strength in a particular final state, assuming pure gluon 
fusion production, can then be obtained as
\begin{equation}
 \mu^{gg}_X = \frac{r_X \,r_{gg}}{r_\text{tot}} \,,
\end{equation}
where $r_\text{tot}=\Gamma_h/\Gamma_h^\text{SM}$ is the modification of the 
total width.

We use ATLAS and CMS measurements to constrain the signal strength. In the case 
of ATLAS, the
gluon fusion result is given explicitly. In the case of CMS, we use the ``0/1 
jet'' result for $WW$, $ZZ$, and $\tau^+\tau^-$, and the ``untagged'' result 
for 
$\gamma\gamma$. We naively combine the ATLAS and CMS results for each final 
state, using the PDG prescription to enlarge the error in the case of poor 
agreement. The resulting constraints are listed in table~\ref{tab:exp}.
We neglect the correlations between individual measurements. Since the $h\to 
b\bar b$ signal strength is only measured in the case of vector boson 
associated 
production, we do not include it in our numerical analysis.

\subsubsection{Meson-antimeson mixing} \label{sec:df2}

The meson-antimeson mixing amplitude for the neutral meson $M^0$ ($=K^0$, 
$B_s$, $B_d$, or $D^0$) can be written as
\begin{equation}
M_{12}^M = \frac{1}{2m_M}\langle \bar M^0 |\mathcal H^{\Delta F=2} |M^0\rangle
= \left(M_{12}^M\right)_\text{SM} +
\sum_a C_a^{q_iq_j}(\mu_l) \langle \bar M^0 | Q_a^{q_iq_j}(\mu_l)|M^0\rangle
\label{eq:M12}
\end{equation}
with $q=u$ or $d$.
The loop-induced SM contribution is discussed e.g.\ in \cite{Buras:1998raa}. 
The sum 
contains NP contributions due to tree-level vector resonance exchange that are 
encoded in the Wilson coefficients of the following $\Delta F=2$ operators.
\begin{align}
Q_{VLL}^{q_iq_j} &= (\bar q^i_L\gamma^\mu q^j_L)(\bar q^i_L\gamma^\mu q^j_L) \,,
&
Q_{VRR}^{q_iq_j} &= (\bar q^i_R\gamma^\mu q^j_R)(\bar q^i_R\gamma^\mu q^j_R) \,,
\label{eq:df2op1}
\\
Q_{VLR}^{q_iq_j} &= (\bar q^i_L\gamma^\mu q^j_L)(\bar q^i_R\gamma^\mu q^j_R) \,,
&
Q_{SLR}^{q_iq_j} &= (\bar q^i_R q^j_L)(\bar q^i_L q^j_R) \,,
\label{eq:df2op2}
\end{align}
that can be written in terms of the Feynman rules defined in 
figure~\ref{fig:feyn} as
\begin{align}
C_{VLL}^{q_iq_j} &=
-\frac{1}{2}
\sum_i
\left(\frac{g^L_{\rho^0_i q_j q_k}}{m_{\rho^0_i}}\right)^2
-
\frac{1}{6}
\left(\frac{g^L_{\rho_G q_j q_k}}{m_{\rho_G}}\right)^2
\,,
\label{eq:cvll}
\\
C_{VRR}^{q_iq_j} &=\left. C_{VLL}^{q_iq_j} \right|_{L\to R}
\,,\\
C_{VLR}^{q_iq_j} &=-
\sum_i
\frac{g^L_{\rho^0_i q_j q_k}g^R_{\rho^0_i q_j q_k}}{m_{\rho^0_i}^2}
+
\frac{1}{6}
\frac{g^L_{\rho_G q_j q_k}g^R_{\rho_G q_j q_k}}{m_{\rho_G}^2}
\,,\\
C_{SLR}^{q_iq_j} &=
\frac{g^L_{\rho_G q_j q_k}g^R_{\rho_G q_j q_k}}{m_{\rho_G}^2}
\,.
\end{align}
These expressions are valid at the matching scale of NP and SM, while 
\eqref{eq:M12} depends on the values at the hadronic scale $\mu_l$ that is 
chosen conventionally as $m_b$ for $B_{d,s}$ mixing, 3~GeV for $D^0$ mixing, 
and 2~GeV for $K^0$ mixing. In principle, the correct matching scale is set by 
the mass scale of the heavy resonances. However, in our numerical scan, we 
often encounter vastly different scales for the different resonances. 
Consequently, we have decided to simply match the tree-level Wilson 
coefficients to the SM at the scale $m_t$ and to neglect the RG evolution above 
$m_t$. A more complete treatment, including intermediate thresholds, is beyond 
the scope of our present work. We note that the RG evolution typically makes 
the NP effects larger at low scales. In that sense, our treatment leads to more 
conservative bounds. All relevant anomalous dimensions for the evolution below 
$m_t$ can be found in \cite{Buras:2001ra}.

The matrix elements in \eqref{eq:M12} depend on meson decay constants and bag 
parameters, both of which can be determined from lattice QCD. They can be 
written as
\begin{equation}
\langle \bar M^0 | Q_a^{q_iq_j}(\mu_l)|M^0\rangle =
m_M f_M^2 \mathcal B_a^M(\mu_l) \,,
\end{equation}
where
\begin{align}
 \mathcal B_{VLL}^M &= \mathcal B_{VRR}^M = \frac{1}{3}B_1^M(\mu_l) \,,
 \\
 \mathcal B_{VLR}^M &= -\frac{1}{6} \left(\frac{m_M}{m_{q_i}+m_{q_j}}\right)^2 
B_5^M(\mu_l)\,,
 &
 \mathcal B_{SLR}^M &= \frac{1}{4} \left(\frac{m_M}{m_{q_i}+m_{q_j}}\right)^2 
B_4^M(\mu_l)\,.
\end{align}
For the lattice predictions of the decay constants as well as the bag 
parameters $B_i$ for $B_d$ and $B_s$ mixing, we use ref.~\cite{Aoki:2013ldr}, 
for the kaon bag parameters we use ref.~\cite{Bertone:2012cu}, and for the 
charm bag parameters ref.~\cite{Carrasco:2014uya}.

We use the following observables sensitive to NP in the meson-antimeson mixing 
amplitude.
\begin{itemize}
\item The mass differences in the $B_d$ and $B_s$ systems,
\begin{equation}
\Delta M_{d,s}=2|M_{12}^{B_{d,s}}| \,.
\end{equation}
For the theoretical uncertainties, which are dominated by lattice 
uncertainties, we take $10.2\%$ relative uncertainty for $\Delta M_d$ and 
$7.6\%$ for $\Delta M_s$. Note that we do not have to account for 
uncertainties due to CKM elements as these are allowed to vary in our scan. We 
further take these lattice uncertainties to be correlated with a coefficient of 
$0.17$, since the ratio of the relevant lattice parameters is known more 
precisely than for the individual systems.
\item The mixing-induced CP asymmetry in $B_d\to J/\psi \, K_S$, 
\begin{equation}
S_{\psi K_S} = \sin\left(\arg\!\left(M_{12}^{B_d}\right)\right) \,,
\end{equation}
which in the SM measures $\sin2\beta$. We add a theory uncertainty of $0.01$ to 
account for possible penguin pollution \cite{Frings:2015eva}.
\item The sine of the $B_s$ mixing phase as obtained from an average of the 
mixing-induced CP asymmetries in $B_s\to J/\psi \, K^+K^-$ and $B_s\to 
J/\psi \, \pi^+\pi^-$ decays\footnote{%
Here we have used  the average performed by the LHCb collaboration. Very 
recently, a measurement with comparable precision has been presented by CMS 
\cite{Khachatryan:2015nza}. The observable has also been measured by ATLAS  
\cite{Aad:2014cqa}.
},
\begin{equation}
\sin\phi_s = \sin\left(\arg\!\left(M_{12}^{B_s}\right)\right) \,.
\end{equation}
In this case, we add a theory uncertainty due to penguin pollution of $0.017$
\cite{Frings:2015eva}.
\item The parameter for indirect CP violation in $K^0$ 
mixing,
\begin{equation}
|\epsilon_K| = \kappa_\epsilon 
\frac{\text{Im}\!\left(M_{12}^{K^0}\right)}
{\sqrt{2}\Delta M_K}
\end{equation}
where the experimental value for $\Delta M_K$ can be used. For the (non-CKM) 
theory uncertainty on $|\epsilon_K|$, we take a relative uncertainty of 11\%.
\item The mass difference in $K^0$ mixing,
\begin{equation}
\Delta M_K = 2 \text{Re}\!\left(M_{12}^{K^0}\right) \,.
\end{equation}
The SM contribution to $\Delta M_K$ is plagued by large uncertainties due to 
long-distance contributions. 
Although first results are available from lattice calculations 
\cite{Bai:2014cva}, these are still for unphysical kinematics.
Thus we conservatively allow the NP contribution to saturate the 
experimental central value at $1\sigma$ (i.e.\ at $3\sigma$, we allow points 
where the NP contribution is three times the experimental central value, 
implying a necessary cancellation with the SM contribution).
\end{itemize}
We do not impose $D^0$ mixing observables as constraints, as they are expected 
to receive small NP contributions in models with minimally broken $U(2)^3$ 
\cite{Barbieri:2012uh}, but we will discuss predictions for them in 
section~\ref{sec:num} below.

\subsubsection{Rare $B$ decays}\label{sec:df1}

The $b\to s\gamma$ transition arises first at the one-loop level; approximate 
analytical results as well as generic formulae that can be used in a numerical 
analysis have been presented in \cite{Konig:2014iqa}. We include the constraint 
from the $B\to X_s\gamma$ branching ratio, that agrees well between SM 
prediction \cite{Misiak:2015xwa} and experimental world average 
\cite{Amhis:2014hma},
\begin{align}
\text{BR}(B\to X_s\gamma)_\text{SM} &= (3.36\pm0.23)\times 10^{-4} \,,
\\
\text{BR}(B\to X_s\gamma)_\text{WA} &= (3.43\pm0.22)\times 10^{-4} \,.
\end{align}
For the NP contribution, we use the following formula (cf.~\cite{Buras:2011zb}),
\begin{equation}
\frac{\text{BR}(B\to X_s\gamma)}{\text{BR}(B\to X_s\gamma)_\text{SM}} =
\frac{1}{\left(|C_7^\text{eff}(m_b)^\text{SM}|^2+N_\gamma\right)}\left(
|C_7^\text{eff}(m_b)|^2+|C_7'(m_b)|^2 + N_\gamma)
\right) \,,
\end{equation}
where $N_\gamma=3.6\times10^{-3}$.

The imaginary part of the Wilson coefficients and the relative size of the 
left- 
and right-handed Wilson coefficients can be constrained by other processes, 
most 
notably $B\to K^*\mu^+\mu^-$ angular observables. We do not impose these 
additional observables as constraints, but will discuss predictions for them in 
section~\ref{sec:num}.

NP contributions to semi-leptonic FCNC decays of $B$ and $K$ mesons arise 
already at tree level, mediated by the $Z$ boson that can obtain 
flavour-changing couplings to quarks at tree level as well as by heavy neutral 
vector resonances (for a thorough discussion of these effects in composite 
Higgs 
models, see \cite{Straub:2013zca}. Similar effects are obtained in models with 
a 
warped extra dimension \cite{Blanke:2008yr,Bauer:2009cf}). Writing the 
four-fermion operators as
\begin{align}
Q_{VAB}^{d_id_j\ell\ell} &= (\bar d^j_A\gamma^\mu d^i_A)(\bar \ell_B\gamma^\mu 
\ell_B) \,,
\end{align}
with $A,B=L,R$, the Wilson coefficients are obtained in analogy with 
section~\ref{sec:df2} as
\begin{align}
C_{VAB}^{d_id_j\ell\ell} &=
- \frac{g^A_{Z d_j d_k}g^B_{Z \ell\ell}}{m_{Z}^2}
-\sum_i
\frac{g^A_{\rho^0_i d_j d_k}g^B_{\rho^0_i \ell\ell}}{m_{\rho^0_i}^2}
\,.
\label{eq:wcsl}
\end{align}
Here we have explicitly included the $Z$ contribution as it contributes 
formally at the same level in $v/f$ as the heavy resonance exchanges. The 
smallness of the flavour-changing coupling (which only arises after EWSB and 
is of order $v^2/f^2$) is compensated by the absence of the 
suppression by the heavy resonance mass in the propagator.

One can map the coefficients \eqref{eq:wcsl} onto the traditional operator 
basis for $d_i\to 
d_j\ell^+\ell^-$ transitions as
\begin{align}
C_9^{d_id_j} &= 
\Lambda_{ij}^2\left(C_{VLR}^{d_id_j\ell\ell}+C_{VLL}^{d_id_j\ell\ell}\right) \,,
&
C_{10}^{d_id_j} &= 
\Lambda_{ij}^2\left(C_{VLR}^{d_id_j\ell\ell}-C_{VLL}^{d_id_j\ell\ell}\right) \,,
\\
C_9^{\prime \, d_id_j} &= 
\Lambda_{ij}^2\left(C_{VRR}^{d_id_j\ell\ell}+C_{VRL}^{d_id_j\ell\ell}\right) \,,
&
C_{10}^{\prime \, d_id_j} &= 
\Lambda_{ij}^2\left(C_{VRR}^{d_id_j\ell\ell}-C_{VRL}^{d_id_j\ell\ell}\right) \,,
\end{align}
where
\begin{equation}
\Lambda_{ij}^2 = \frac{\pi}{\sqrt{2}G_F\alpha_\text{em}V_{ti}V_{tj}^*} \,.
\end{equation}
The primed coefficients are only generated at a very suppressed level in the 
flavour-symmetric models we consider.
Since we are assuming leptons to be elementary, all Wilson coefficients are 
lepton flavour universal. Relaxing this assumption, the recent hint for lepton 
flavour non-universality can potentially be explained as well
\cite{Niehoff:2015bfa}, but we 
are not considering this possibility here.
Since the lepton-$Z$ couplings are SM-like to an excellent precision in our 
setup, the $Z$-mediated contributions fulfill the well-known relation
$C_9=(4s_w^2-1)C_{10}$, i.e.\ they mostly contribute to $C_{10}$.

Concerning the resonance-mediated contributions, as mentioned above, they 
formally contribute at the same order in $v/f$ as the $Z$ contributions. Their 
couplings to elementary leptons however only arise through mixing of the 
composite and elementary vectors, so the resonance-mediated contributions are 
expected to be parametrically suppressed compared to the $Z$-mediated ones by a 
factor $g_\text{el}^2/g_\text{co}^2$, where $g_\text{el,co}$ are generic 
elementary and composite gauge couplings. Still, there are parts of parameter 
space where these contributions can be relevant. To understand their impact, it 
is instructive to work in a basis where instead of the three electrically 
neutral electroweak resonances $\rho_L^0$, $\rho_R^0$, and $\rho_X^0$, one 
works with 
three linear combinations that, before EWSB, couple to the same quantum 
numbers as the $Z$, the photon, and one which does not couple to the leptons at 
all. The first two states are analogous to the KK $Z$ and the KK photon in 
Randall-Sundrum models (cf. \cite{Agashe:2007ki,Albrecht:2009xr}). This basis 
is relevant because the ``KK $Z$'' contribution leads to $C_9=(4s_w^2-1)C_{10}$ 
just as the $Z$-mediated one, while the ``KK photon'' contribution affects only 
$C_9$ and not $C_{10}$.
In addition, the part of the ``KK $Z$'' contribution that involves the 
composite-elementary mixings $\Delta_{u_L}$ is forbidden by the same custodial 
protection that protects the $Zb_L\bar b_L$ coupling, while a similar 
protection is absent for the ``KK photon'' contribution. This is particularly 
relevant in $U(3)^3_\text{RC}$, where the 
correction involving $\Delta_{d_L}$ is flavour diagonal in the mass basis 
\cite{Barbieri:2012tu}, cf.\ \eqref{eq:u3rc-deltadL}.
As a consequence, the $Z$-mediated as well as the ``KK $Z$'' contribution to 
the $\Delta F=1$ operators vanish, while the ``KK photon'' contribution can be 
nonzero.

Recently, a number of tensions between measurements and SM expectations have 
appeared in several observables in rare $b\to s$ decays. This includes in 
particular
\begin{itemize}
 \item A suppression of the angular observable $P_5'$ in $B\to K^*\mu^+\mu^-$ 
\cite{LHCb:2015dla,Descotes-Genon:2014uoa,Straub:2015ica};
 \item A suppression of the branching ratio of $B_s\to \phi\mu^+\mu^-$ 
\cite{Aaij:2015esa,Straub:2015ica};
 \item A suppression of $R_K$, the ratio of branching ratios of $B^+\to 
K^+\mu^+\mu^-$ and $B^+\to K^+e^+e^-$ \cite{Aaij:2014ora}.
\end{itemize}
While the first two of these anomalies could be due to unexpected hadronic 
effects (see e.g.\ 
\cite{Lyon:2014hpa,Jager:2014rwa}) and the last one due to a statistical 
fluctuation, all of them could be explained consistently by a negative NP 
contribution to the Wilson coefficient $C_9^{bs}$ (a positive contribution to 
$C_{10}^{bs}$ is allowed in addition) with muons only 
\cite{Descotes-Genon:2013wba,Altmannshofer:2013foa,
Beaujean:2013soa,Hiller:2014yaa,Ghosh:2014awa,Altmannshofer:2014rta}. 
In composite Higgs models, such lepton flavour non-universal contribution was 
shown by us to arise if muons carry a significant degree of compositeness 
\cite{Niehoff:2015bfa}\footnote{Another possibility is to introduce composite 
leptoquarks \cite{Gripaios:2014tna}.}.
In the present setup, since we are considering elementary leptons only, all 
effects are lepton flavour universal. Although in this case, 
the deviation in $R_K$ cannot be explained,  the overall agreement with the 
data could still be significantly improved compared to the SM if there are 
(lepton flavour universal) NP contributions in $C_9^{bs}$ (and possibly 
$C_{10}^{bs}$), which can resolve the tensions in $B\to K^*\mu^+\mu^-$ angular 
observables and various branching ratios and give a good fit to the data 
\cite{Altmannshofer:2014rta}.
As discussed above, such contribution can arise from ``KK photon''-like 
resonance exchanges.

In view of these tensions, we do not impose semi-leptonic $b\to s$ transitions 
as constraints in our numerical analysis, but rather consider the predictions 
for them {\em a posteriori}.

We do however include the branching ratio of $B_s\to\mu^+\mu^-$ as a 
constraint. This branching ratio, which has reduced theoretical uncertainties 
compared to the semi-leptonic decays, was
recently observed by LHCb and CMS \cite{CMS:2014xfa} in agreement with the
SM expectation
\cite{Bobeth:2013uxa},
\begin{align}
\text{BR}(B_s\to\mu^+\mu^-)_\text{SM} &= (3.65\pm0.23)\times 10^{-9} \,,
\\
\text{BR}(B_s\to\mu^+\mu^-)_\text{exp} &= (2.8^{+0.7}_{-0.6})\times 10^{-9} \,.
\end{align}
In the presence of new physics, the branching ratio is modified as
\begin{equation}
\frac{\text{BR}(B_s\to\mu^+\mu^-)}{\text{BR}(B_s\to\mu^+\mu^-)_\text{SM}}
= \frac{|C_{10}^{bs}-C_{10}^{\prime\,bs}|^2}{|(C_{10}^{bs})_\text{SM}|^2} \,.
\end{equation}
Again, the imaginary parts and the chirality structure can be constrained
by other observables in exclusive and inclusive semi-leptonic decays.

\subsubsection{Contact interactions}\label{sec:ci}

Four-quark contact interactions are constrained by measurements of the
dijet angular distribution at LHC. These constraints become relevant when
some of the first-generation quark fields have a significant degree of
compositeness. This is unavoidable in the $U(3)^3$ models, but also occurs
in part of the parameter space of the $U(2)^3$ models.
The relevant four-quark operators involve only the first generation quarks as 
the contribution from the other generations is PDF-suppressed. The Wilson 
coefficients are computed analogously to the $\Delta F=2$ ones in 
section~\ref{sec:df2}.
Experimental bounds are usually quoted on operators in an $SU(2)\times U(1)_Y$ 
gauge-invariant basis. Using the notation of \cite{Domenech:2012ai}, their 
Wilson coefficients can be related to the ones in the low-energy basis as
\begin{align}
c_{qq}^{(1)} &= C_{VLL}^{uu} + \frac{1}{6}C_{VLL}^{uddu} \,,
&
c_{qq}^{(8)} &= C_{VLL}^{uu} \,,
\\
c_{qu}^{(1)} &= C_{VLR}^{uu} - \frac{1}{6}C_{SLR}^{uu} \,,
&
c_{qu}^{(8)} &= - \frac{1}{6}C_{SLR}^{uu} \,,
\\
c_{uudd}^{(1)} &= C_{VRR}^{uudd}+\frac{1}{3}C_{VRR}^{uddu} \,,
&
c_{uudd}^{(8)} &= 2C_{VRR}^{uddu} \,,
\\
c_{uu}^{(1)} &= C_{VRR}^{uu} \,,
&
c_{dd}^{(1)} &= C_{VRR}^{dd} \,,
\end{align}
and with the appropriate replacement $u\to d$ for $c_{qd}^{(1,8)}$. In addition 
to the operators
in \eqref{eq:df2op1}, \eqref{eq:df2op2} and the ones with $d\to u$, we have 
defined
\begin{align}
Q_{VLL}^{uddu} &= (\bar u_L\gamma^\mu d_L)(\bar d_L\gamma^\mu u_L) \,,
&
Q_{VLL}^{uudd} &= (\bar u_L\gamma^\mu u_L)(\bar d_L\gamma^\mu d_L) \,,
\end{align}
as well as $L\to R$.

The Wilson coefficients of the four-quark operators in the low-energy basis can 
be computed analogously to the $\Delta F=2$ Wilson coefficients in 
section~\ref{sec:df2}. However, an important difference is that the measurement 
of the dijet angular distribution at LHC involves processes at much higher 
energies compared to meson decays. The EFT approach is only valid if the 
exchanged resonances are much heavier than the typical energy scale of the 
process in question. In \cite{deVries:2014apa}, it has been shown that for 
resonance masses below about 5~TeV, the contact interaction bounds become much 
weaker than a naive application of the EFT would suggest. To account for this 
fact in an approximate way, we follow a prescription advocated in this paper 
and multiply every individual contribution to the four-quark operators arising 
from exchange of a resonance with mass $m_{\rho_i}$ by a correction factor 
$(1+C^2/m_{\rho_i}^2)^{-2}$, adopting $C=1.3$~TeV as a rough estimate based on 
a numerical analysis of the full mass dependence in two 
benchmark scenarios \cite{deVries:2014apa}.

ATLAS and CMS have presented constraints on contact interactions using the full 
run-1 data set. However, the constraints are only quoted for a single operator 
(in the case of ATLAS) or for individual operators, but only allowing one at a 
time (in the case of CMS). In our case, multiple operators might be present 
simultaneously, and the operators with right-handed quarks typically differ for 
up- and down-type quarks. The full dependence of the dijet angular distribution 
on all operators has been discussed in \cite{Domenech:2012ai} and simple 
formulae for the impact of the operators in specific rapidity bins have been 
presented there for the 7~TeV LHC. We use these results, updated to the 8~TeV 
LHC, to obtain the relative contributions of individual operators to the 
differential cross section, while we use the bound on the Wilson coefficient 
$c_{qq}^{(1)}$ quoted by the experimentalists for the normalization. Details on 
the procedure are discussed in appendix~\ref{sec:dijetapp}.

\subsection{Direct searches}\label{sec:direct}

In addition to the {\em indirect} searches, i.e.\ precision constraints from 
flavour, electroweak, and Higgs physics, composite Higgs models are also 
subject to increasingly strong {\em direct} constraints from searches for 
composite resonances at the LHC. Since we are focusing on a model with a 
minimal Higgs sector and we are ignoring partial compositeness of leptons, in 
our case the relevant searches are the ones for quark partners, to be 
discussed in section~\ref{sec:spin1/2}, and for spin-1 resonances, to be 
discussed in section~\ref{sec:spin1}.

\subsubsection{Quark partners}\label{sec:spin1/2}

\begin{table}[tbp]
\centering
\renewcommand{\arraystretch}{1.15}
\begin{tabular}{llllll}
\hline
Decay		&Experiment	&$\sqrt{s}$ [TeV]	&Luminosity 
[fb$^{-1}$]&Analysis			&	\\
\hline
$Q \to tW$	&	CMS		&	7		&	5	
	
	&	B2G-12-004		&	\cite{Chatrchyan:2012af}	
\\
\hline
\multirow{2}{*}{$Q \to jW$}	&	ATLAS	&	7		&	
1.04		
	
&	EXOT-2011-28		&	\cite{Aad:2012bt}	\\
			&	CDF		&	1.96		&	
4.6 & &\cite{CDF-PUB-TOP-PUBLIC-10110}	\\
\hline
$Q \to qW$	&	CMS		&	8		&	19.7	
	
	&	B2G-12-017		&	\cite{CMS:2014dka}	\\
\hline
$Q \to jZ$ &	CDF		&	1.96		&	
1.055 & &\cite{CDFnote8590}	\\
\hline
\multirow{2}{*}{$U \to tZ$}
			&	CMS		&	7		&	
5			&	B2G-12-004		&	
\cite{Chatrchyan:2012af}	\\
			&	CMS		&	7		&	
1.1			&	EXO-11-005		&	
\cite{Chatrchyan:2011ay}
\\\hline
\multirow{4}{*}{$D \to bH$}
			&	ATLAS	&	8		&	20.3	
	
	&	CONF-2015-012	&	\cite{ATLAS:2015dka}	\\
			&	CMS		&	8		&	
19.8			&	B2G-12-019		&	
\cite{CMS:2012hfa}	\\
			&	CMS		&	8		&	
19.5			&	B2G-13-003		&	
\cite{CMS:2013una}	\\
			&	CMS		&	8		&	
19.7			&	B2G-14-001		&	
\cite{CMS:2014bfa}	\\
\hline
\multirow{4}{*}{$D \to bZ$}
			&	CMS		&	7		&	
5			&	EXO-11-066		&	
\cite{CMS:2012jwa}	\\
			&	CMS		&	8		&	
19.8			&	B2G-12-019		&	
\cite{CMS:2012hfa}	\\
			&	CMS		&	8		&	
19.5			&	B2G-13-003		&	
\cite{CMS:2013una}	\\
			&	CMS		&	8		&	
19.6			&	B2G-12-021		&	
\cite{CMS:2013zea}	\\
\hline
\multirow{4}{*}{$D \to tW$}
			&	ATLAS	&	8		&	20.3	
	
	&	EXOT-2013-16		&	\cite{Aad:2015gdg}	\\
			&	CMS		&	8		&	
19.8			&	B2G-12-019		&	
\cite{CMS:2012hfa}	\\
			&	CMS		&	8		&	
19.5			&	B2G-13-003		&	
\cite{CMS:2013una}	\\
			&	CDF		&	1.96		&	
2.7			&			&	
\cite{Aaltonen:2009nr}	\\
\hline
\multirow{5}{*}{$Q \to bW$}
			&	CMS		&	7		&	
5			&	EXO-11-050		&	
\cite{CMS:2012ab}	\\
			&	CMS		&	7		&	
5			&	EXO-11-099		&	
\cite{Chatrchyan:2012vu}	\\
			&	ATLAS	&	7		&	4.7	
	
	&	EXOT-2012-07		&	\cite{ATLAS:2012qe}	\\
			&	ATLAS	&	8		&	20.3	
	
	&	CONF-2015-012	&	\cite{ATLAS:2015dka}	\\
			&	CMS		&	8		&	
19.7			&	B2G-12-017		&	
\cite{CMS:2014dka}	\\
\hline
\multirow{3}{*}{$Q_{5/3} \to tW$}
			&	ATLAS	&	8		&	20.3	
	
	&	EXOT-2013-16		&	\cite{Aad:2015gdg}	\\
			&	ATLAS	&	8		&	20.3	
	
	&	EXOT-2014-17		&	\cite{Aad:2015mba}	\\
			&	CMS		&	8		&	
19.6			&	B2G-12-012		&	\cite{CMS:vwa}	
\\
\hline
$U \to tH$		&	CMS		&	8		&	
19.7			&	B2G-12-004		&	
\cite{CMS:2014rda}	\\

\hline
\end{tabular}
\caption{Experimental analyses included in our numerics for heavy quark 
partner decay. $Q$ stands for any quark partner where the decay in question is 
allowed by electric charges, $j$ stands for a light quark or $b$ jet, and $q$ 
for a light quark jet.}
\label{tab:exp_quark_res}
\end{table}

Pair production of heavy quarks and subsequent decay to SM quarks and weak 
bosons has been searched for at Tevatron and LHC. Recently, also final states 
involving the Higgs have been included in the searches. In the simplest case 
where only decays to third generation quarks and a $W$, $Z$ or Higgs are 
allowed, these channels can be combined to obtain stringent bounds on the 
masses 
of vector-like 3rd generation quark partners. In our numerical analysis, we aim 
to be more general since in principle, a quark partner can have several 
relevant 
decay modes involving SM or partner quarks, 3rd or light generation quarks. To 
this end, we compute the production cross section times branching ratio of each 
quark partner in each experimentally relevant decay mode, and compare it 
directly to the upper bounds on this quantity provided in the experimental 
analysis.

For the pair production cross section, we simply take the model-independent 
NNLO 
QCD production cross section for a heavy quark computed in Hathor 
\cite{Aliev:2010zk}. This means 
we 
neglect
\begin{itemize}
\item Single production, that is relevant for quarks that have a significant 
degree of compositeness, and in this case dominates at higher masses 
\cite{DeSimone:2012fs,Li:2013xba,Redi:2013eaa,Azatov:2013hya,
Delaunay:2013pwa,Backovic:2014uma};
\item Contributions to the pair production cross section from heavy resonance 
exchange \cite{Azatov:2015xqa,Araque:2015cna}.
\end{itemize}
Taking into account these two effects is beyond the scope of our study, as it 
cannot be implemented efficiently in a fast parameter scan. The bounds we 
obtain 
should thus be considered conservative.

Since the experimental analyses typically quote bounds on the pair production 
cross section assuming a 100\% branching ratio to the desired final state, we 
correct for the branching ratio $\text{BR}(Q\to f)$ of the quark partner to 
final state $f$ by multiplying the production cross section with
\begin{itemize}
\item $\text{BR}(Q\to f)^2$ in case the experimental analysis requires both 
partners to decay to~$f$;
\item $\left(1-\left(1-\text{BR}(Q\to f)\right)^2\right)$ in case the 
experimental analysis requires one or both of the partners to decay 
to~$f$.
\end{itemize}

In the M4DCHM5, there are in total 24 heavy charge-$2/3$ quarks (denoted with 
$U$ in the following), 24 charge-$(-1/3)$ quarks ($D$), as well as 6 
exotic charge-$5/3$ quarks ($Q_{5/3}$) and 6 charge-$(-4/3)$ quarks 
($Q_{-4/3}$). The decay modes always involve one SM quark or quark resonance 
plus one $W$, $Z$, Higgs, or vector resonance. In our numerical analysis, we 
can only impose constraints on decays involving SM particles only. This is not 
a strong restriction since the lightest quark partners are always required to 
decay to SM states for kinematic reasons.
In table~\ref{tab:exp_quark_res}, we list all the experimental searches that
we include in our numerical analysis for the individual decay modes. In this 
table, $Q$ stands for any quark partner where the decay in question is allowed 
by electric charges, $j$ stands for a light quark or $b$ jet, and $q$ for a 
light quark jet. Note that there are no dedicated searches for the $Q_{-4/3}$, 
but searches of the type $Q\to (bW,jW,qW)$ are also sensitive to these states.

\begin{figure}
\includegraphics[width=0.48\textwidth]{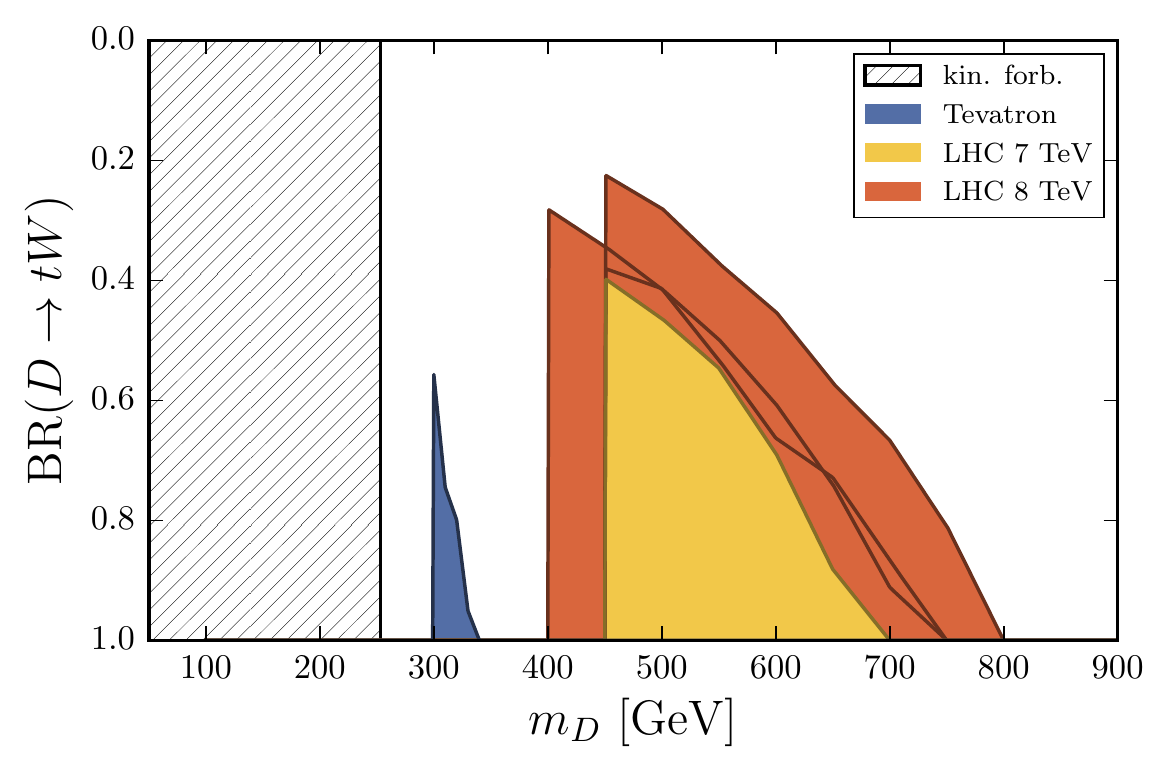}
\includegraphics[width=0.48\textwidth]{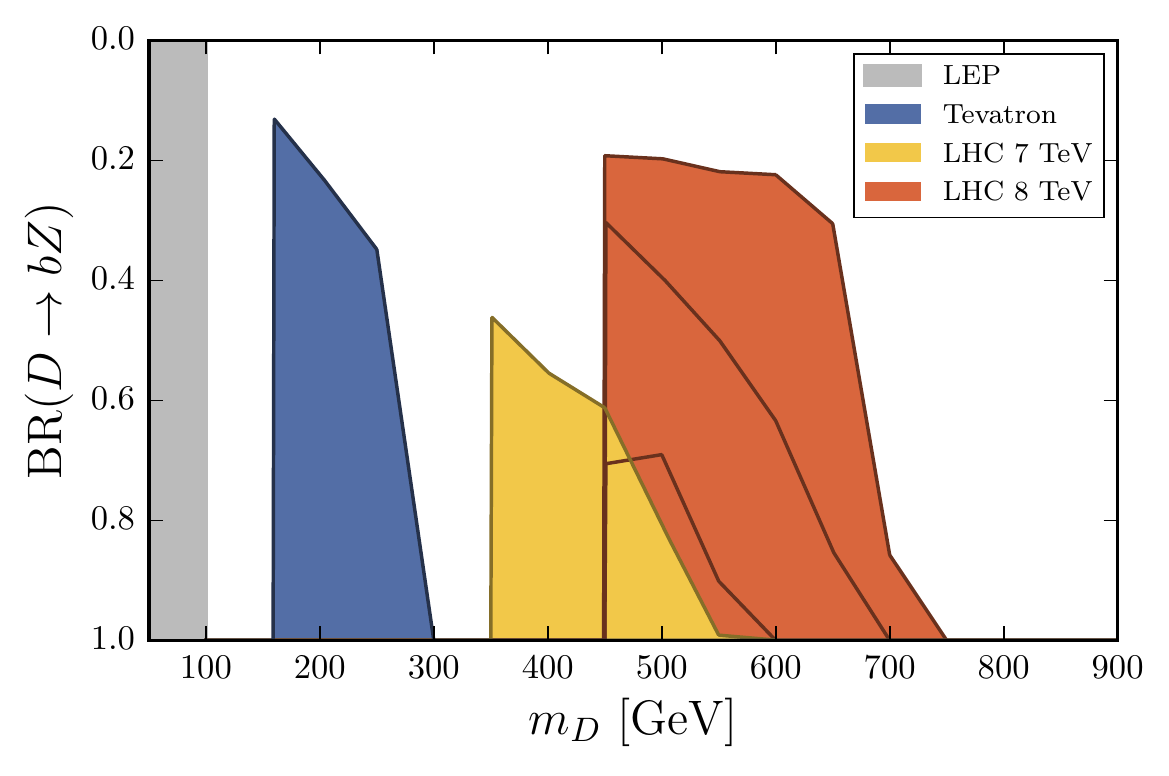}

\includegraphics[width=0.48\textwidth]{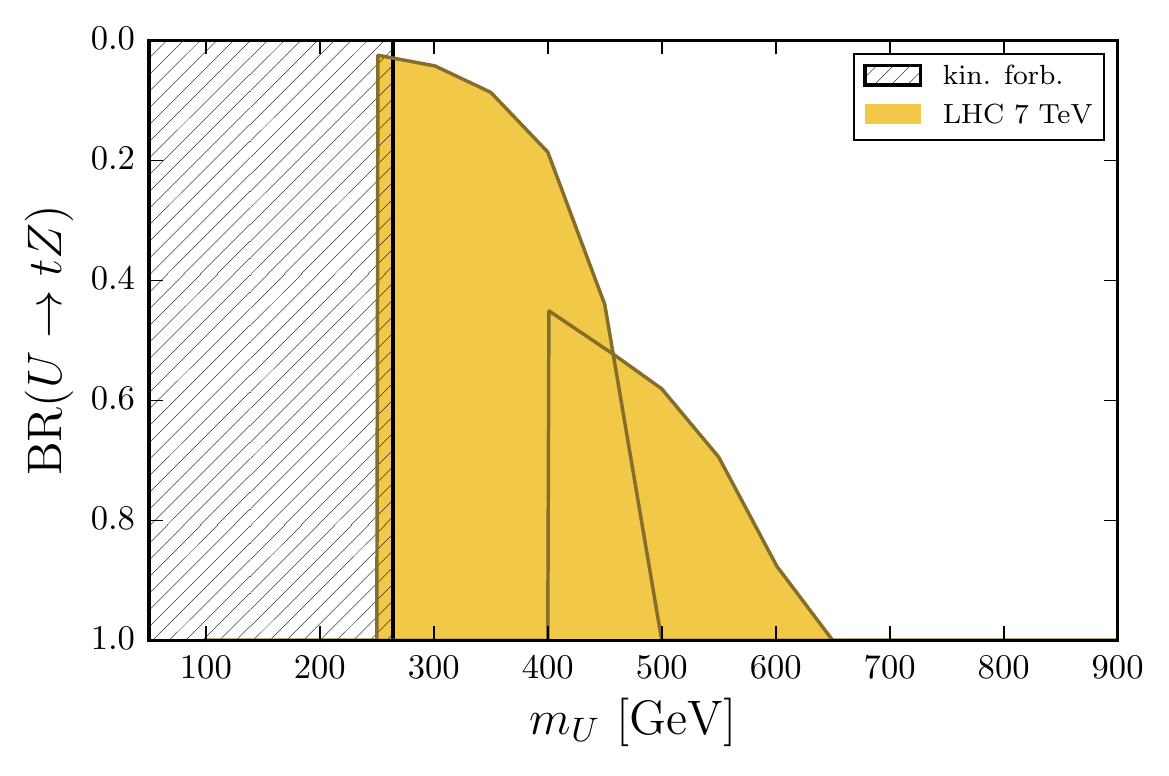}
\includegraphics[width=0.48\textwidth]{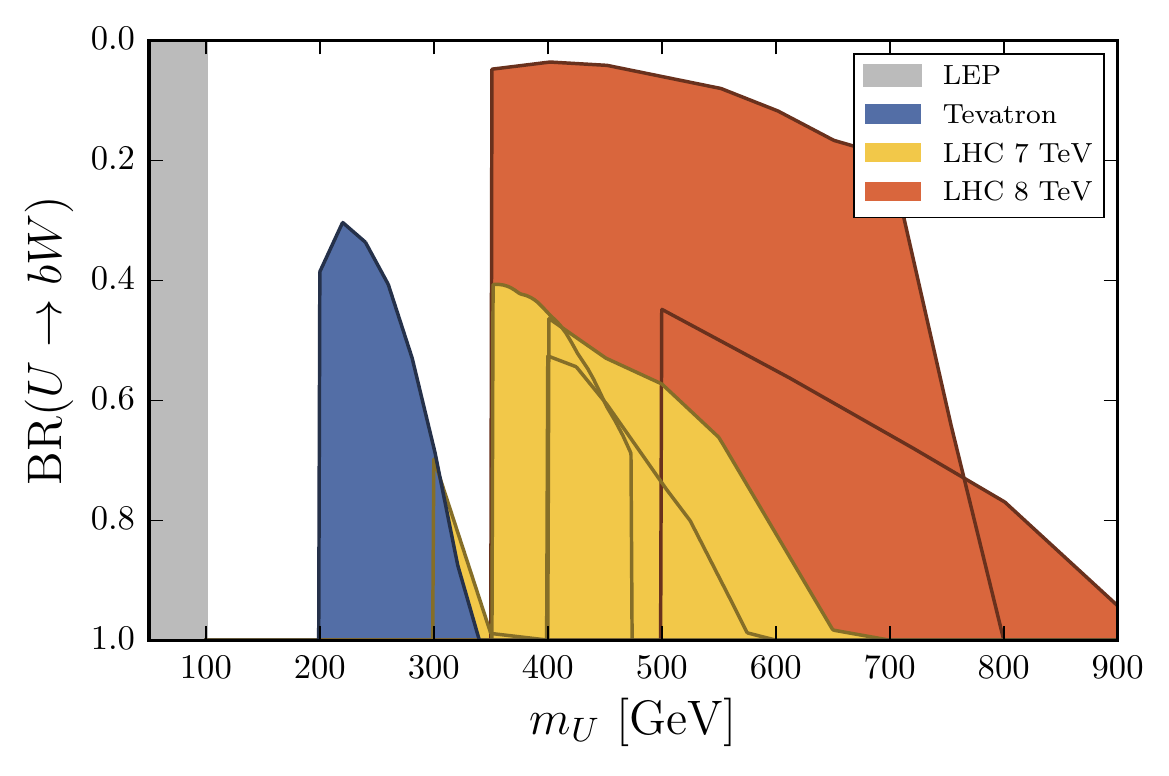}

\includegraphics[width=0.48\textwidth]{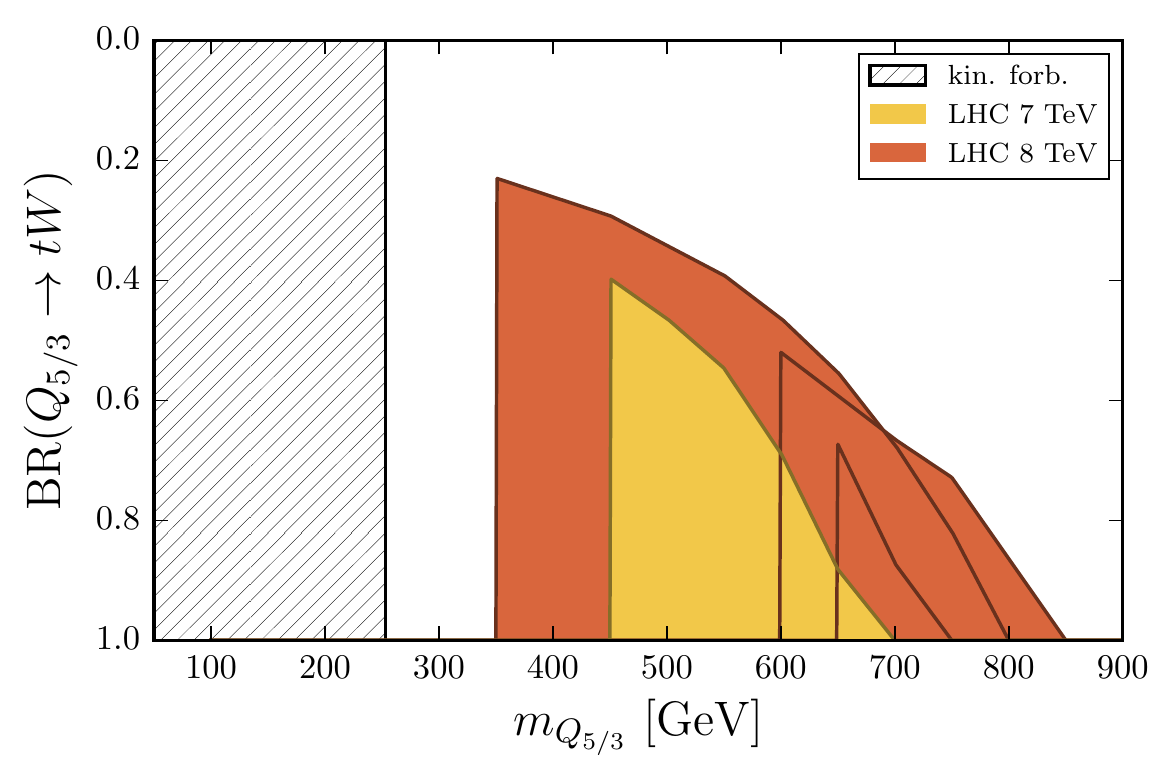}
\includegraphics[width=0.48\textwidth]{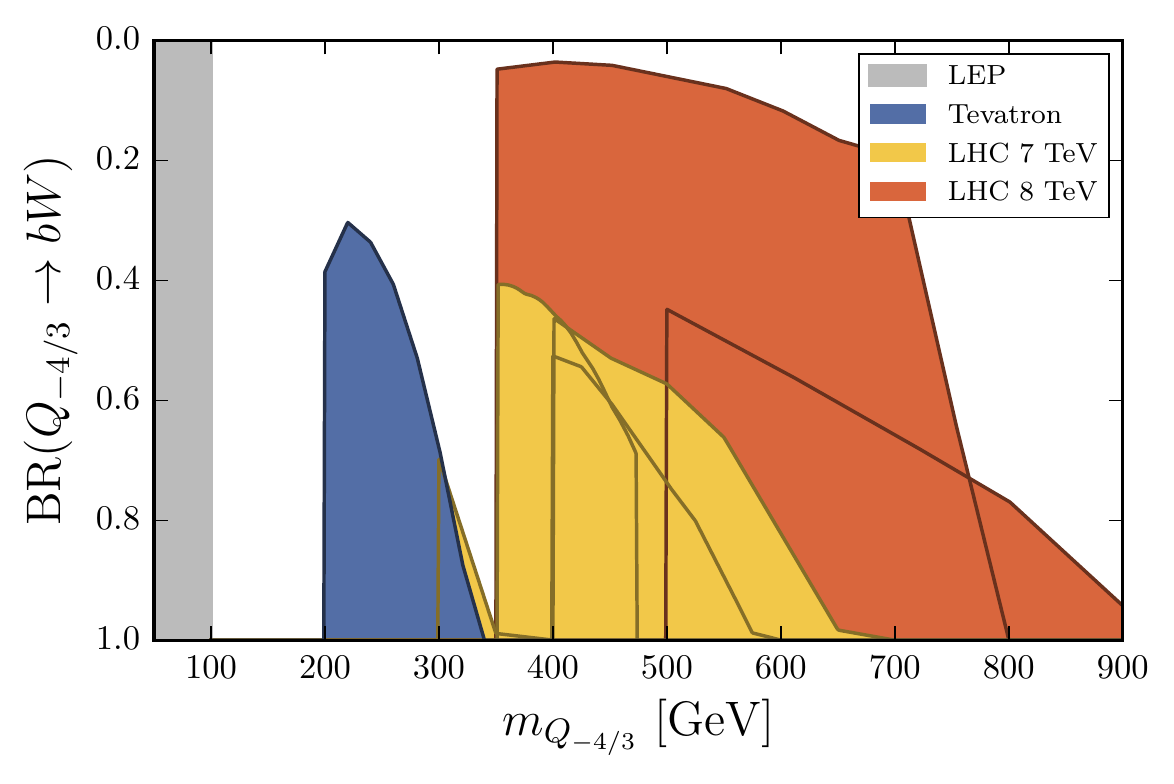}

\includegraphics[width=0.48\textwidth]{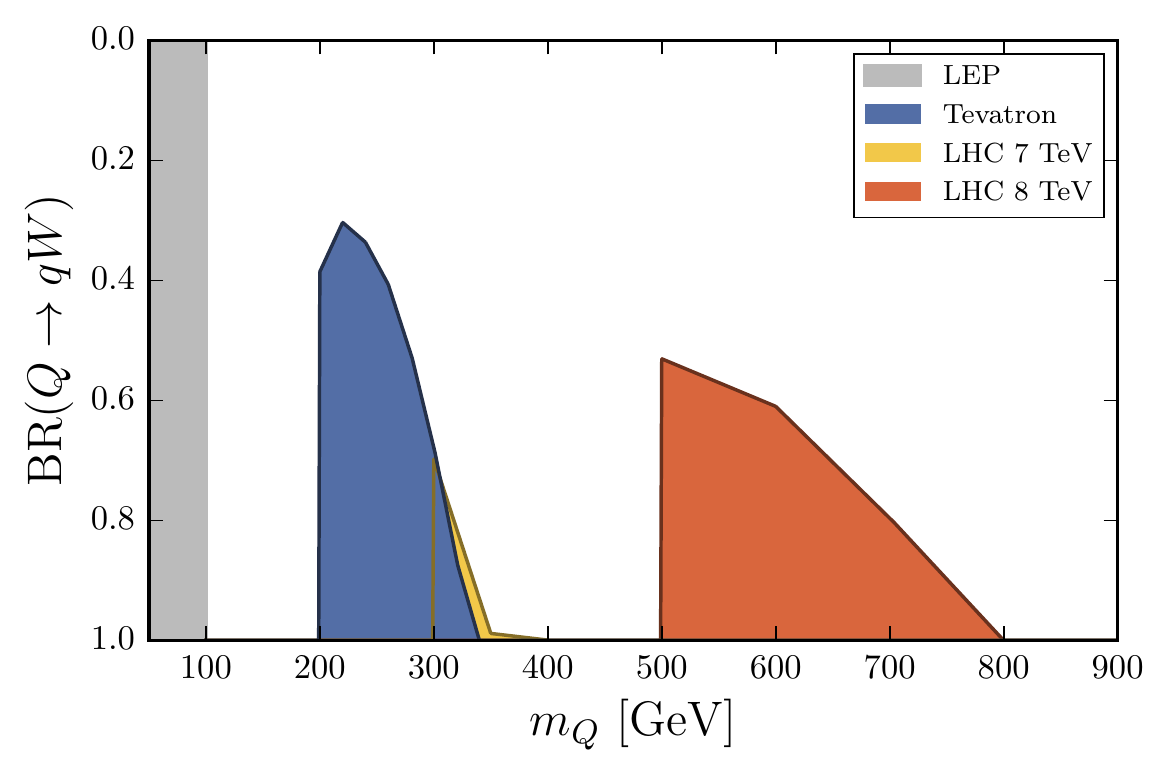}
\includegraphics[width=0.48\textwidth]{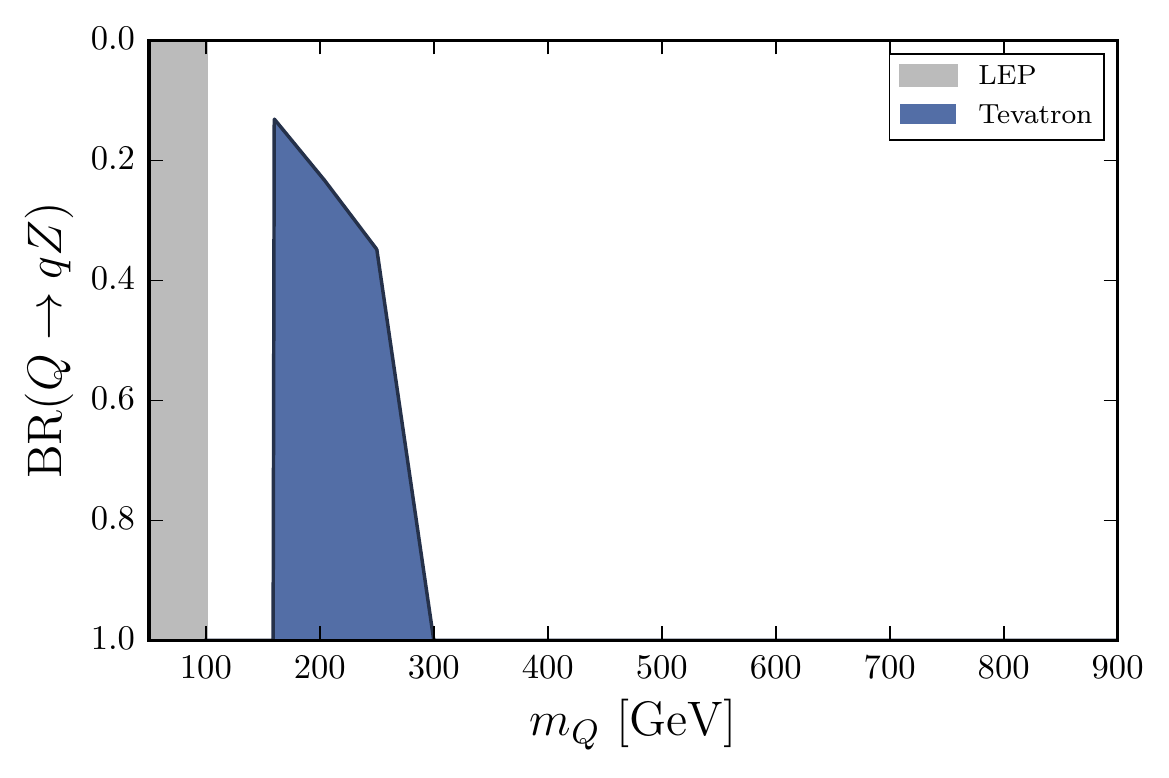}
\caption{Upper bounds on the branching ratios of quark partners decaying to SM 
states from individual Tevatron and LHC searches. QCD pair production is 
assumed. $Q$ stands for any quark partner where the decay is allowed by the 
electric charges. $q$ stands for a light quark (excluding the $b$).}
\label{fig:fermion_br}
\end{figure}

An important point concerning the experimental coverage of parameter space is 
that the experiments typically employ a hard $p_T$ cut to reduce backgrounds. 
This maximizes the sensitivity to heavy states, but misses out on the low end 
of 
the mass spectrum. In fact, combining all existing analyses in 
table~\ref{tab:exp_quark_res}, we have identified 
a number of gaps in the coverage of quark partner masses. This is illustrated 
by 
the plots in figure~\ref{fig:fermion_br}. They show the 95\% C.L.\ upper bound 
on the branching 
ratio in the decay modes to $W$ or $Z$ bosons as a function of the quark 
partner mass. We make the following observations.
\begin{itemize}
\item When kinematically allowed, there is a gap between the LEP bound of 
100~GeV and the lower end of the Tevatron bounds. This should however not be 
taken seriously as quark partners with mass comparable to the top quark would 
very likely have shown up already.
\item More seriously, there are gaps between the upper end of the Tevatron 
exclusions and the lower end of the LHC exclusions. This leaves a window 
between about 300 and 350--500~GeV not covered by existing searches.\footnote{%
The presence of a gap between Tevatron and LHC 7~TeV searches has already been
noticed in \cite{Gillioz:2012se}.}
The only 
exception is the mode $U\to tZ$, but quark partners around 300~GeV would have a 
very small phase space to decay to $tZ$, making it plausible that the branching 
ratio is smaller than in the other channels.
\item Bounds on light quark partners are weak, with no existing LHC search for 
the decay mode involving the $Z$ boson. This is problematic since,  depending 
on their quantum numbers, some of the light generation partners have very small 
branching ratio into $qW$ as will be demonstrated in 
section~\ref{sec:prospects-quark} below.
\end{itemize}
Concerning the gaps mentioned in the second item,
they could very likely be closed by a reanalysis of existing run-1 data (this 
is also indicated by recasting of some existing new physics searches, see e.g.\
\cite{Anandakrishnan:2015yfa}). We
call upon the experimental collaborations to perform such a reanalysis, given 
the importance of the partner mass scale for naturalness and radiative 
electroweak symmetry breaking in the models at hand.
In our numerical analysis, in order not to be biased by these low-mass regions 
for quark partners, we have imposed a hard lower bound of 500~GeV on all quark 
partner masses, in addition to the LHC 7 and 8~TeV searches that are 
sensitive to higher masses (while the Tevatron searches become irrelevant).

\subsubsection{Spin-1 partners}\label{sec:spin1}

\begin{table}[tbp]
\centering
\renewcommand{\arraystretch}{1.15}
\begin{tabular}{llllll}
\hline
Decay		&Experiment	&$\sqrt{s}$ [TeV]	&Lum.
[fb$^{-1}$]&Analysis			&	\\
\hline
$\rho^{\pm} \to \ell^{\pm}\nu$	&	ATLAS	&	7		&	
4.7		
&	EXOT-2012-02		&	\cite{Aad:2012dm}	\\
\hline
\multirow{2}{*}{$\rho^{\pm} \to W^{\pm}h$}
			&	ATLAS	&	8		&	20.3	
	
&	EXOT-2013-23		&	\cite{Aad:2015yza}	\\
			&	CMS		&	8		&	
19.7		&	EXO-14-010		&	\cite{CMS:2015gla}	
\\
\hline
\multirow{4}{*}{$\rho^{\pm} \to W^{\pm}Z$}
			&	ATLAS	&	8		&	20.3	
	
&	EXOT-2013-01		&	\cite{Aad:2015ufa}	\\
			&	ATLAS	&	8		&	20.3	
	
&	EXOT-2013-07		&	\cite{Aad:2014pha}	\\
			&	ATLAS	&	8		&	20.3	
	
&	EXOT-2013-08		&	\cite{Aad:2015owa}	\\
			&	CMS		&	8		&	
19.7		&	EXO-12-024		&	
\cite{Khachatryan:2014hpa}	\\
\hline
$\rho^{\pm} \to tb$		&	CMS		&	8		
&	
19.5		&	B2G-12-010		&	
\cite{Chatrchyan:2014koa}	\\
\hline
\multirow{2}{*}{$\rho^0 \to W^+W^-$}
			&	ATLAS	&	8		&	20.3	
	
&	EXOT-2013-01		&	\cite{Aad:2015ufa}	\\
			&	CMS		&	8		&	
19.7		&	EXO-13-009		&	
\cite{Khachatryan:2014gha}	\\
\hline
\multirow{2}{*}{$\rho^0 \to Zh$}
			&	ATLAS	&	8		&	20.3	
	
&	EXOT-2013-23		&	\cite{Aad:2015yza}	\\
			&	CMS		&	8		&	
19.7		&	EXO-13-007		&	
\cite{Khachatryan:2015ywa}	\\
\hline
\multirow{2}{*}{$\rho^0 \to \ell^+\ell^-$}
			&	ATLAS	&	8		&	20.3	
	
&	EXOT-2012-23		&	\cite{Aad:2014cka}	\\
			&	CMS		&	8		&	
20.6
&	EXO12061		&	
\cite{Khachatryan:2014fba}	\\
\hline
\multirow{2}{*}{$\rho^0/\rho_G \to t\bar t$}
			&	ATLAS	&	8		&	20.3	
	
&	CONF-2015-009	&	\cite{ATLAS:2015aka}	\\
			&	CMS		&	8		&	
19.5		&	B2G-12-008		&	\cite{CMS:2013gqa}	
\\

\hline
\end{tabular}
\caption{Experimental analyses included in our numerics for heavy vector 
resonance decay.}
\label{tab:exp_vector_res}
\end{table}

Spin-1 resonances can be pair-produced in a Drell-Yan like process. If narrow 
enough, they would show up as a peak in the dilepton, dijet, $t\bar t$, $VV$, 
or 
$Vh$ final state, depending on the branching ratios. In the models considered 
by 
us, the spin-1 and spin-1/2 resonances are strongly coupled to each other. 
Consequently, if the decay to two fermion resonances is kinematically allowed, 
the resonances become very broad and are not captured by the experimental 
analyses anymore\footnote{They would however still constitute a contribution to 
the 
pair production cross section of the fermion resonances, see the comment in 
section~\ref{sec:spin1/2}}. Still, we expect that they are sufficiently narrow 
in part of the parameter space, so we include the experimental constraints.

The hadronic production cross section of a spin-1 resonance can be written as
\begin{equation}
\sigma(pp\to \rho+X) = \sum_{q_1q_2} \frac{\Gamma(\rho\to q_1\bar 
q_2)}{s\,m_\rho}
\frac{4\pi^2}{3} \mathcal L_{q_1\bar q_2} c_\rho \,,
\label{eq:prodV}
\end{equation}
where $q_{1,2}=u$ or $d$,
$c_\rho$ is the colour multiplicity of the resonance ($c_\rho=1$ for electroweak
resonances and $c_\rho=8$ for $\rho_G$), $s$ is the hadronic center-of-mass 
energy
squared and $\mathcal L_{i_1i_2}$ is the parton-parton luminosity of the 
relevant
initial state defined as
\begin{equation}
\mathcal L_{q_1\bar q_2}(s,\hat s)=
\int_{\hat s/s}^1 \frac{dx}{x}
f_{q_1}(x,\mu)
f_{\bar q_2}\!\left(\frac{\hat s}{xs},\mu\right)
\,.
\end{equation}
For the Tevatron, \eqref{eq:prodV} is valid with the appropriate
replacements.

The M4DCHM5 contains 3 charged and 5 neutral electroweak vector resonances, 
plus the colour-octet gluon resonance. As in the spin-$\frac{1}{2}$ case, we 
can only impose constraints on decays to SM states. 
We include all the analyses listed in table~\ref{tab:exp_vector_res}.
The only relevant final state that we have not included is the decay to dijets.
The reason is that the experimental bounds on the dijet resonance cross section 
depend on an acceptance factor that is model dependent and that is not easy to 
take into account in a parameter scan.

In our numerical analysis, 
we employ a cut of 5\% on the relative total width $\Gamma_\rho/m_\rho$, above 
which all bounds are ignored for an individual resonance. While several of the 
analyses are actually sensitive to broader resonances, it is not possible to 
include this dependence in a parameter scan. As in the case of fermion 
resonances, our bounds should thus be considered as conservative.

\subsubsection{LHC excesses}\label{sec:excess}

Interestingly, several of the searches for spin-1 resonances we include as 
constraints contain an excess of events around 2~TeV. The most significant 
deviation is in the ATLAS search for $\rho^\pm \to WZ$, corresponding to a 
local significance of $3.4\sigma$, but an excess around the same mass appears 
also in the corresponding CMS search, and, to a lesser extent, in searches for 
$WW$ or $Wh$ final states. This is particularly interesting in the context of 
composite Higgs models as the resonances associated to $SU(2)_L$ (denoted 
$\rho_{L\mu}$ in the gauge eigenstate basis in section~\ref{sec:model}) 
form a triplet of a charged and a neutral vector that are almost degenerate,
have approximately equal branching ratios to $WZ$, $WW$, $Wh$, and $Zh$ final 
states, and can have a production cross section in the right ballpark to 
explain these excesses 
\cite{Pappadopulo:2014qza,Lane:2014vca,Thamm:2015csa,Carmona:2015xaa,
Bian:2015ota, Lane:2015fza,Low:2015uha}.

\section{Numerical analysis and predictions}\label{sec:num}

This section contains the main results of our paper.
After describing the 
numerical analysis procedure in section~\ref{sec:strategy}, 
we will discuss fine-tuning in all models in section~\ref{sec:tuning},
the numerical results of signals in indirect searches in
$U(2)^3_\text{LC}$ in section~\ref{sec:unlc},
for indirect searches in
$U(2)^3_\text{RC}$ and $U(3)^3_\text{RC}$ in section~\ref{sec:unrc},
and for direct LHC searches in all models in
section~\ref{sec:num-direct}.

\subsection{Strategy}\label{sec:strategy}
\subsubsection{Scanning procedure}

Our aim is to sample the parameter space of the M4DCHM with four different 
flavour structures while satisfying all the experimental constraints discussed 
in section~\ref{sec:exp}. This is particularly challenging because partial 
compositeness implies that all SM masses and couplings are relatively 
complicated functions of the model parameters and the additional requirement of 
correct radiative EWSB is even harder to control analytically. To cope with 
these challenges, we have improved a method first employed in 
\cite{Straub:2013zca}. We construct a $\chi^2$ function depending on all the 
theoretical predictions for all constraints discussed in section~\ref{sec:exp} 
as well as the corresponding experimental measurements. We then proceed in 
four steps.
\begin{enumerate}
\item We randomly generate a set of input parameters that only fulfills the 
most rudimentary consistency conditions (e.g. composite gauge couplings being 
greater than 1, effective potential possesses a minimum away from zero).
\item We use brute-force numerical minimization (with \texttt{NLopt} 
\cite{johnsonnlopt}) to 
``burn-in'' into a region of parameter space not too far from viability.
\item We use a Markov Chain Monte Carlo (MCMC, with \texttt{pypmc} 
\cite{beaujean_2015_20045}) to sample 
the viable parameter space.
\item We filter the Markov chains so that only points remain where each 
individual constraint is satisfied at the $3\sigma$ level.
\end{enumerate}
After the burn-in phase, the generation of viable parameter points is very 
efficient, as the MCMC is adaptive and has an acceptance rate around 23\%. The 
downside of the method is that adjacent points have high autocorrelation, 
implying that very long chains are needed to obtain a reasonable coverage of 
the parameter space. Furthermore, the parameter space can contain several 
disconnected minima. For these reasons, we run a large number of chains (around 
500 for each model) starting at different (random) initial points.

We stress that we do not interpret the outcome of the Markov Chain
statistically, in the sense of a posterior probability distribution for the
model parameters. Apart from the problem of reaching sufficient coverage of
the parameter space, this would be problematic due to dependence on the
choice of priors. Instead, we use the MCMC algorithm as a shortcut to generate
a sufficient number of valid points.
In our final sample, these points
approximately follow a normal
distribution peaked around 30--40,
for 48 individual contributions to the $\chi^2$.
We also find that model-independently for
nearly all points only $\leq 5$ individual constraints are violated by more than
$2 \sigma$ at a time.
Consequently, the hard $3\sigma$ cut only removes a small fraction of extreme
points.
The largest deviations are typically found for 
$m_t$, $\text{BR}(B \to X_s \gamma)$, the inclusive values of
$V_{ub}$ and $V_{cb}$, and for the Higgs signal strengths.

\subsubsection{Model parameters}\label{sec:param}

Below, we specify the model parameters and any relations we have imposed among 
them in our scan.
\begin{itemize}
\item $f$, $f_1$, $f_X$, $f_G$.

We have imposed $1 < f_1/f < \sqrt{3}$, where the lower bound corresponds to a 
decoupling of the axial resonances and the upper bound is motivated by the 
partial unitarization of Goldstone boson scattering \cite{Marzocca:2012zn}. We 
have not assumed $f_1$, $f_X$, and $f_G$ to be degenerate, but we have 
restricted them to be within a factor of two, i.e.\ $\frac{1}{2}\leq 
f_{X,G}/f_1\leq2$, to prevent the fit from completely decoupling one of the 
resonances.
\item $g_\rho$, $g_X$, $g_G$.

We have varied these couplings completely independently, only imposing
$1<g_{\rho,X,G}\leq 4\pi$ to have a strong but semi-perturbative coupling (in 
the case of the gluon resonance, we also imposed $g_G > g_{s0}$). We further 
imposed $f_{1,X,G}\,g_{\rho,X,G}/\sqrt{2} < 4\pi f$, to not have resonance 
masses 
above the (naive) cutoff of the theory.
\item $g_0$, $g_0'$, $g_{s0}$.

These parameters are fixed by the known gauge couplings once the composite 
gauge couplings are specified.
\item $m_{Q}$, $m_{\widetilde{Q}}$, $m_{Y_Q}$, $m_{Y_Q} + Y_Q$ where 
$Q=U$~or~$D$, in the case of $U(2)^3$ different for the first two and the third 
generation.

In our numerical analysis described above, we have treated the {\em logarithms} 
of these parameters as scan parameters, in order not to get a bias towards 
heavier masses. The only relation (apart from the ones forced by the flavour 
symmetries) we have imposed is that all these parameters are $< 4\pi f$.
Note that this implies that the above parameters can only take positive 
values in our scan, but this can always be arranged by a suitable choice of 
phases for the fermion fields.  
\item Quantities parametrizing the composite-elementary mixings, see 
appendix~\ref{sec:deltas} for details.

Again, we have scanned dimensionful parameters logarithmically and require them 
to be $< 4\pi f$, but otherwise we impose no restrictions (many relations among 
these parameters are fixed by the requirement to have the correct quark masses 
and CKM mixing).
\end{itemize}
With these assumptions, the total number of real parameters or phases is
44 for $U(2)^3_\text{LC}$ and $U(2)^3_\text{RC}$
and
30 for $U(3)^3_\text{LC}$ and $U(3)^3_\text{RC}$.
To compare these parameters to the SM, it should be noted that we do not treat 
lepton masses as free parameters, have massless neutrinos and set the QCD 
$\bar\theta$ term to zero, but the Higgs mass and VEV are predictions rather 
than inputs.

\subsection{General fit results and fine-tuning}

\subsubsection{Failure of $U(3)^3_\text{LC}$}\label{sec:u3lc}

In the case of $U(3)^3_\text{LC}$, our scans have not been able to find a 
single viable parameter point, even for a large number of chains. This is not 
surprising as already a qualitative analysis of the relevant constraints on 
$U(3)^3_\text{LC}$ \cite{Barbieri:2012tu} has revealed that there are extremely 
strong constraints on 
the model from electroweak precision test and CKM unitarity. It seems to be 
impossible to reconcile these constraints with the need for correct radiative 
EWSB. We will thus not consider $U(3)^3_\text{LC}$ any further.

\subsubsection{Fine-tuning}\label{sec:tuning}

\begin{figure}[tbp]
\centering
\includegraphics[width=0.75\textwidth]{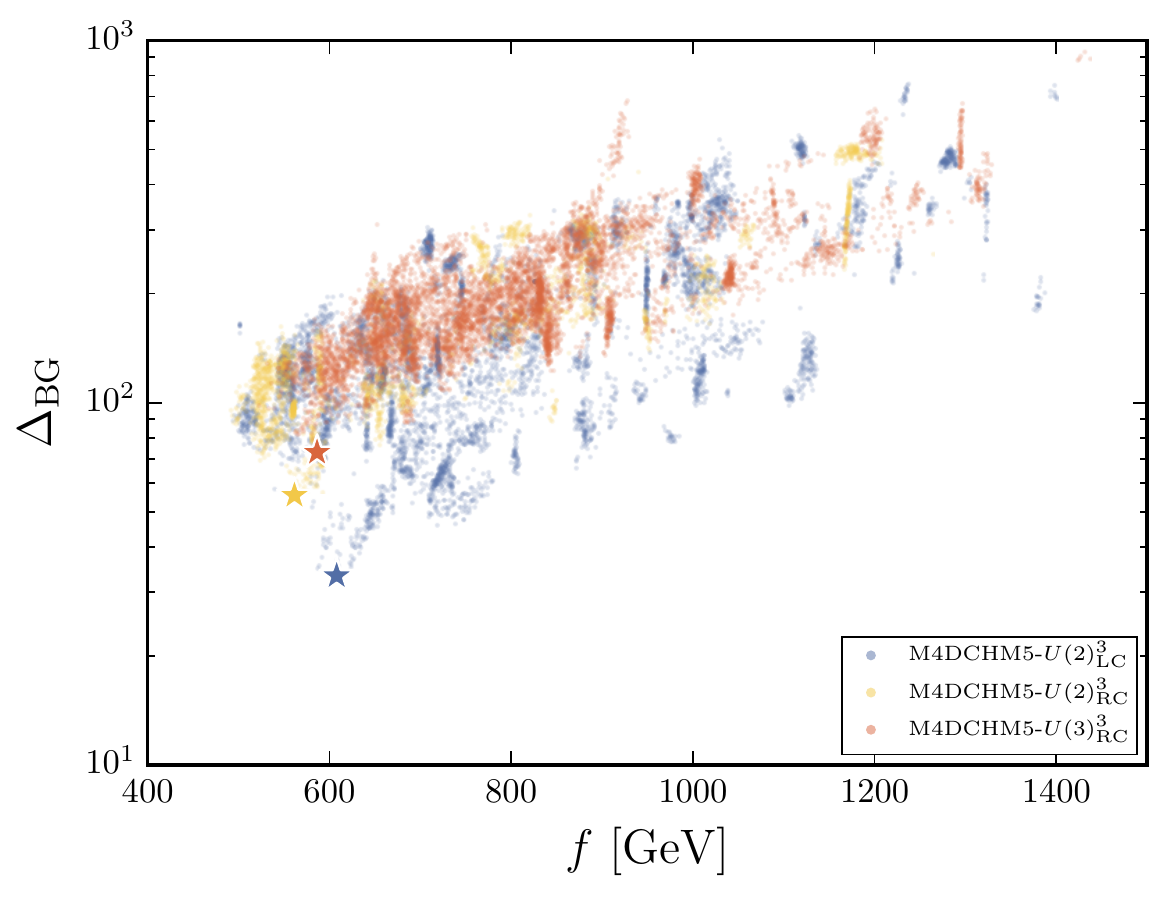}
\caption{Barbieri-Giudice fine-tuning measure vs.\ $f$ for the three viable 
models for a thinned-out sample of all our viable parameter points. The stars 
show the points with lowest fine-tuning measure for each model.}
\label{fig:tuning}
\end{figure}

Before discussing predictions for physical observables, let us address the 
degree to which the viable model points we have found can address the 
electroweak hierarchy problem. To this end, we have computed the 
Barbieri-Giudice fine-tuning measure \cite{Barbieri:1987fn}
\begin{equation}
\Delta_\text{BG} = \max_i \left|
\frac{\partial \ln m_Z}{\partial \ln x_i}
\right|
\end{equation}
that quantifies the sensitivity of the weak scale to variations in the model 
parameters $x_i$. In composite Higgs models, $\Delta_\text{BG}$ can be obtained 
directly from derivatives of the potential \cite{Panico:2012uw}. Still, given 
the large parameter space, the computation turns out to be more time-consuming 
than for the physical observables, so we have computed $\Delta_\text{BG}$ only 
for 
a subset (2\%) of all our points. The results for the three viable models are 
shown in fig.~\ref{fig:tuning}.\footnote{Note that the individual ``speckles''  
-- visible in many of the scatter plots -- correspond to different Markov 
chains.}
Not surprisingly, the lowest $\Delta_\text{BG}$ is obtained for low $f$ and a 
sub-percent fine-tuning is possible in all models as long as $f\lesssim1$~TeV.
This is compatible with earlier analyses in similar models 
\cite{Panico:2012uw,Barnard:2015ryq}.
The points with the lowest tuning measure, highlighted by stars in the plot, 
have
\begin{itemize}
\item $\Delta_\text{BG}^\text{min}=33$ for $U(2)^3_\text{LC}$,
\item $\Delta_\text{BG}^\text{min}=55$ for $U(2)^3_\text{RC}$,
\item $\Delta_\text{BG}^\text{min}=73$ for $U(3)^3_\text{RC}$.
\end{itemize}
Two comments are in order here. First,
we reiterate that we define viability for a point as fulfilling all individual 
constraints at $3\sigma$. Since we assume a 5\% relative uncertainty on $m_t$ 
and $m_h$ (see section~\ref{sec:sm}), the known tendency of the model to have a 
too heavy Higgs and a too 
light top means that the points with lowest tuning typically have the Higgs and 
top mass $15\%$ away from their central values.
Second, we stress that there are variants of the model considered by us that 
have lower fine-tuning since the M4DCHM5 suffers from a ``double tuning'' by
an accidental enhancement of the Higgs mass due to the structure of the 
potential, see \cite{Panico:2012uw} for a discussion and alternatives.

To get a better understanding of the tuning in the Higgs potential, we adopted 
an expansion of the potential in terms of $s_h$ (as also used e.g. in 
\cite{Marzocca:2012zn}),
\begin{equation} \label{eq:Veff_approx}
 V_\mathrm{eff} \approx - \gamma \, f^4 \, 
s_h^2 + \beta \, f^4 \, s_h^4,
\end{equation}
where we have explicitly defined the $\gamma$ and $\beta$ parameters as 
dimensionless 
by factoring out their typical scale $f^4$ and we have neglected terms of 
$\mathcal{O}(s_h^6)$. In this notation, the Goldstone VEV and the Higgs mass 
are 
given as
\begin{equation}
   s_h^* = \sqrt{\frac{\gamma}{2 \beta}}, \qquad m_h^2 = 4 \gamma \left( 1 - 
\frac{\gamma}{2 \beta}  \right) \, f^2. 
\end{equation}
The requirement to fulfill EWPT's (and therefore to have a not too large 
$s_h^*$) together with the hierarchy $m_h \ll f$ forces $\gamma$ to take a 
rather small value. As already mentioned in \cite{Marzocca:2012zn, 
Panico:2012uw}, this requires a cancellation between the fermion and gauge 
contributions to a large degree. 

In our framework, we can extract the $\gamma$ and $\beta$ parameters for each 
contributing field by simply fitting the numerical values of (\ref{eq:Veff}) 
to the parametrization (\ref{eq:Veff_approx}). We indeed find that the fit 
prefers highly correlated gauge and fermion contributions
that are large individually but almost completely compensate each other.
We also find large cancellations between 
the up- and down-quark sector and also between individual contributions in each 
sector.

\subsection{Left-handed compositeness: indirect searches}\label{sec:unlc}

As discussed in section~\ref{sec:u3lc}, we have not found any 
viable points for $U(3)^3_\text{LC}$. We will thus restrict ourselves to
$U(2)^3_\text{LC}$ in this section.

\subsubsection{Light quark compositeness}

\begin{figure}[tbp]
\centering
\includegraphics[width=0.48\textwidth]{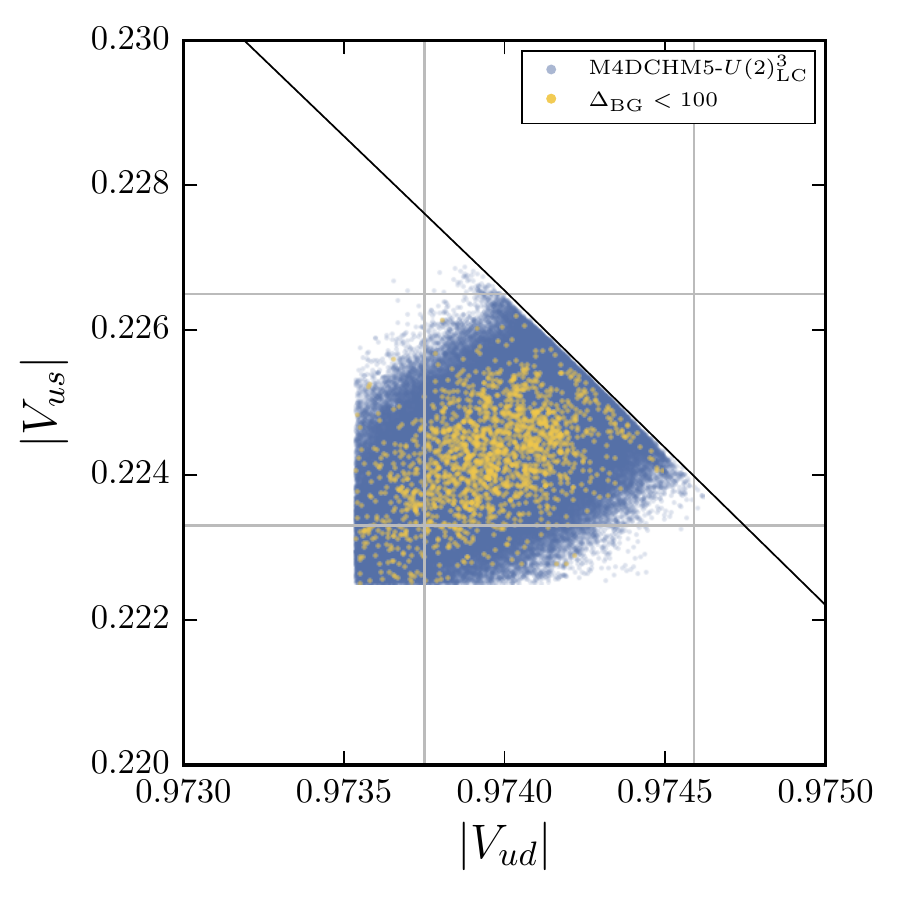}
\includegraphics[width=0.48\textwidth]{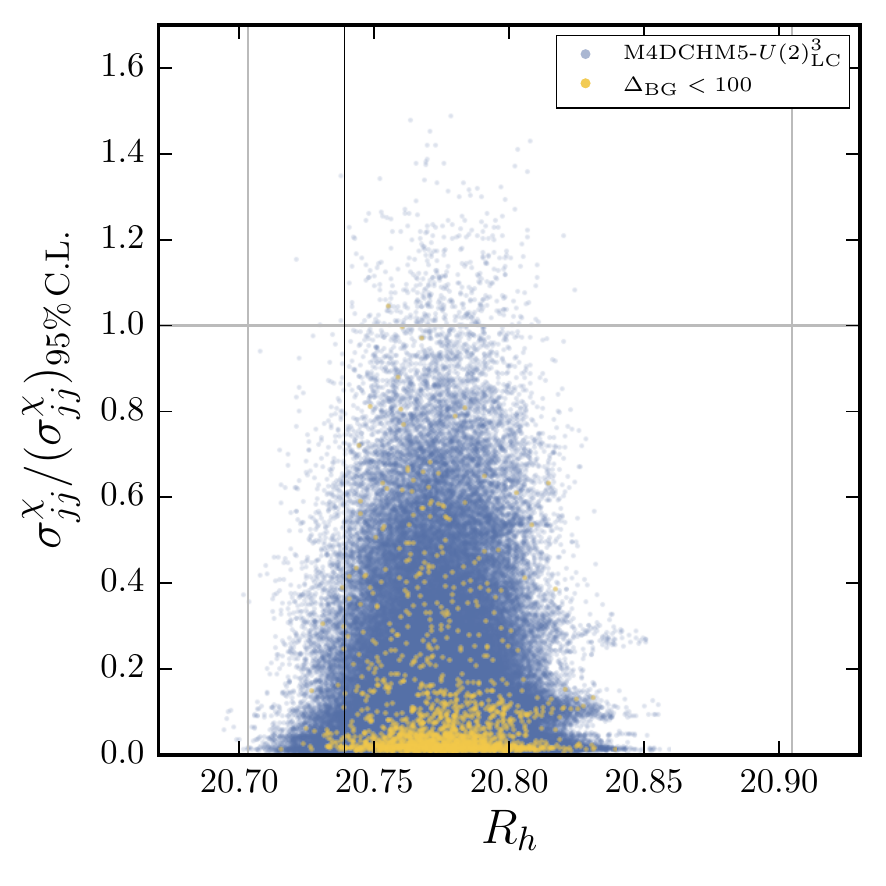}
\caption{Observables sensitive to light quark compositeness in 
$U(2)^3_\text{LC}$. Left: first-row effective
CKM elements. The black line corresponds to the SM limit of a unitary CKM 
matrix. Right: hadronic $Z$ width (normalized to $Z\to e\bar e$) vs.\ the 
$pp\to jj$ cross section in the rapidity bin described in 
appendix~\ref{sec:dijetapp}, normalized 
to the 95\% C.L. limit extracted from the ATLAS analysis. The black line 
corresponds to the central value of the SM prediction.}
\label{fig:u2lc-ckm}
\end{figure}

Compositeness of the first two generation quarks is mainly constrained by
\begin{itemize}
\item First-row CKM unitarity, see section~\ref{sec:ckm};
\item The hadronic $Z$ width, see section~\ref{sec:z};
\item The dijet angular distribution at LHC, see 
section~\ref{sec:ci}.
\end{itemize}
The predictions for these quantities are shown in figure~\ref{fig:u2lc-ckm}.
The left-hand plot shows the effective CKM elements $|V_{us}|$ vs.\ $|V_{ud}|$ 
and demonstrates that large deviations from the SM relation 
$|V_{us}|^2+|V_{ud}^2|\approx 1$ (shown as a black line) are possible.
The solid gray lines show the current experimental $2\sigma$ bounds. At 
$3\sigma$, the points stop because of our procedure described above.
On the
right, we show the predictions for the hadronic $Z$ width as defined in 
\eqref{eq:RZ} (the SM central value is shown as a black line) vs.\ the $pp\to 
jj$ cross section in the bin described in
appendix~\ref{sec:dijetapp}, normalized to 
the 95\% C.L. limit extracted from the ATLAS analysis. This plot shows that 
sizable effects are possible in these observables as well, but almost all 
points lie within the $2\sigma$ region for both observables (shown as solid 
gray lines), demonstrating that 
CKM unitarity is by far the strongest constraint on light quark compositeness 
in $U(2)^3_\text{LC}$ at present.

In these plots (as in almost all the plots of this section), on top of all the 
viable points in blue, we show points that have a fine-tuning measure 
$\Delta_\text{BG}<100$ in yellow. The rationale is to demonstrate in which part 
of the viable space for the observables in question the most ``natural'' 
parameter points lie. We warn the reader however that these points do {\em not} 
correspond to {\em all} viable points with $\Delta_\text{BG}<100$ -- as 
mentioned in section~\ref{sec:tuning}, we have only computed $\Delta_\text{BG}$ 
for a subset of the points. One should also keep in mind that, simply due to 
their smaller number, these points typically cluster in the region with the 
highest point density.

\subsubsection{Higgs production and decay}\label{sec:u2lc-h}

The left-hand plot in figure~\ref{fig:u2lc-h-st} shows the signal strengths of 
the Higgs produced in gluon fusion and decaying to $ZZ$ (which equals the one 
to $WW$ due to custodial 
symmetry), $\gamma\gamma$, and $b\bar b$. Most of the points lie on the curves 
that are expected from analytical considerations of coupling modifications (see 
e.g.\ \cite{Giudice:2007fh}). This is even true for $h\to\gamma\gamma$ and 
$gg\to h$, since 
in pNGB Higgs models, the loop contribution of heavy resonances to these 
processes almost entirely cancels with the coupling modification of the top 
quark, leaving the Higgs non-linearities as the dominant effect 
\cite{Low:2010mr}.

However the plot also shows deviations from these relations. These can be 
understood to be caused by light quark compositeness, spoiling the 
above mentioned cancellation \cite{Delaunay:2013iia}. In this way, the signal 
strength can be closer to (or further away from) their SM values than naively 
expected for small $f$. 
Nevertheless, we find this effect to be mild, given the strong constraints on 
light quark compositeness discussed in the previous section.\footnote{%
The fact that almost all points lie on the same curve and only a few individual 
Markov chains have strayed away from it explains the frayed appearance of the 
plot.}

Concerning the $h\to b\bar b$ signal strength, we note that the figure shows 
the signal strength in the case of gluon fusion production, while the 
experimental bounds are currently based on the associated production with 
vector bosons, that we do not include in our analysis.

\subsubsection{Oblique parameters}\label{sec:u2lc-st}

\begin{figure}[tbp]
\centering
\includegraphics[width=0.48\textwidth]{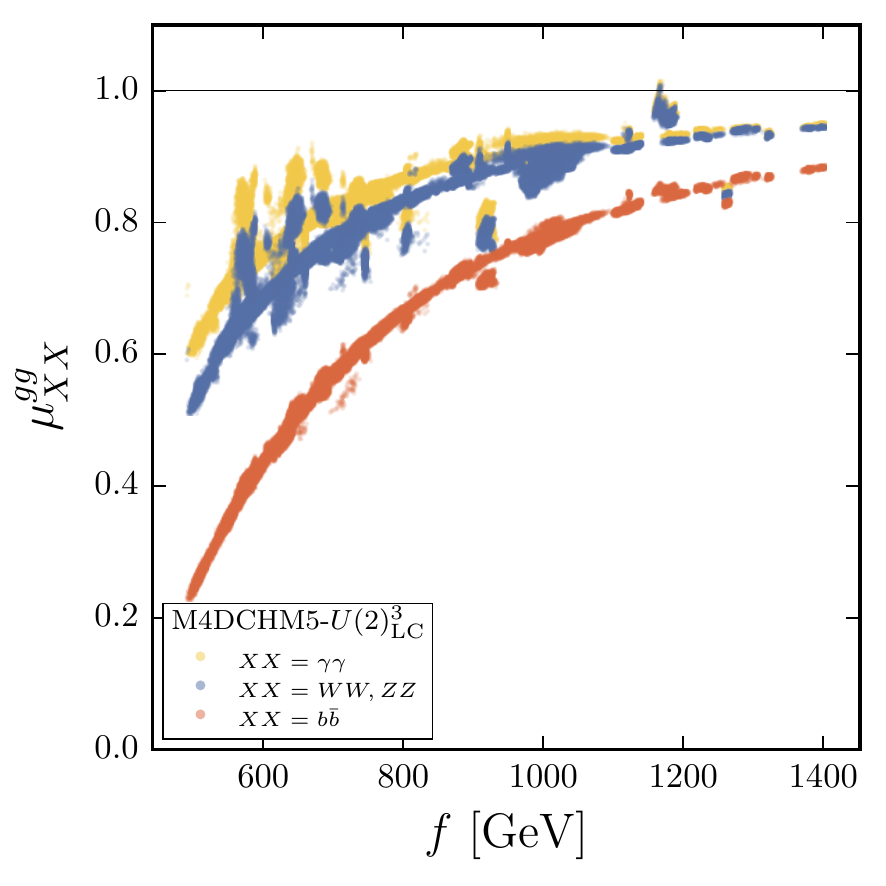}
\includegraphics[width=0.48\textwidth]{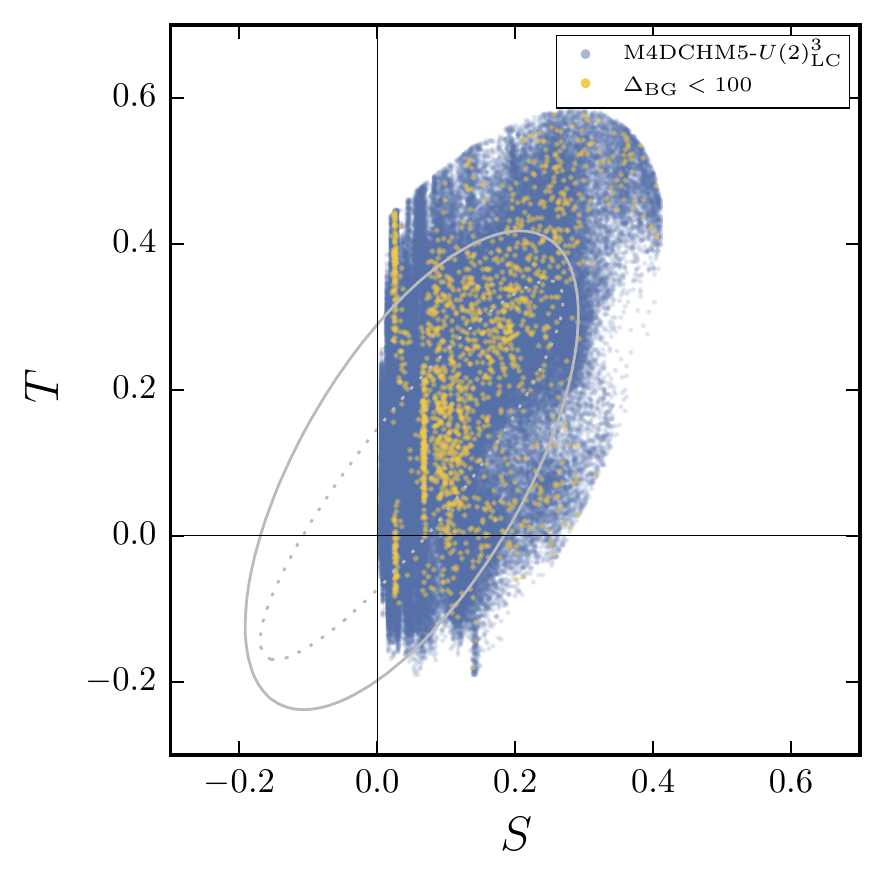}
\caption{%
Left: Higgs signal strength for gluon fusion production and decay to final 
states $ZZ$ (equal to $WW$ by custodial symmetry),  $\gamma\gamma$, 
and $b\bar b$ in $U(2)^3_\text{LC}$. The SM corresponds to $\mu=1$, shown as a 
horizontal line.
Right: Oblique 
parameters $S$ and $T$ in $U(2)^3_\text{LC}$, defined to be 0 in the SM.}
\label{fig:u2lc-h-st}
\end{figure}

The right-hand plot in figure~\ref{fig:u2lc-h-st} shows the predictions for the 
$S$ and $T$ parameters.
We show the region allowed by experiment at $2\sigma$ as a gray dashed ellipse, 
while the gray solid ellipse takes into account also the additional theory 
uncertainty discussed in section~\ref{sec:ST}.
The tree-level contribution to $S$ is strictly 
positive, while the fermionic loop contribution to $T$ can have either sign, 
but 
is preferred to be positive by experiment for positive values of $S$ and indeed 
large positive contributions are possible for our choice of fermion 
representations, which is important as it helps to alleviate the bound from $S$.

\subsubsection{Meson-antimeson mixing}\label{sec:u2lc-df2}

\begin{figure}[tbp]
\centering
\includegraphics[width=0.45\textwidth]{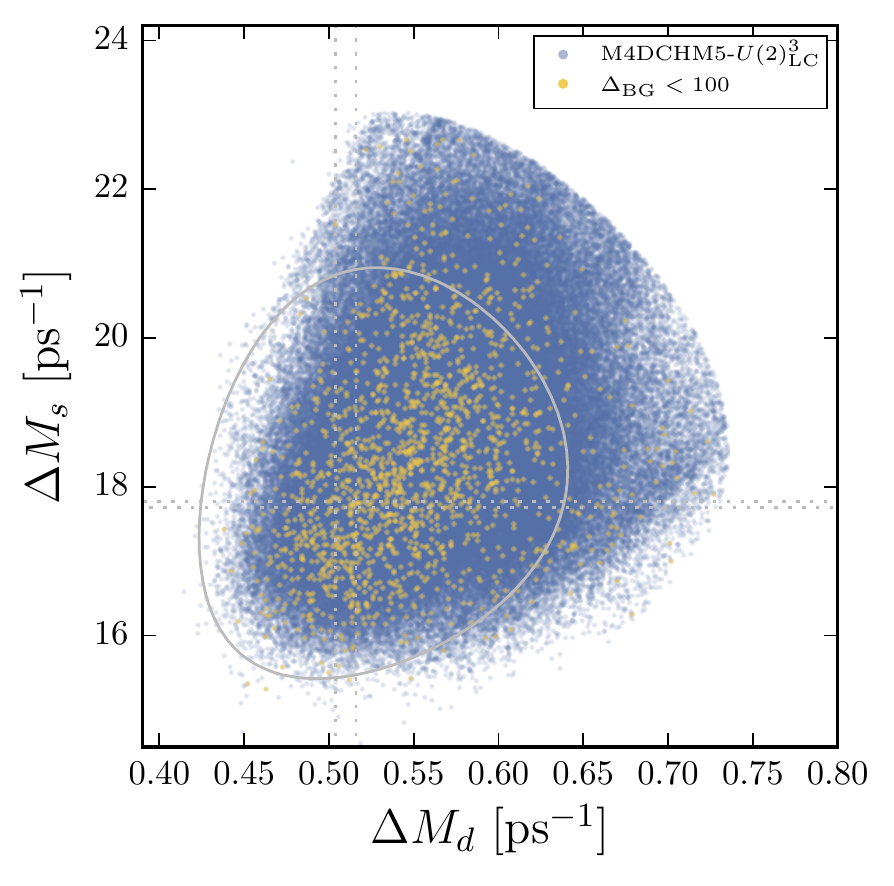}
\includegraphics[width=0.45\textwidth]{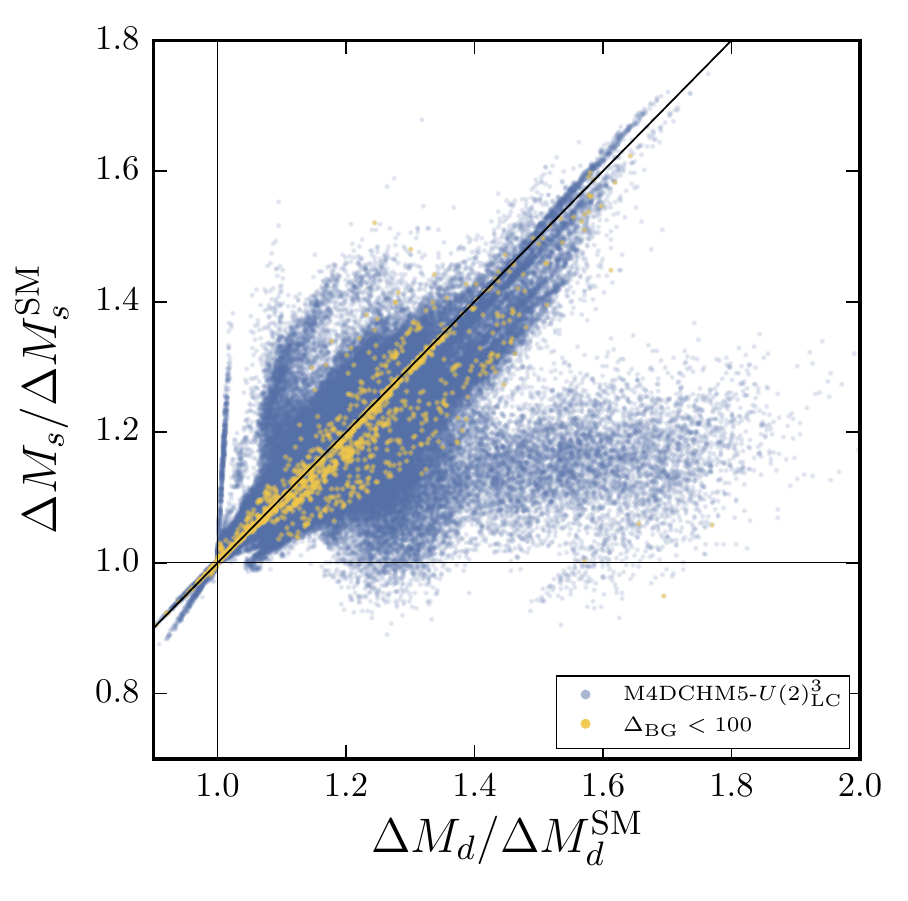}

\includegraphics[width=0.45\textwidth]{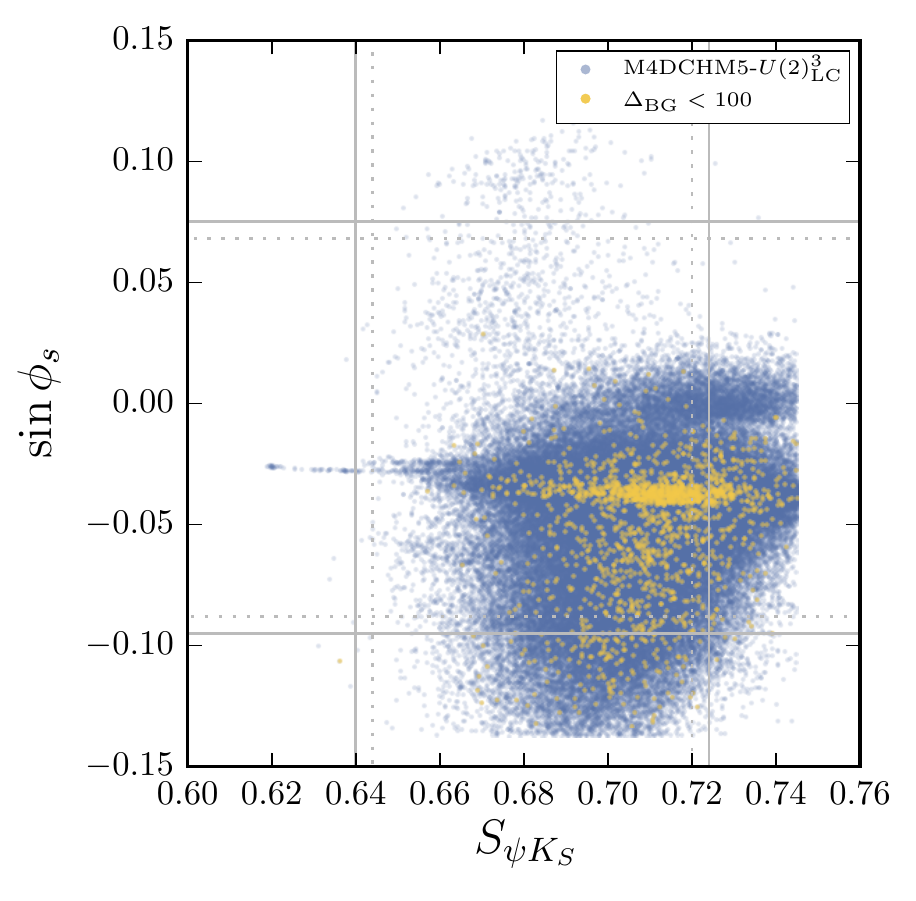}
\includegraphics[width=0.45\textwidth]{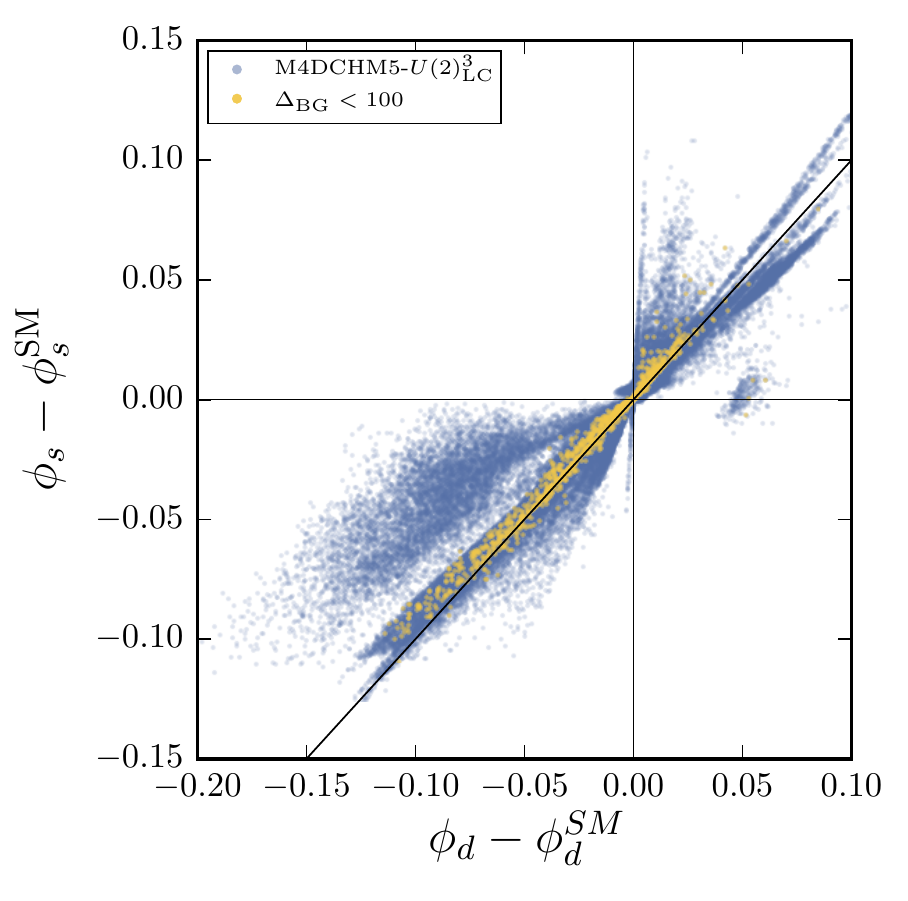}

\includegraphics[width=0.45\textwidth]{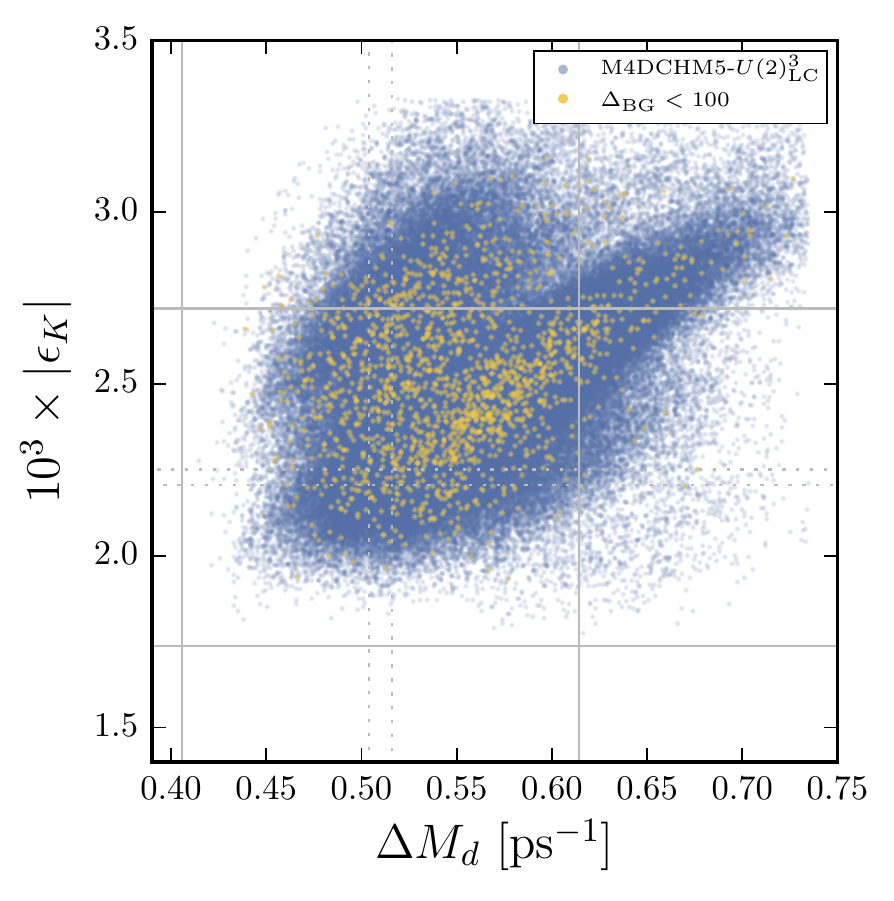}
\includegraphics[width=0.45\textwidth]{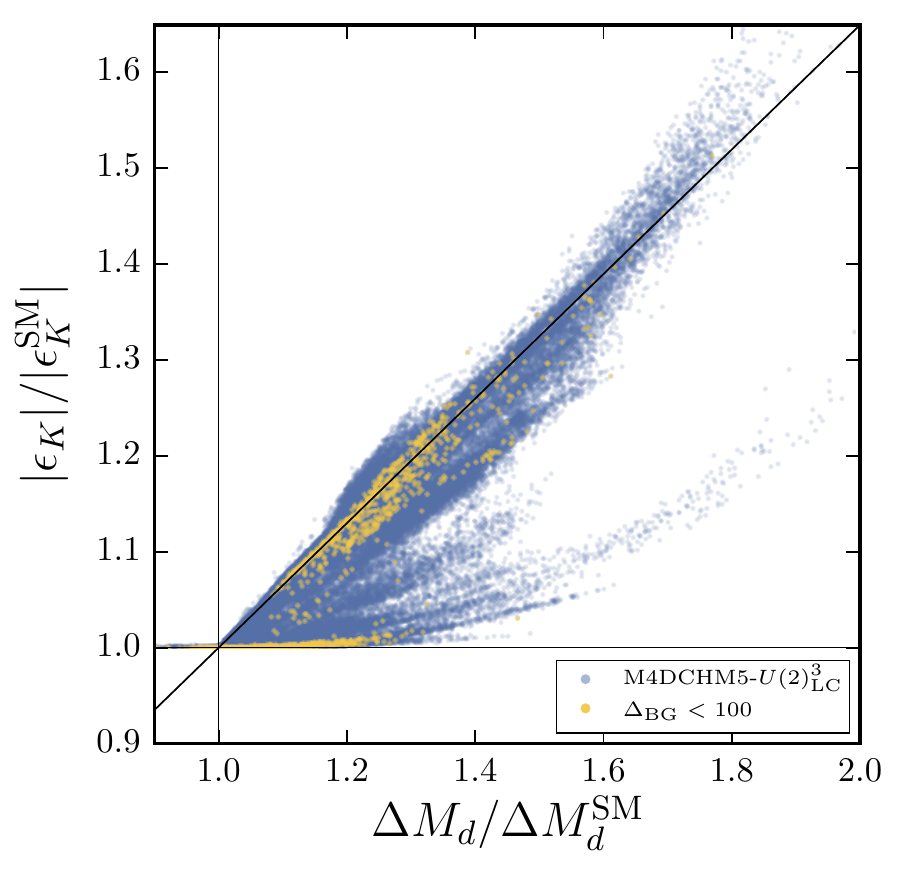}
\caption{$\Delta F=2$ observables in $U(2)^3_\text{LC}$.}
\label{fig:u2lc-df2}
\end{figure}

Figure~\ref{fig:u2lc-df2} shows the predictions for $\Delta F=2$ observables in 
$U(2)^3_\text{LC}$. In the left-hand column, we directly show the correlation 
between observables. In this case, it is important to notice that the CKM 
parameters themselves are varied in our fit and are not fixed to their SM
central values (as is often done in parameter scans of, e.g., SUSY models). As 
a consequence, any correlations among (tree-level) NP contributions are washed 
out by the spread in the allowed values for the CKM parameters.
The dashed gray lines in these plots show the allowed regions with merely the
experimental $2\sigma$ uncertainties, while the solid gray contours take into 
account additionally the (correlated) theory uncertainties at $2\sigma$. As 
discussed in section~\ref{sec:df2}, we only need to account for the non-CKM 
(i.e., lattice) theory uncertainties, as for a given point, the CKM parameters 
are predictions.

While these plots are more directly related to the experimental measurements, 
the variation of CKM parameters obscures the relation to the NP contributions. 
This is why
in the right-hand column of figure~\ref{fig:u2lc-df2} we show the ratios (or 
phase differences) of the observables and the SM $W$-loop contribution for each 
parameter point. In this way, the correlations valid at leading order in the 
$U(2)^3$ spurion expansion,
$\Delta M_d/\Delta M_d^\text{SM} = \Delta M_s/\Delta M_s^\text{SM}$ and
$\phi_s-\phi_s^\text{SM}=\phi_d-\phi_d^\text{SM}$ (where $S_{\psi 
K_S}=\sin\phi_d$), shown by solid black lines in the plots, become apparent.
In the bottom right plot, the black line corresponds to the MFV limit of equal 
relative modification in the $B_d$ and $K^0$ mixing amplitudes, while no such 
correlation is expected in $U(2)^3$.
We make the following observations.
\begin{itemize}
\item The mass differences in the $B_d$ and $B_s$ systems can receive 
corrections up to $+60\%$, but negative NP contributions are disfavoured. This 
can be understood from the fact that the tree-level Wilson coefficient of 
$Q_{VLL}^{d_ib}$, cf.\  \eqref{eq:cvll}, involves the square of a coupling that 
carries a small phase.
\item The $B_s$ mixing phase can saturate the experimental lower bound, but 
positive values for $\sin\phi_s$ are only predicted for a small number of 
points. This is due to the 
correlation with $\phi_d$ and the preference for a negative NP contribution to 
the latter, that is also visible in global CKM fits 
\cite{Barbieri:2014tja,Bevan:2014cya,Charles:2015gya}.
\item Both for the mass differences and for the phases in the $B_d$ and $B_s$ 
systems, the leading-order 
$U(2)^3$ correlations are broken for a significant fraction of the points, seen 
as a deviation from the black diagonal lines. This is due to non-negligible 
contributions from left-right operators. We have identified two reasons for why 
these effects are larger than expected from a general EFT analysis 
\cite{Barbieri:2012uh}.
\begin{enumerate}
 \item The Wilson coefficients of these operators are RG-enhanced;
 \item Due to partial compositeness and the possibility to have a hierarchy 
even among the (diagonal) left-handed composite-elementary mixings, the spurion 
hierarchies in the right-handed mixings can be milder than the Yukawa 
hierarchies, effectively enhancing subleading terms in the spurion expansion.
\end{enumerate}
A similar effect has already been noted in the context of the MSSM with a 
$U(2)^3$ symmetry \cite{Barbieri:2014tja} (where it was mostly due to an 
accidental enhancement of a loop function) and we find the effect to be even 
more 
pronounced in the composite Higgs case. We stress nevertheless that the 
majority of parameter points does fulfill the $U(2)^3$ relations to a good 
precision, corresponding to a large density of points around the black lines in 
the plots.
\item The relative modification of $\epsilon_K$ compared to the SM is always 
equal\footnote{With a small correction factor stemming from the SM charm 
contribution.} to or smaller than the relative modification of $\Delta M_d$. 
This confirms the general expectation for $U(2)^3_\text{LC}$ in 
\cite{Barbieri:2012tu}.
\end{itemize}

So far, we have not discussed $D^0$-$\bar D^0$ mixing. On the one hand, the 
$D^0$ system is plagued by large theoretical uncertainties due to poorly known 
long-distance contributions; on the other, the effects in $U(2)^3$ models are 
expected to be small on general grounds \cite{Barbieri:2012uh}. To 
investigate whether this expectation is correct, we have computed the 
tree-level NP contribution to the $D^0$ mixing amplitude in $U(2)^3_\text{LC}$. 
Since the SM contribution is expected to be real to a good accuracy, the most 
promising NP effect would be a CP violating one. A global fit to data from the 
$D$ system \cite{Bevan:2014tha} allows to directly constrain the absolute value 
and the phase of the mixing amplitude. At $2\sigma$, this constrains the 
imaginary part of the mixing amplitude to be
\begin{equation}
-0.5~\text{ns}^{-1} \gtrsim \text{Im}\,M_{12}^D \lesssim 1.6~\text{ns}^{-1} \,.
\end{equation}
Numerically, we have found that the NP contributions to $\text{Im}\,M_{12}^D$ 
are always negative in $U(2)^3_\text{LC}$ and can reach at most 
$-0.5~\text{ns}^{-1}$. We conclude that CP violation in $D^0$ mixing is 
currently not a relevant constraint on the model, but future improvements of 
the bound by factors of a few would start to cut into its parameter space.
We have further found that the NP contributions to $\text{Im}\,M_{12}^D$ are 
strongly anticorrelated with $\epsilon_K$: sizable NP contributions to the 
former never occur simultaneously with sizable NP contributions to the latter. 
However, the NP contributions to both observables can be small simultaneously.

\subsubsection{Rare $B$ decays}\label{sec:u2lc-df1}

The Wilson coefficient $C_7^{bs}$ of the electromagnetic dipole operator, cf. 
sec.~\ref{sec:df1}, receives NP contributions, but only to the extent that is 
allowed by the strong constraint from the branching ratio of $B\to X_s\gamma$. 
We find these contributions to be aligned in phase with the SM to a high 
degree, such that CP violating effects, e.g.\ in the direct CP asymmetry in 
$B\to K^*\gamma$, are expected to be small. Contributions to the 
chirality-flipped coefficient $C_7^{\prime bs}$ are small by $U(2)^3$ symmetry.

\begin{figure}[tbp]
\centering
\includegraphics[width=0.48\textwidth]{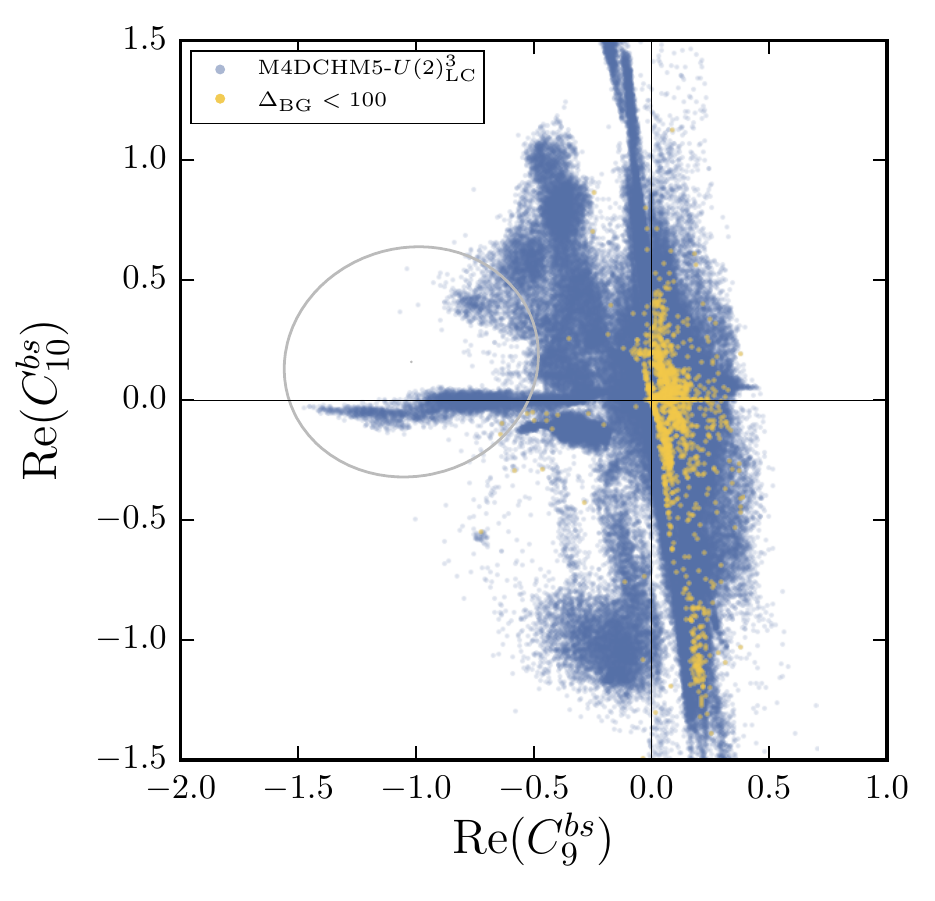}
\includegraphics[width=0.48\textwidth]{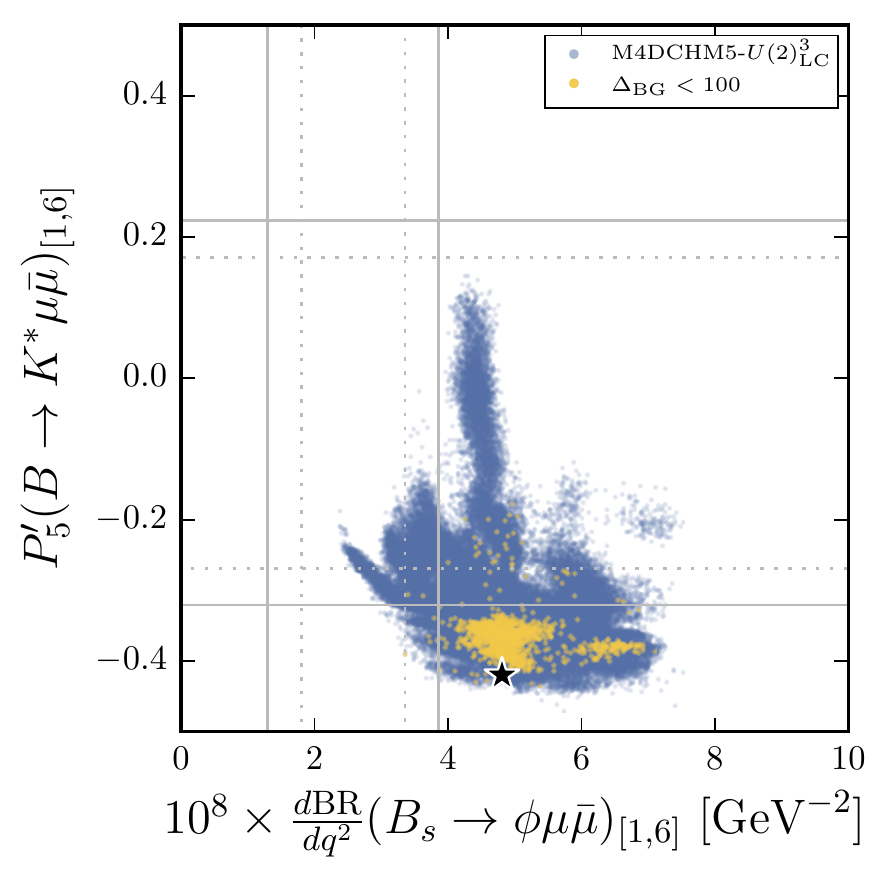}
\caption{%
Left: new physics contributions to the Wilson coefficients $C_{9}^{bs}$ and 
$C_{10}^{bs}$ in $U(2)^3_\text{LC}$.
Right: predictions for the angular observable $P_5'$ in $B\to 
K^*\mu^+\mu^-$ and the branching ratio of $B_s\to\phi\mu^+\mu^-$, both in the 
low-$q^2$ bin from 1 to 6~GeV$^2$. The star corresponds to the central values 
of the SM predictions.}
\label{fig:u2lc-c910}
\end{figure}

The most interesting effects in rare $B$ decays stem from the tree-level 
contributions to the semi-leptonic Wilson coefficients $C_9^{bs}$ and 
$C_{10}^{bs}$. As discussed in sec.~\ref{sec:df1}, there are $Z$-mediated and 
resonance-mediated effects that dominantly contribute to $C_{10}^{bs}$, but 
also resonance-mediated effects that contribute only to $C_9^{bs}$. We remind 
the reader that in our numerical analysis, the only observable sensitive to 
these Wilson coefficients that we have imposed as a constraint is the branching 
ratio of $B_s\to\mu^+\mu^-$, essentially limiting the absolute value of 
$C_{10}^{bs}$. All points passing this constraint are shown in the left-hand 
plot of fig.~\ref{fig:u2lc-c910}. We observe that large NP effects in 
$C_{10}^{bs}$ -- saturating the experimental bound on $B_s\to\mu^+\mu^-$ --  
are 
possible, but also sizable effects in $C_{9}^{bs}$. Interestingly, the largest 
effects allowed in $C_{9}^{bs}$ correspond to a negative sign that is preferred 
by the anomalies in $B\to K^*\mu^+\mu^-$ angular observables discussed in 
sec.~\ref{sec:df1}.
The gray ellipse in figure~\ref{fig:u2lc-c910} left corresponds to the 
$2\sigma$ preferred 
region in a global fit to $b\to s\mu^+\mu^-$ observables 
\cite{Altmannshofer:2014rta}, which shows a clear 
tension with the SM point $(0,0)$. The figure clearly shows that 
if these tensions are due to NP, the M4DCHM5-$U(2)^3_\text{LC}$ can 
explain them.
This is also demonstrated by the right-hand plot in fig.~\ref{fig:u2lc-c910}, 
which shows the predictions for two of the observables that currently show the 
biggest tensions with the SM, namely the low-$q^2$ branching ratio of 
$B_s\to\phi\mu^+\mu^-$  and the angular observable $P_5'$ in $B\to 
K^*\mu^+\mu^-$. In this plot, the black star shows the central value of the SM 
predictions (taken from \cite{Altmannshofer:2014rta,Straub:2015ica}), the gray 
dashed line the values allowed at $2\sigma$ by experiment, and the gray solid 
lines the $2\sigma$ allowed values taking into account also the theoretical 
uncertainties.

We have found that all of the points that have $C_{9}^{bs}\lesssim-0.5$ --
and could thus account for the tensions in angular observables and branching 
ratios --
correspond to a small value (between 1 and 2) of the composite coupling $g_X$ 
and a correspondingly small mass (below 1~TeV) of the mass eigenstate that 
is 
dominantly the $\rho_X$ resonance. This can be understood from the discussion 
in 
section~\ref{sec:df1}: in the limit $g_X\ll g_\rho$, the ``KK photon''-like 
state is dominantly the field $\rho_X$. In addition, since $g_X$ is not much 
larger 
than the elementary gauge coupling, the parametric suppression of the 
resonance-mediated contribution is lifted.
It is also important to notice that this linear combination of gauge fields 
does not contribute to the $S$ parameter and thus is allowed to be lighter than 
the other vector resonances.
Sizable contributions to $B_s$-$\bar B_s$ mixing are also generated by the 
exchange of the light resonance, but we find that the shift in $\Delta M_s$ 
is below 20\% relative to the SM.
We have also computed the LHC production cross section and decay branching 
ratios of the light resonance for the points with sizable NP effects in $C_9$. 
For most of the points, the dominant decay mode is $t\bar t$ and the 
cross-section is just below the ATLAS and CMS searches for resonances in this 
mode that we have imposed as a constraint in our scan (see 
section~\ref{sec:spin1}). The width of the resonance is small enough to show up 
in a ``bump hunt''. Prospects for vector resonances will be discussed in more 
detail in section~\ref{sec:prospects-vector} below.

\begin{figure}[tbp]
\centering
\includegraphics[width=0.48\textwidth]{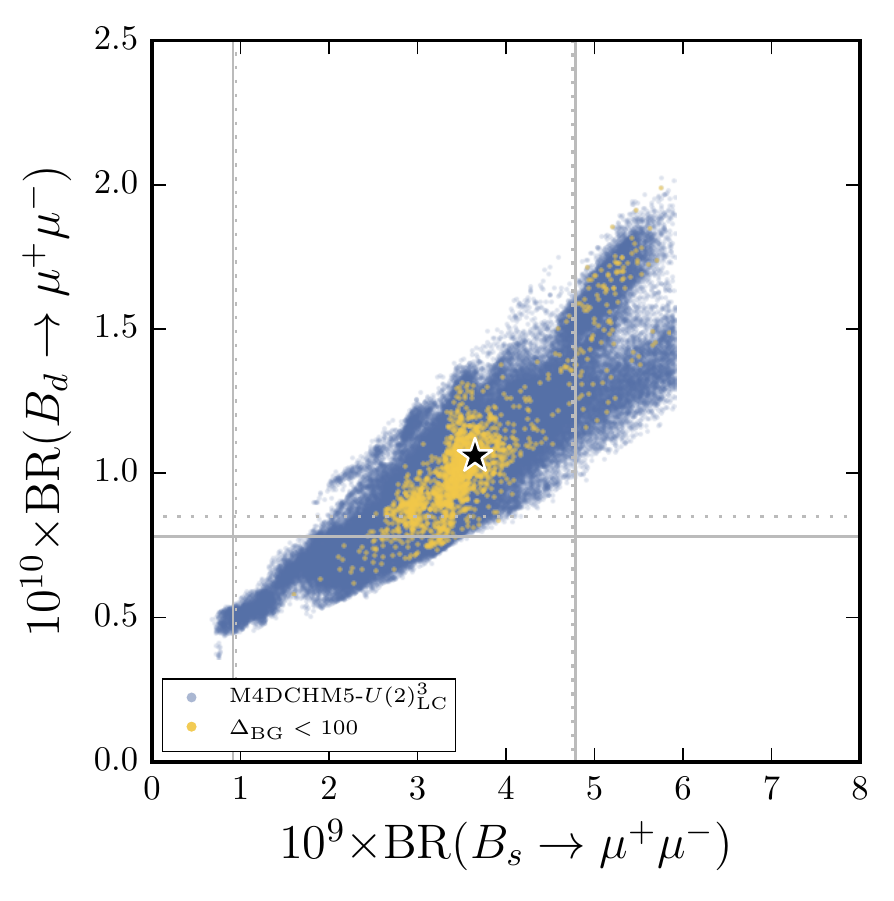}
\includegraphics[width=0.48\textwidth]{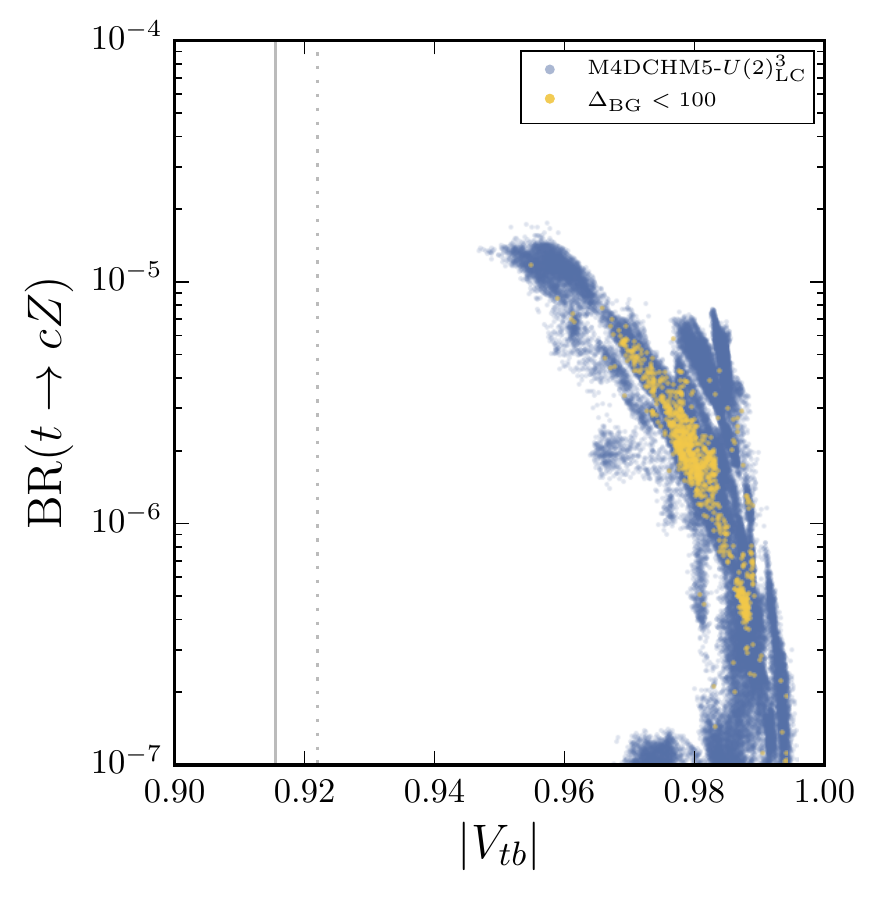}
\caption{%
Left:
predictions for the branching ratios of
$B_s\to\mu^+\mu^-$ and $B_d\to\mu^+\mu^-$ in $U(2)^3_\text{LC}$.
Right:
predictions for the deviation of the effective CKM element $V_{tb}$ from 1 vs.\ 
the branching ratio of the FCNC top decay $t\to cZ$ in $U(2)^3_\text{LC}$.}
\label{fig:u2lc-bmumu-top}
\end{figure}

In the left-hand plot of fig.~\ref{fig:u2lc-bmumu-top}, we finally show the 
predictions for the correlation between the branching ratios of 
$B_s\to\mu^+\mu^-$ and $B_d\to\mu^+\mu^-$, which is fixed by $U(2)^3$ to be 
equal relative to the respective SM predictions (but is again slightly washed 
out by the variation in CKM elements). The current $3\sigma$ upper bound on 
$B_s\to\mu^+\mu^-$ can be saturated, but also a significant 
suppression can occur. This is in contrast to, e.g., the Littlest Higgs model 
with T-parity, where this branching ratio can only be enhanced with respect to 
the SM \cite{Blanke:2015wba}.

\subsubsection{Top decays}

A significant degree of compositeness of the left-handed top quark can lead to  
a reduction of the single top production cross section at LHC, corresponding to 
a reduced value for the effective CKM element $V_{tb}$ as discussed in
section~\ref{sec:ckm}. In addition, in this case there can be sizable 
flavour-changing couplings of the top quark to the $Z$ boson, since the 
left-handed couplings are not custodially protected, in contrast to the 
right-handed ones.

These two effects manifest themselves in a correlation between the deviation of 
$V_{tb}$ from 1 and the branching ratio of the FCNC top decay $t\to cZ$ as 
shown in the right-hand plot of figure~\ref{fig:u2lc-bmumu-top}.
Both effects are quite moderate after imposing all 
the bounds. The deviation in $V_{tb}$ is always within the current $2\sigma$ 
experimental constraint and percent-level experimental accuracy will be 
necessary to find a significant deviation. The branching ratio of the FCNC top 
decay can reach at most $10^{-5}$, which will be challenging to see at the LHC 
\cite{Agashe:2013hma,Azatov:2014lha}.

\subsubsection{Other processes}\label{sec:u2lc-other}

So far, we have not discussed rare $K$ decays. While these processes are 
important constraints on many NP models, we find the effects in 
$U(2)^3_\text{LC}$ to be rather small. For instance, the branching ratios of 
$K^+\to \pi^+\nu\bar\nu$ and $K_L\to \pi^0\nu\bar\nu$ are modified by at most 
$\pm20\%$ with respect to the SM (and are perfectly correlated due to $U(2)^3$). 
The short-distance contribution to the branching ratio of $K_L\to\mu^+\mu^-$ is 
always below $2\times10^{-9}$.

\subsection{Right-handed compositeness: indirect searches}\label{sec:unrc}

In contrast to $U(3)^3_\text{LC}$, we do find a viable parameter space for the
$U(3)^3_\text{RC}$ model. Since $U(3)^3_\text{RC}$ is a limiting case of the 
more general $U(2)^3_\text{RC}$ (the limit in which the composite sector mass 
parameters and the right-handed composite-elementary mixings for the first two 
and the third generation coincide), it is natural to discuss them together.
We will proceed as in the case of left-handed compositeness in 
section~\ref{sec:unlc}.

\subsubsection{Light quark compositeness}

\begin{figure}[tbp]
\centering
\includegraphics[width=0.48\textwidth]{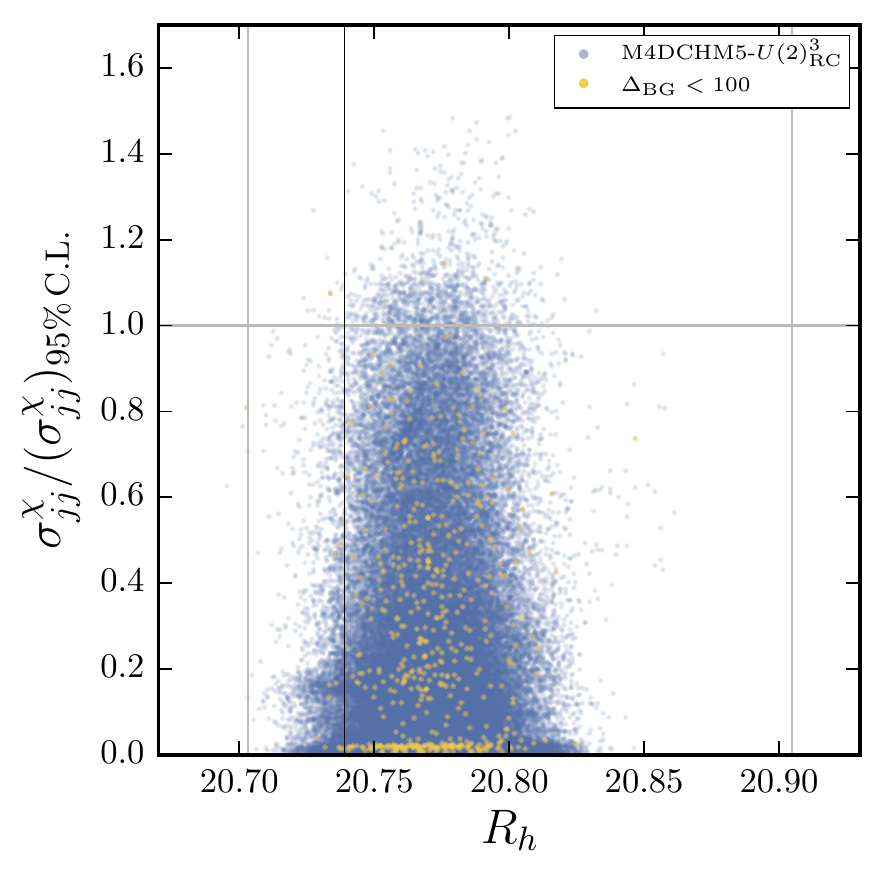}
\includegraphics[width=0.48\textwidth]{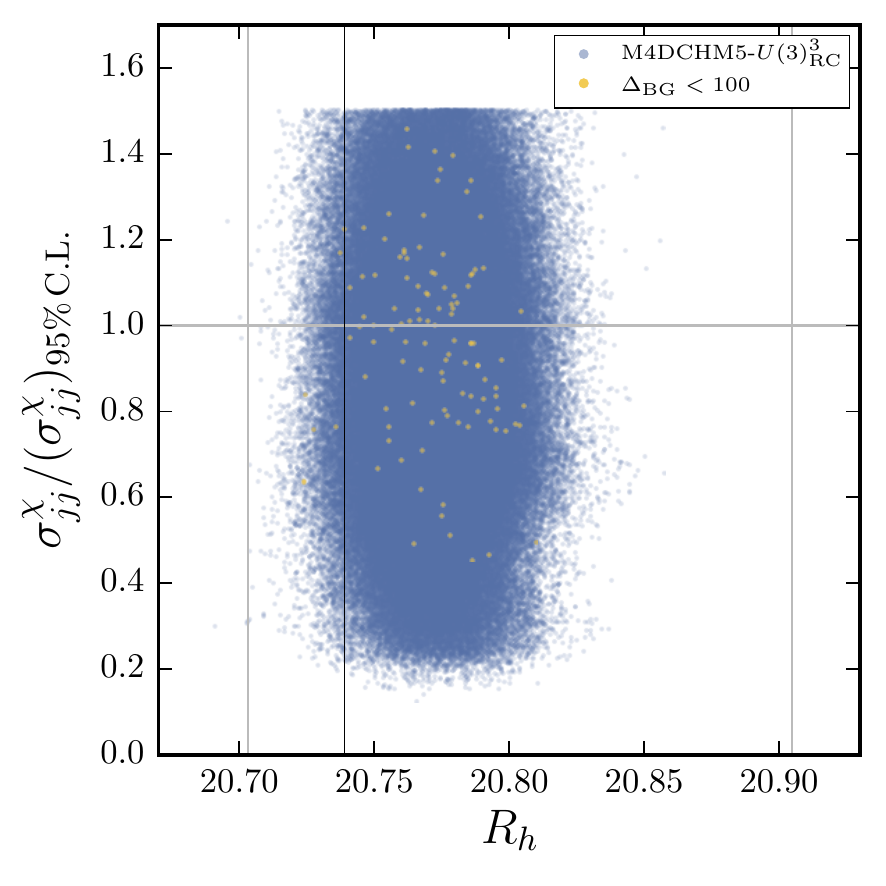}
\caption{Hadronic $Z$ width (normalized to $Z\to e\bar e$) vs.\ the 
$pp\to jj$ cross section in the rapidity bin described in the text, normalized 
to the 95\% C.L. limit extracted from the ATLAS analysis. The black line 
corresponds to the central value of the SM prediction.
Left: $U(2)^3_\text{RC}$, right: $U(3)^3_\text{RC}$.}
\label{fig:unrc-jj}
\end{figure}

In the case of right-handed compositeness, it is typically the right-handed 
light quarks that can carry a sizable degree of compositeness. Consequently, in 
contrast to left-handed compositeness, first-row CKM unitarity does not 
constitute a relevant constraint and the main constraint is given by the 
hadronic $Z$ width and the dijet angular distribution.
The predictions for these quantities are shown in figure~\ref{fig:unrc-jj} 
that is the analogue of figure~\ref{fig:u2lc-ckm} right. We make the following 
observations.
\begin{itemize}
\item In both models, $R_h$ is within the $2\sigma$ bounds for almost all 
the points.
\item In both models\footnote{The alert reader may have noticed that in 
$U(3)^3_\text{RC}$, many points saturate the experimental upper bound, while 
in $U(2)^3_\text{RC}$, this does not seem to be the case, even though we have 
stated that $U(3)^3_\text{RC}$ is a subset of the $U(2)^3_\text{RC}$ model. 
The reason is a volume effect in the high-dimensional parameter space: one 
would need a huge number of points in the $U(2)^3_\text{RC}$ model to get a 
reasonable coverage of the subspace corresponding to $U(3)^3_\text{RC}$. This 
effect is visible in many of the plots in this section and justifies the 
separate analysis of $U(3)^3_\text{RC}$.}, large effects relative to the 
experimental constraints are obtained in the dijet angular distribution. This 
is 
the strongest bound on light-quark compositeness in the right-handed 
compositeness models.
\item In the case of $U(3)^3_\text{RC}$, we see that there is even a {\em 
lower} bound on the modification of the dijet angular distribution. Improved 
experimental measurements in the future could help to disfavour this scenario.
\end{itemize}

\subsubsection{Higgs production and decay}

\begin{figure}[tbp]
\centering
\includegraphics[width=0.48\textwidth]{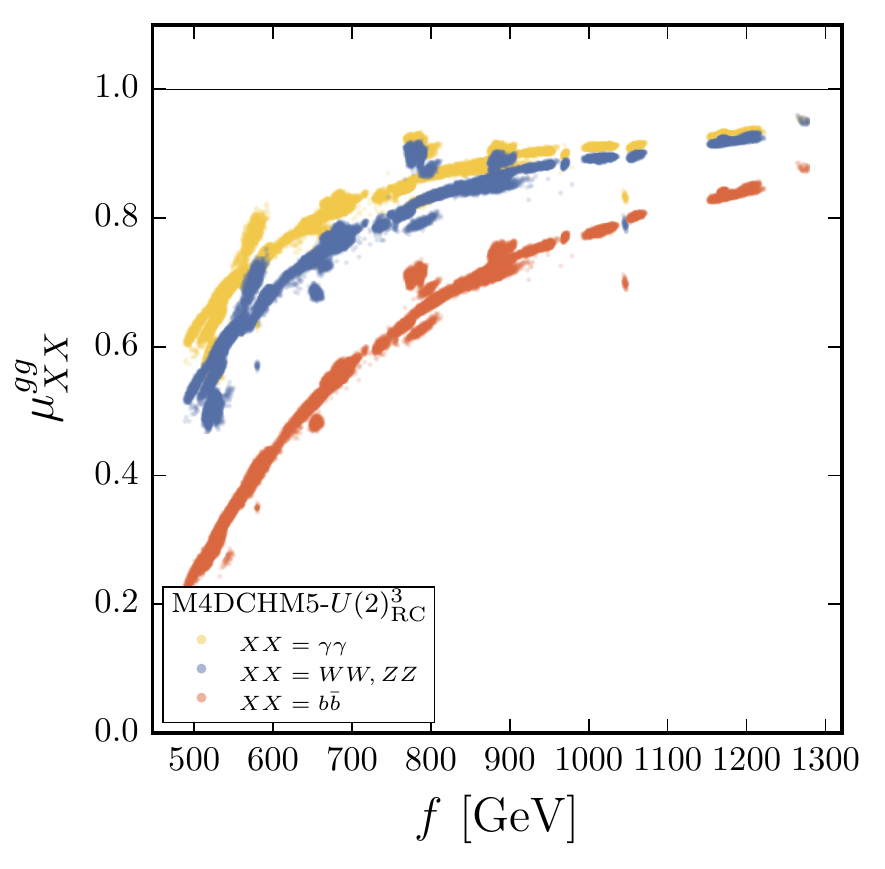}
\includegraphics[width=0.48\textwidth]{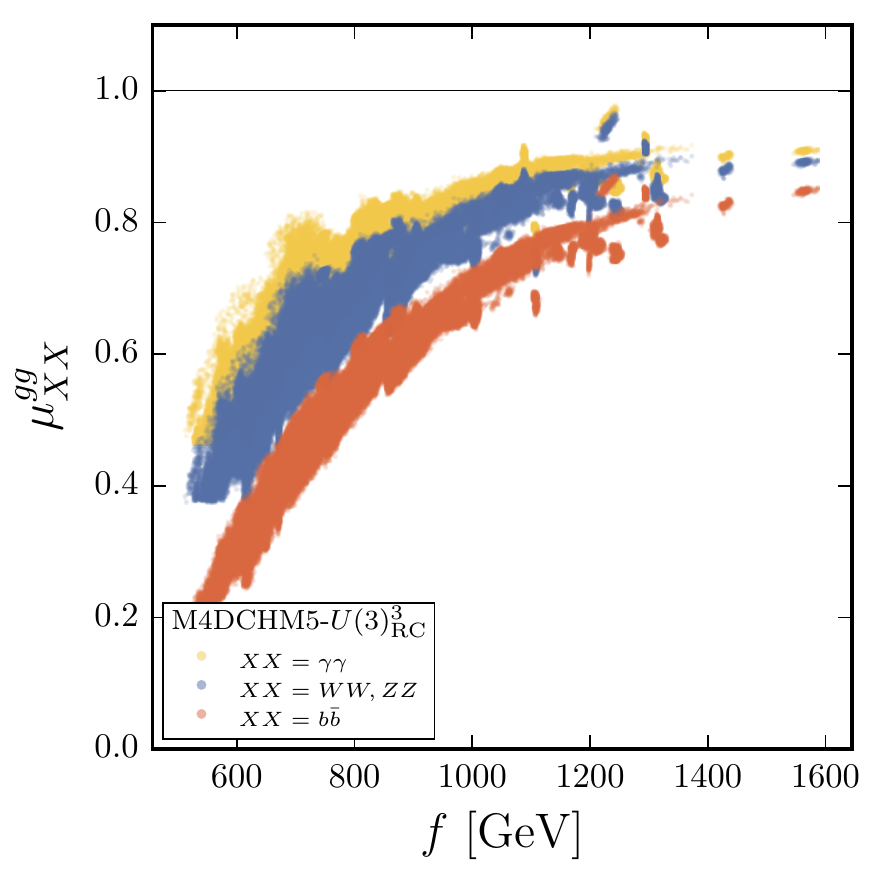}
\caption{%
Higgs signal strength for gluon fusion production and decay to final 
states $ZZ$ (equal to $WW$ by custodial symmetry),  $\gamma\gamma$, 
and $b\bar b$ in $U(2)^3_\text{RC}$ (left) and $U(3)^3_\text{RC}$ (right). 
The SM corresponds to $\mu=1$, shown as a horizontal line.}
\label{fig:unrc-h}
\end{figure}

Figure~\ref{fig:unrc-h} shows the Higgs signal strengths for right-handed 
compositeness in analogy to figure~\ref{fig:u2lc-h-st}. As discussed in 
section~\ref{sec:u2lc-h}, the leading dependence on $f$ is modified by light 
quark compositeness, which is more pronounced in $U(3)^3_\text{RC}$ due to the 
requirement to have a large degree of compositeness for all right-handed 
up-type quarks.

\subsubsection{Meson-antimeson mixing}

\begin{figure}[tbp]
\centering
\includegraphics[width=0.45\textwidth]{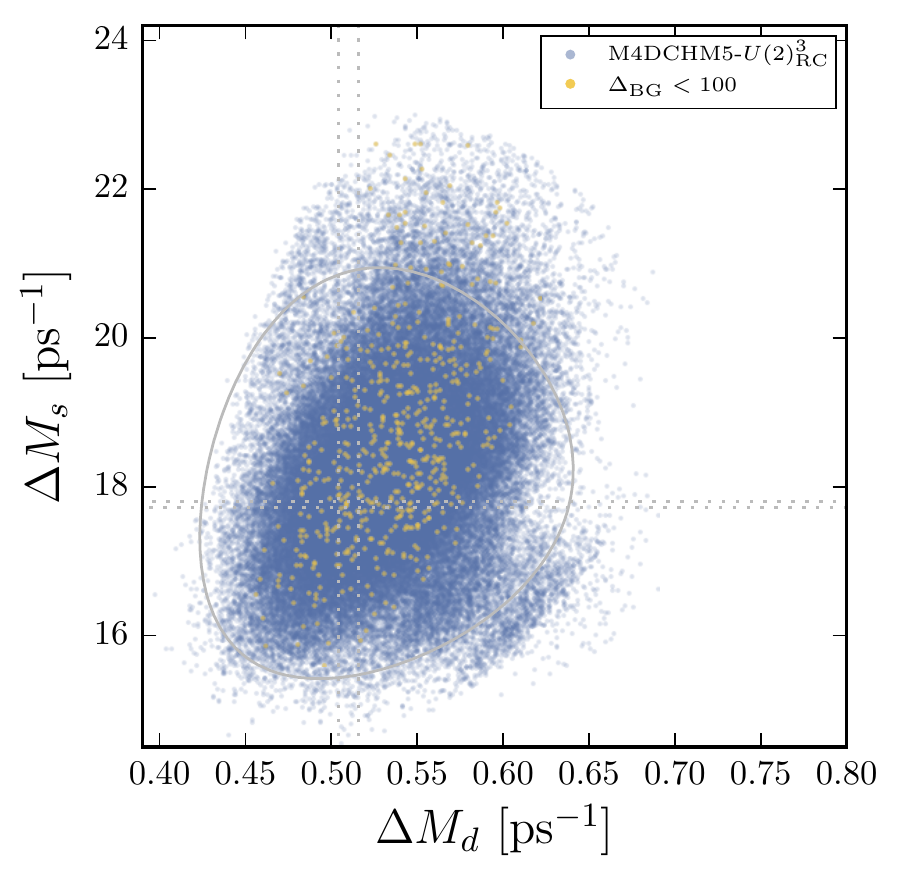}
\includegraphics[width=0.45\textwidth]{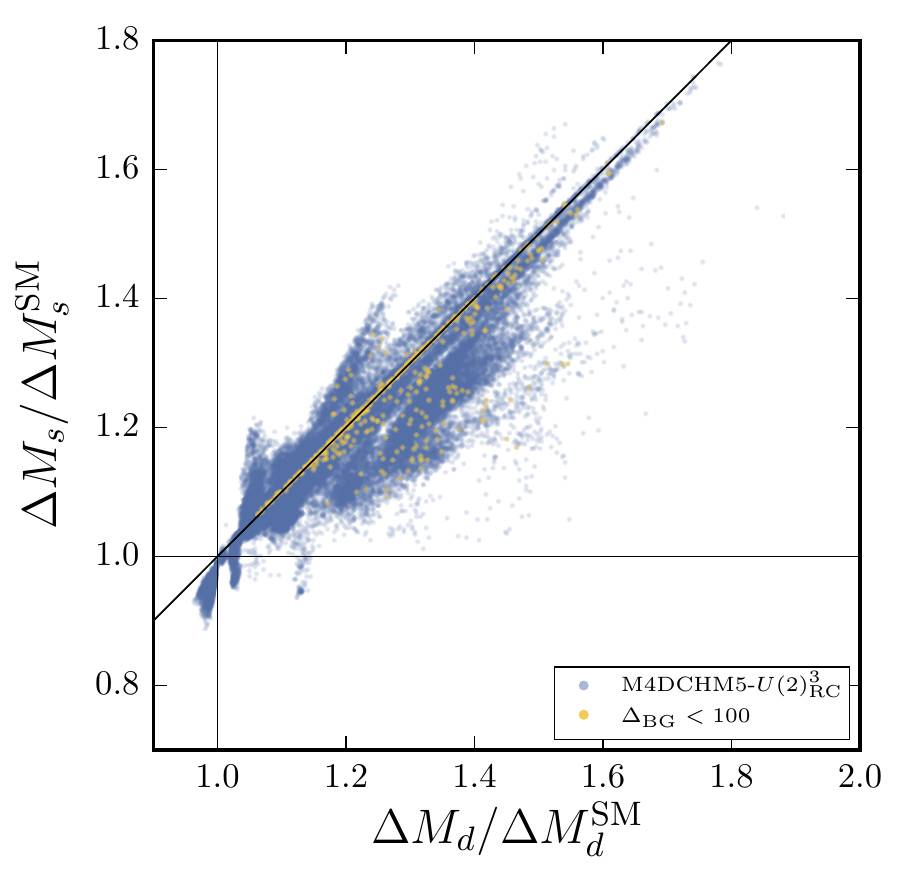}

\includegraphics[width=0.45\textwidth]{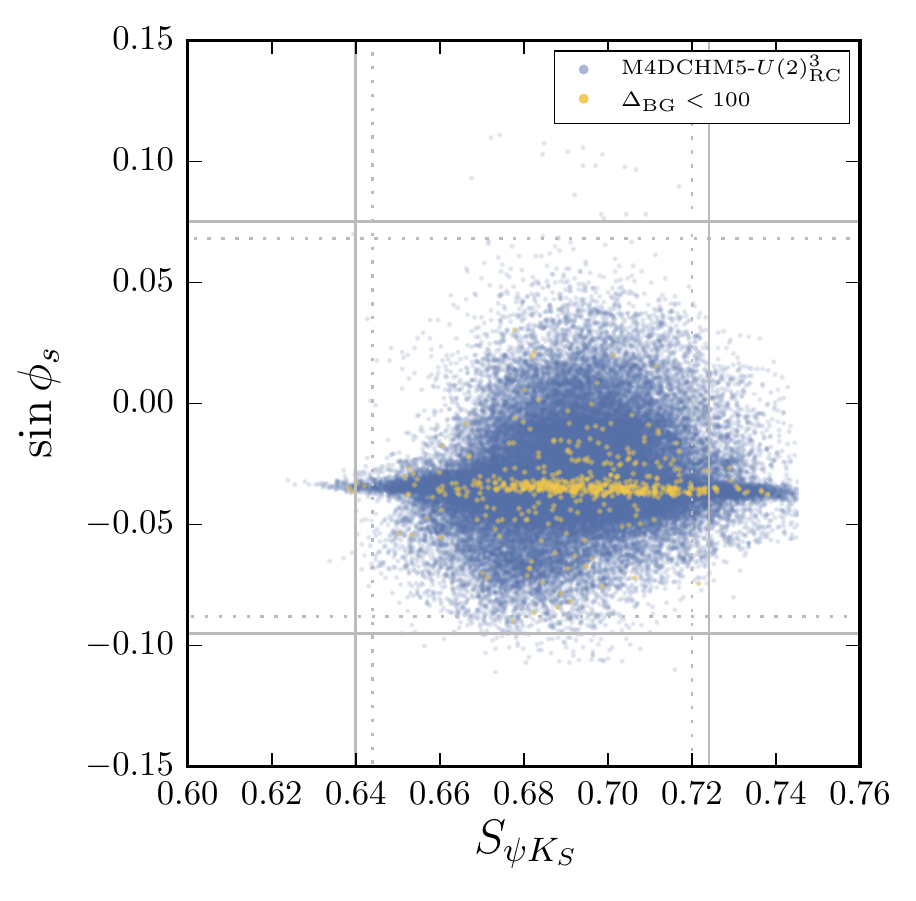}
\includegraphics[width=0.45\textwidth]{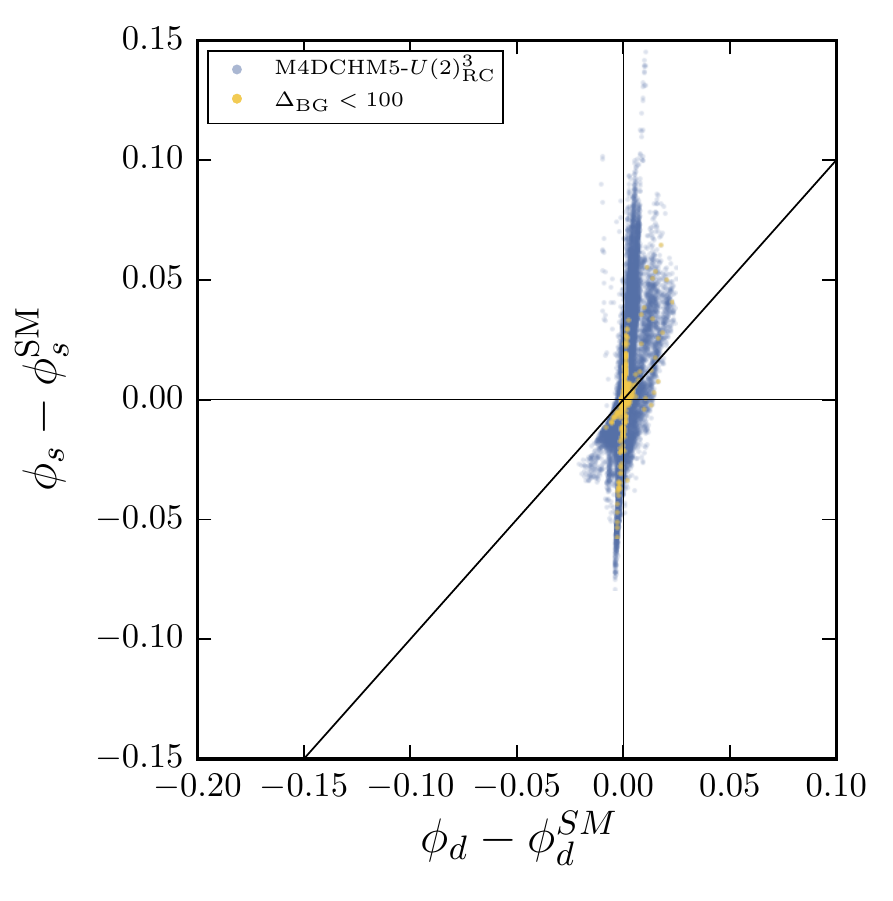}

\includegraphics[width=0.45\textwidth]{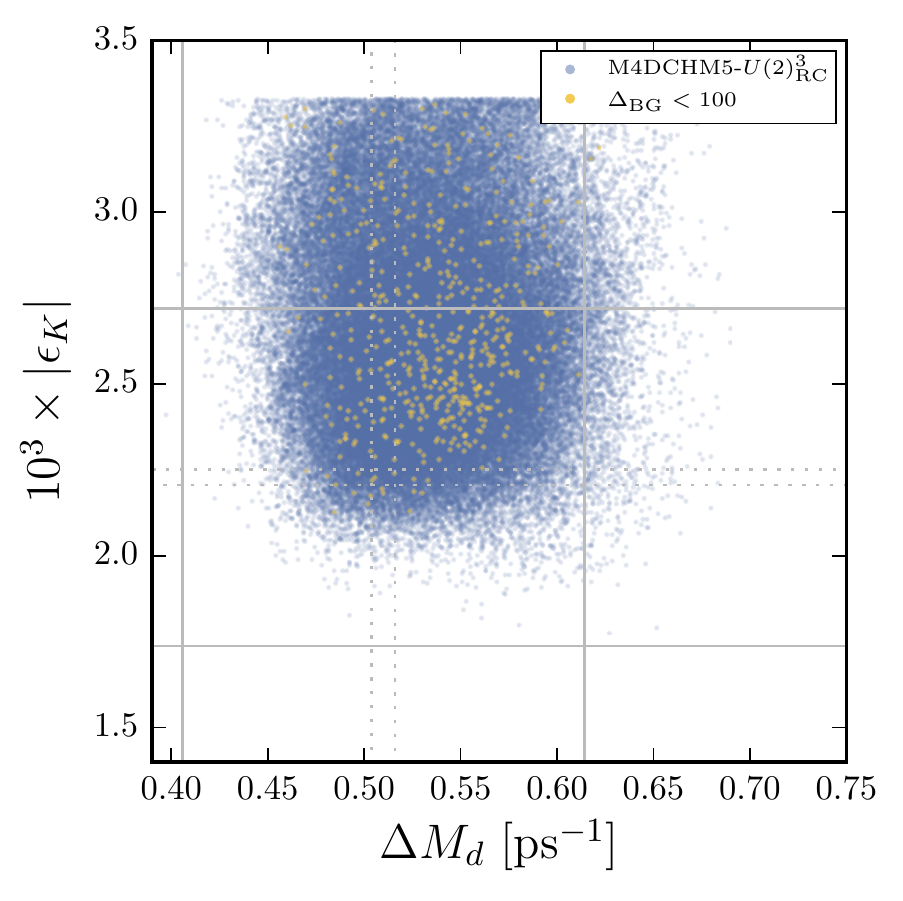}
\includegraphics[width=0.45\textwidth]{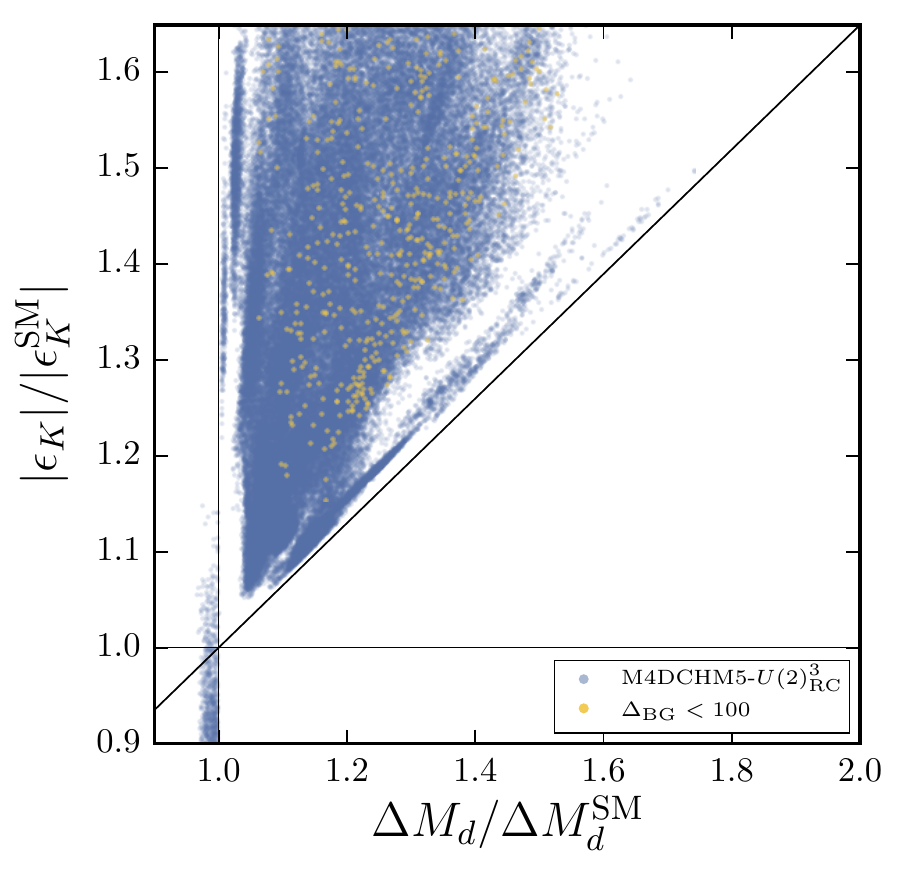}
\caption{$\Delta F=2$ observables in $U(2)^3_\text{RC}$.}
\label{fig:u2rc-df2}
\end{figure}

Figure~\ref{fig:u2rc-df2} shows the predictions for $\Delta F=2$ observables in 
$U(2)^3_\text{RC}$, in analogy to figure~\ref{fig:u2lc-df2}. We first point out 
the similar features,
\begin{itemize}
\item Sizable enhancements at the level of 60\% with respect to the SM are 
possible in the $B_d$ and $B_s$ mass differences, but a suppression is strongly 
disfavoured.
\item The leading-order $U(2)^3$ relation between $\Delta M_d$ and $\Delta M_s$ 
(shown as a black line) is violated by LR operators.
\end{itemize}
But there are also important differences between $U(2)^3_\text{LC}$ and 
$U(2)^3_\text{RC}$.
\begin{itemize}
\item There is no new phase in $B_d$ mixing, as was already pointed out in 
\cite{Barbieri:2012tu}.
\item In $B_s$ mixing, on the other hand, there can be a new phase roughly at 
the level of the current experimental uncertainties. This phase stems from the 
subleading terms in the spurion expansion and thus violates the leading order 
$U(2)^3$ relation (implying equal phase shifts in $B_d$ and $B_s$ mixing).
\item The enhancement of $\epsilon_K$ relative to the SM is {\em always larger} 
than the one in $\Delta M_d$. This is the opposite of what happened in 
$U(2)^3_\text{LC}$, where the relative enhancement was {\em always smaller} in 
$\epsilon_K$, cf.\ figure~\ref{fig:u2lc-df2} bottom-right. In the future, this 
could serve as a way to distinguish the two models based on $\Delta F=2$ 
observables alone.
\end{itemize}

\begin{figure}[tbp]
\centering
\includegraphics[width=0.48\textwidth]{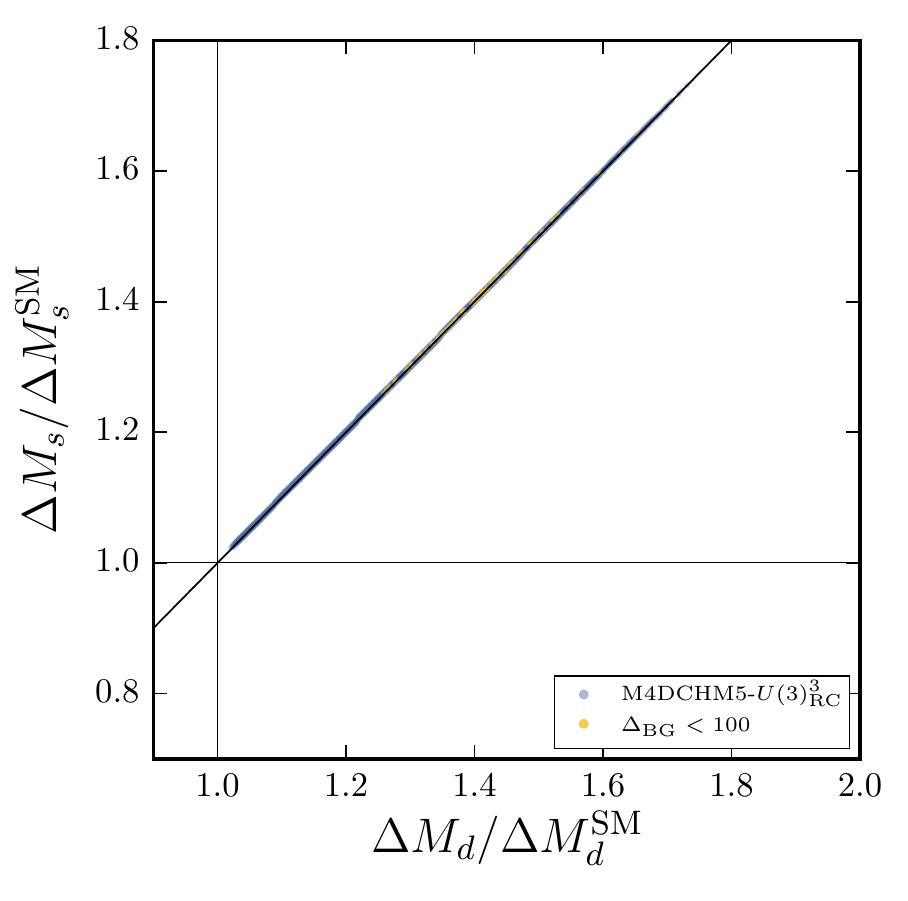}
\includegraphics[width=0.48\textwidth]{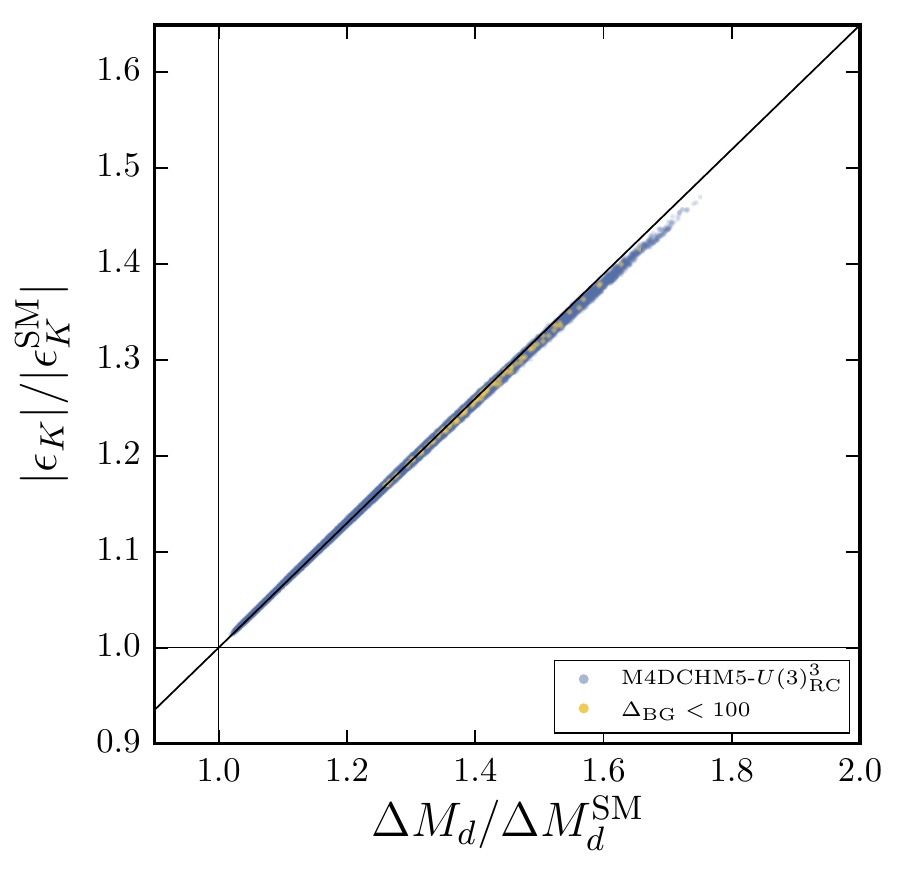}
\caption{$\Delta F=2$ observables in $U(3)^3_\text{RC}$.}
\label{fig:u3rc-df2}
\end{figure}

In $U(3)^3_\text{RC}$, we only directly show the observables 
normalized to their SM values in figure~\ref{fig:u3rc-df2}. In this case, the 
MFV relations, shown by black 
lines, are fulfilled exactly and there is no new phase, neither in $B_d$ nor in 
$B_s$ mixing.

Concerning $D^0$-$\bar D^0$ mixing, in $U(2)^3_\text{RC}$, similarly to 
$U(2)^3_\text{LC}$ discussed at the end of section~\ref{sec:u2lc-df2}, the NP 
effects are quite small and we find that the imaginary part of the mixing 
amplitude is always between $-0.4$ and $+0.2~\text{ns}^{-1}$, which 
is not relevant at the current experimental precision, but will become 
relevant when the experimental bound improves by an order of magnitude.
In  $U(3)^3_\text{RC}$, there is no new phase and thus no NP contribution to 
the imaginary part of the mixing amplitude.

\subsubsection{Rare $B$ decays}\label{sec:unrc-df1}

\begin{figure}[tbp]
\centering
\includegraphics[width=0.48\textwidth]{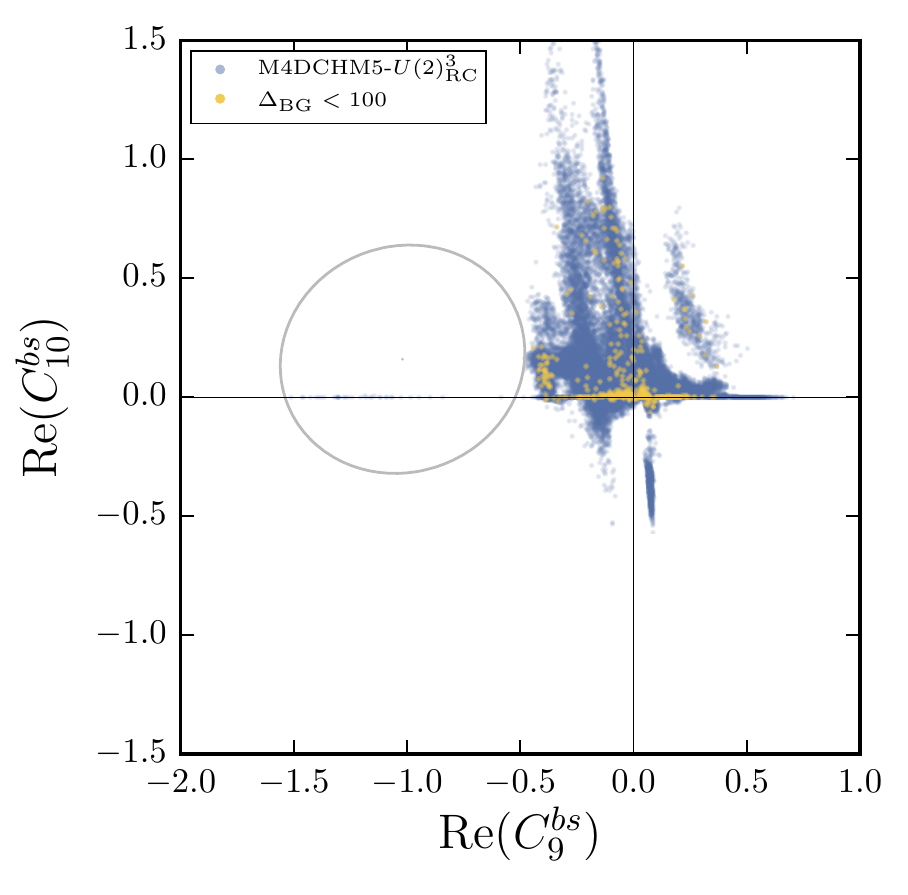}
\includegraphics[width=0.48\textwidth]{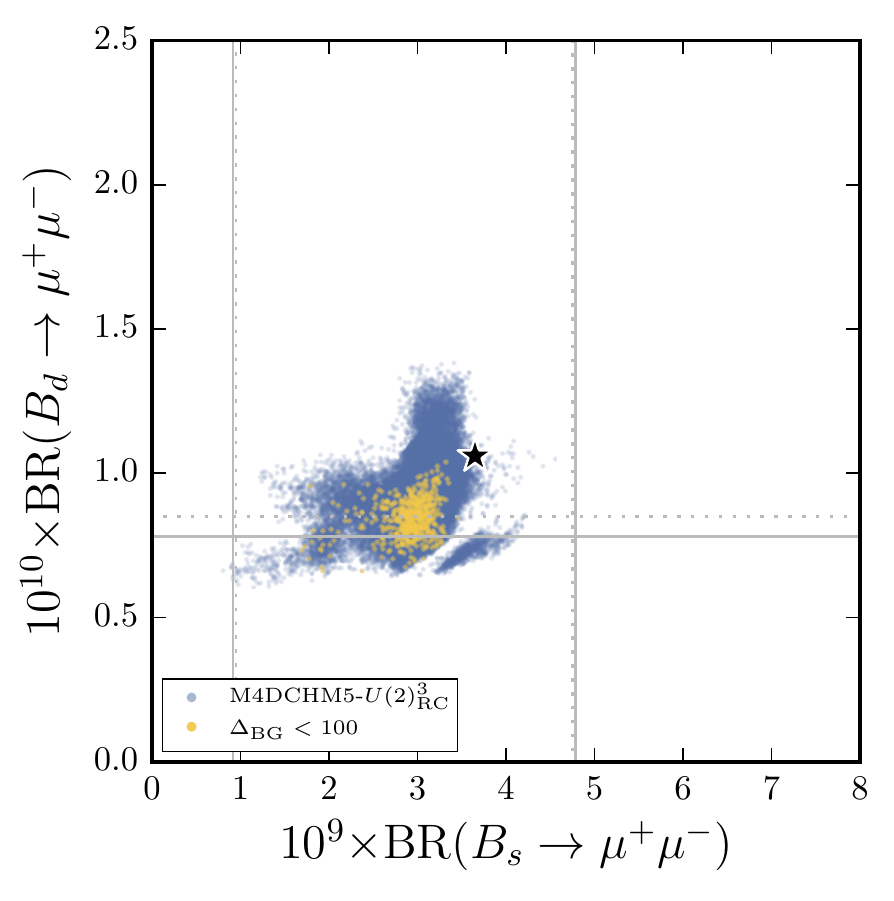}
\caption{%
Left: new physics contributions to the Wilson coefficients $C_{9}^{bs}$ and 
$C_{10}^{bs}$ in $U(2)^3_\text{RC}$.
Right: predictions for the branching ratios of
$B_s\to\mu^+\mu^-$ and $B_d\to\mu^+\mu^-$ in $U(2)^3_\text{RC}$.}
\label{fig:u2rc-c9c10}
\end{figure}

\begin{figure}[tbp]
\centering
\includegraphics[width=0.48\textwidth]{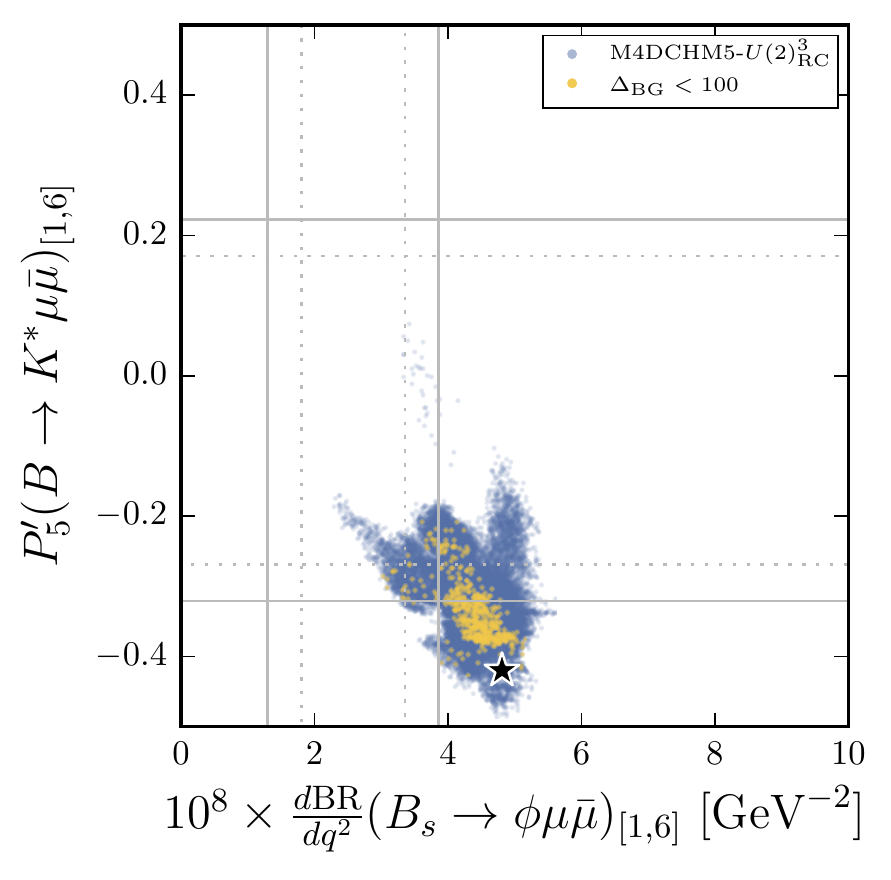}
\includegraphics[width=0.48\textwidth]{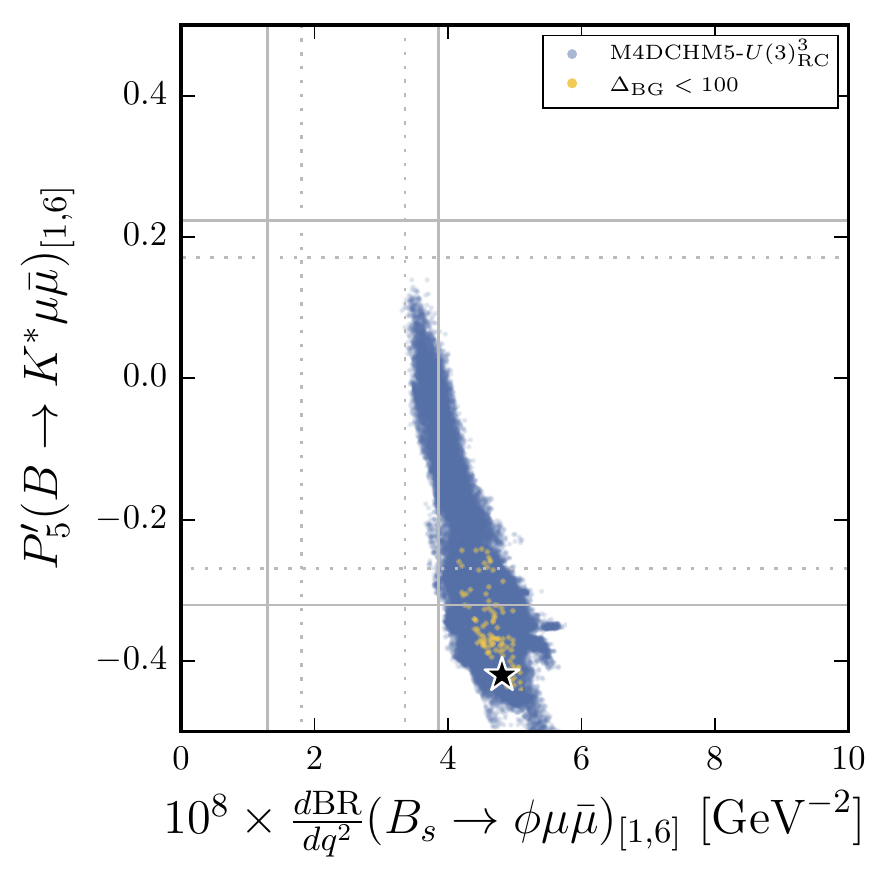}
\caption{%
Predictions for the angular observable $P_5'$ in $B\to 
K^*\mu^+\mu^-$ and the branching ratio of $B_s\to\phi\mu^+\mu^-$, both in the 
low-$q^2$ bin from 1 to 6~GeV$^2$
in $U(2)^3_\text{RC}$ (left) and $U(3)^3_\text{RC}$ (right).
The star corresponds to the central values 
of the SM predictions.}
\label{fig:unrc-p5p}
\end{figure}

In $U(2)^3_\text{RC}$, similarly to the case of left-handed compositeness 
discussed in section~\ref{sec:u2lc-df1}, the largest contribution to the 
Wilson coefficients of the semi-leptonic $b\to s\ell\ell$ transition occurs in 
the Wilson coefficient $C_{10}^{bs}$, but there are also contributions to the 
Wilson coefficient $C_{9}^{bs}$, as shown in figure~\ref{fig:u2rc-c9c10} left. 
In this case, we only find a small number of points with sizably negative 
$C_{9}^{bs}$ that populate the region preferred by a global fit to $b\to 
s\ell\ell$ data, indicated by a gray ellipse. These points then predict a 
significant suppression in absolute value of the angular observable $P_5'$ in 
$B\to K^*\mu^+\mu^-$ at low $q^2$, see the points 
around $P_5'\approx-0.1$ in figure~\ref{fig:unrc-p5p} left.
A distinguishing feature compared to $U(2)^3_\text{LC}$ is that the 
contributions to $C_{10}^{bs}$ are almost always positive, implying that the 
branching ratio of $B_s\to\mu^+\mu^-$ is almost always suppressed, as shown in 
fig.~\ref{fig:u2rc-c9c10} right.

In $U(3)^3_\text{RC}$, the contributions to $C_{10}^{bs}$ are 
forbidden by an interplay between custodial protection and the flavour 
structure as discussed in section~\ref{sec:df1}.
However, the resonance-mediated contributions to $C_{9}^{bs}$ are still 
present and we find viable points in the range $-1.7\lesssim 
C_{9}^{bs}\lesssim0.9$. Consequently, also $U(3)^3_\text{RC}$ can explain the 
anomalies in $b\to s\ell\ell$ angular observables and branching ratios. For the 
observable $P_5'$ and the branching ratio of $B_s\to\phi\mu^+\mu^-$, this is 
illustrated in figure~\ref{fig:unrc-p5p} right. Finally, we remind the reader 
again the $U(3)^3_\text{RC}$ is actually a limiting subset of 
$U(2)^3_\text{RC}$, so the fact that in $U(3)^3_\text{RC}$ there are much more 
points with sizable NP effects in $C_{9}^{bs}$ compared to $U(2)^3_\text{RC}$ 
is simply a statistical effect since the $U(3)^3_\text{RC}$ parameter space is 
more restricted.

As in $U(2)^3_\text{LC}$ discussed in section~\ref{sec:u2lc-df1}, the solution 
of the $B$ physics anomalies by a negative NP contribution to $C_{9}^{bs}$ 
implies the presence of a light, narrow neutral vector resonance below about 
1~TeV. In $U(2)^3_\text{RC}$, the dominant decay mode of this resonance is 
$t\bar t$ or two light quark jets, while in $U(3)^3_\text{RC}$ the dominant 
decay mode is always dijets.

\subsubsection{Other processes}

We have not discussed the oblique parameters as the predictions in both models 
with right-handed compositeness are analogous to the effects in 
$U(2)^3_\text{LC}$ shown in fig.~\ref{fig:u2lc-h-st} right, so the same 
comments as in section~\ref{sec:u2lc-st} apply.

In rare $K$ decays, the effects both in $U(2)^3_\text{RC}$ and in 
$U(3)^3_\text{RC}$ are even smaller than in $U(2)^3_\text{LC}$ discussed in 
section~\ref{sec:u2lc-other}.

In contrast to left-handed compositeness, the branching ratio of the FCNC top 
decay $t\to cZ$ is always below $10^{-6}$ in $U(2)^3_\text{RC}$ and even below 
$10^{-8}$ in $U(3)^3_\text{RC}$ and thus negligible.

\subsection{Direct searches in left- and right-handed 
compositeness}
\label{sec:num-direct}

\subsubsection{Prospects for quark partner searches}\label{sec:prospects-quark}

The direct bounds on quark partner masses discussed in 
section~\ref{sec:spin1/2} are among the most important constraints in our 
analysis. It is thus clear that future searches for quark partners will be 
instrumental in probing these models. Since in our numerical analysis, the 
lightest vector resonances are always found to be heavier than the lightest 
quark partners, which is due to electroweak precision tests and the other 
indirect bounds discussed in section~\ref{sec:indirect}, the lightest
quark partners always decay to SM states. To judge which of the search channels 
will be most promising at run~2 of the LHC, let us first discuss the dominant 
decay modes of the lightest resonances.

\begin{figure}[tbp]
\includegraphics[width=0.48\textwidth]{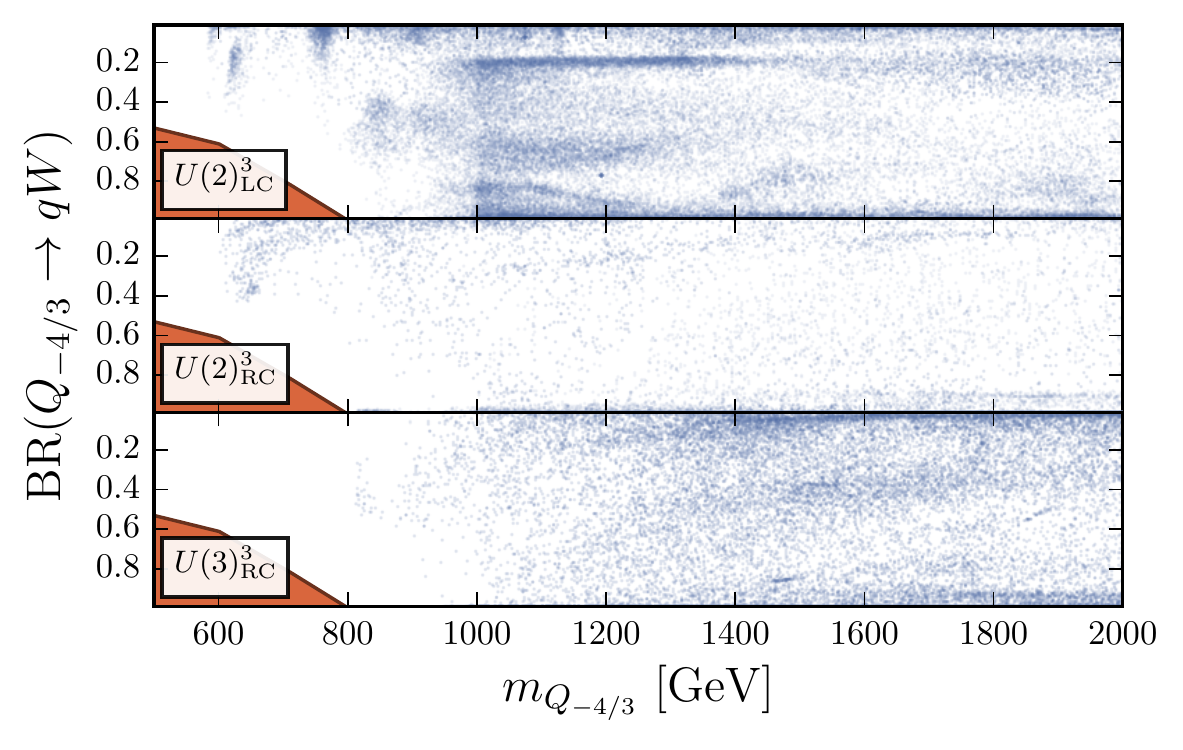}
\includegraphics[width=0.48\textwidth]{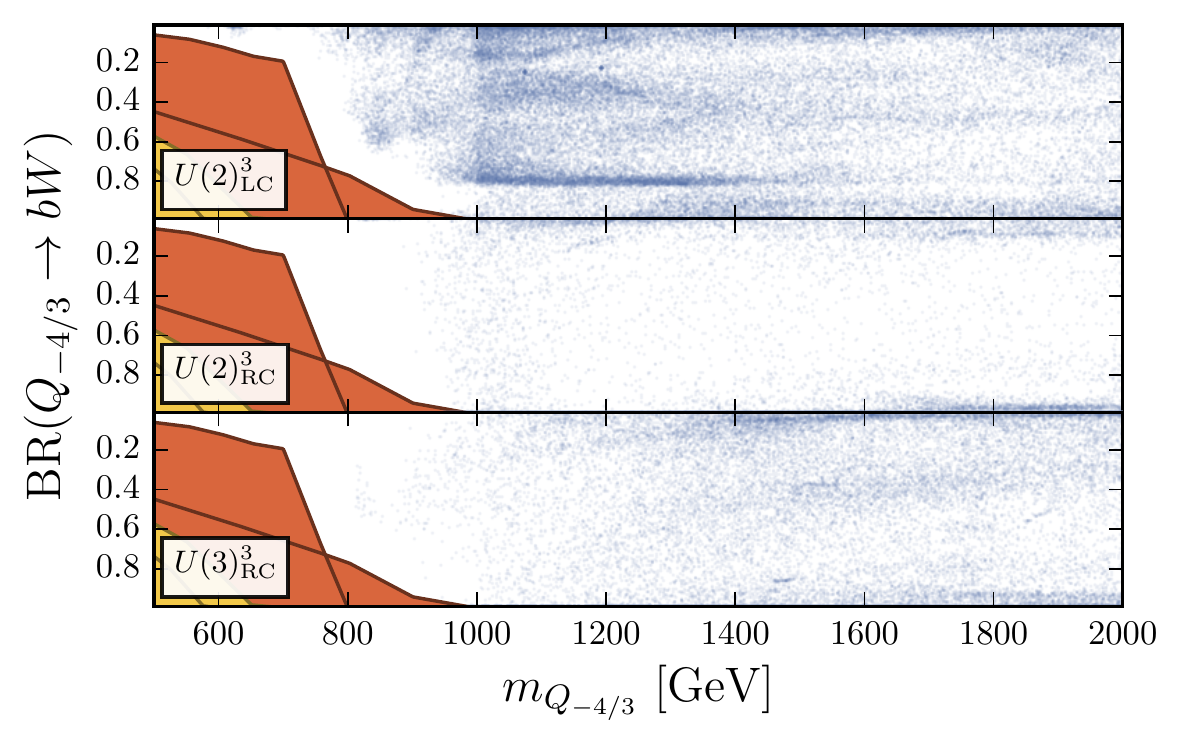}

\includegraphics[width=0.48\textwidth]{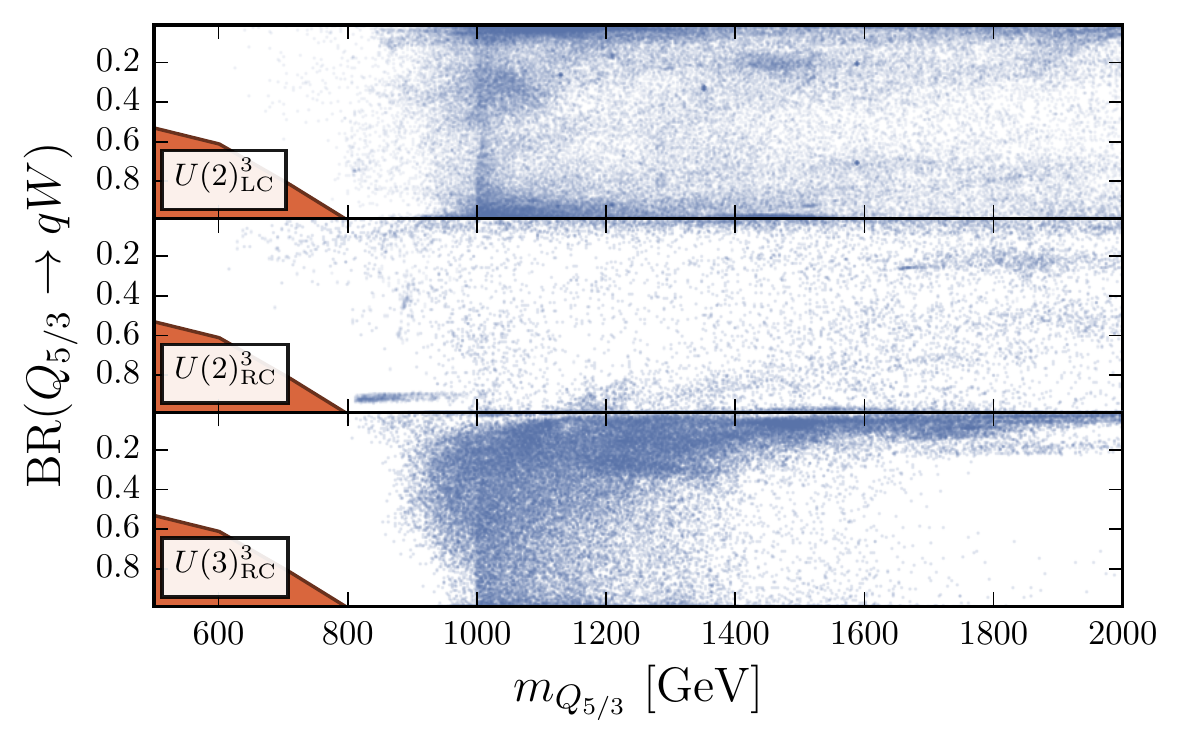}
\includegraphics[width=0.48\textwidth]{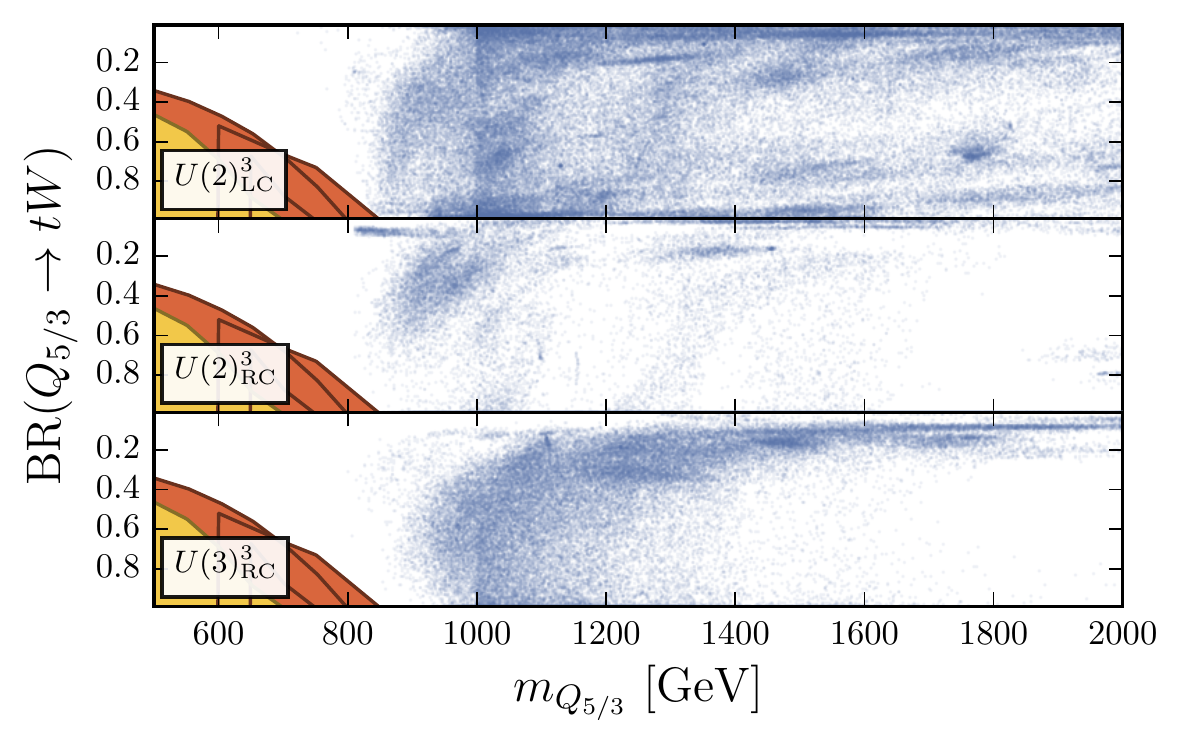}
\caption{Branching ratios of charge-$(-4/3)$ (first row) and charge-$5/3$ (second row) 
quarks to light (left) and third-generation (right) quarks as function of their 
mass for all three models.
The coloured regions are the same as in fig.~\ref{fig:fermion_br}.}
\label{fig:br-predictions-ex}
\end{figure}

\paragraph{Exotic charge quarks} The charge-$5/3$ and charge-$(-4/3)$ quarks 
always decay to a $W$ boson and a SM quark or quark resonance. In 
$U(2)^3_\text{LC}$, we find that there is a significant number of points where 
the $Q_{5/3}$ can decay to both $Wt$ and $Wq$ ($q=u,c$) with significant 
branching ratio, and similarly for the $Q_{-4/3}$ decaying into final states 
with bottoms vs.\ light quarks. In the right-handed compositeness models, we 
find in contrast that for any given exotic quark partner, only the decay to 3rd 
generation {\em or} the one to light quarks is relevant. This can be understood 
from the fact that the decay of the exotic charge quarks always involves 
right-handed composite-elementary mixings, and these are flavour-diagonal in 
right-handed compositeness, but involve flavour mixing in left-handed 
compositeness. In figure~\ref{fig:br-predictions-ex}, we show the predictions 
for the branching ratios as a function of the mass for the exotic charged quark 
partners for a subset of all viable points in all three models.
An interesting feature of these plots is that there is a significant
number of points with branching ratios different from zero or one in a
given channel. Apart from flavour mixing, this is due to the competition
with decays involving a fermion resonance in the final state. For
heavier masses, the branching ratios into SM-only states decrease, as can
be seen from the  plots  as well.

\begin{figure}[tbp]
\includegraphics[width=0.48\textwidth]{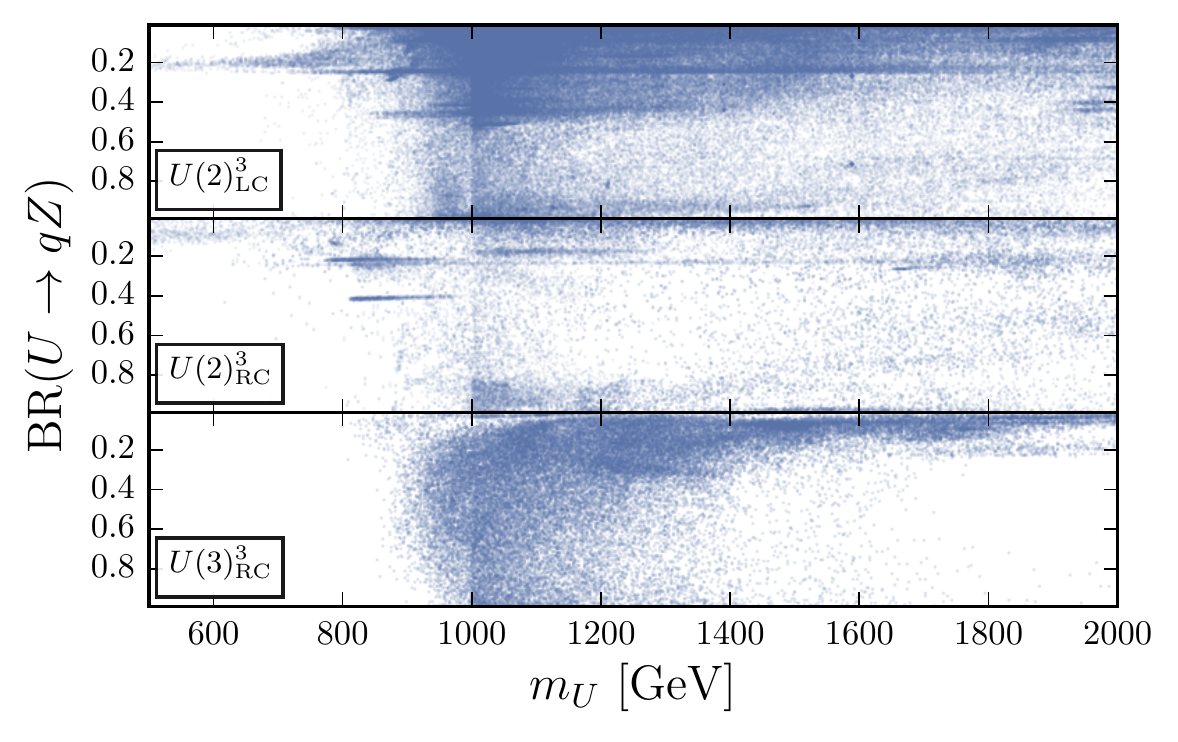}
\includegraphics[width=0.48\textwidth]{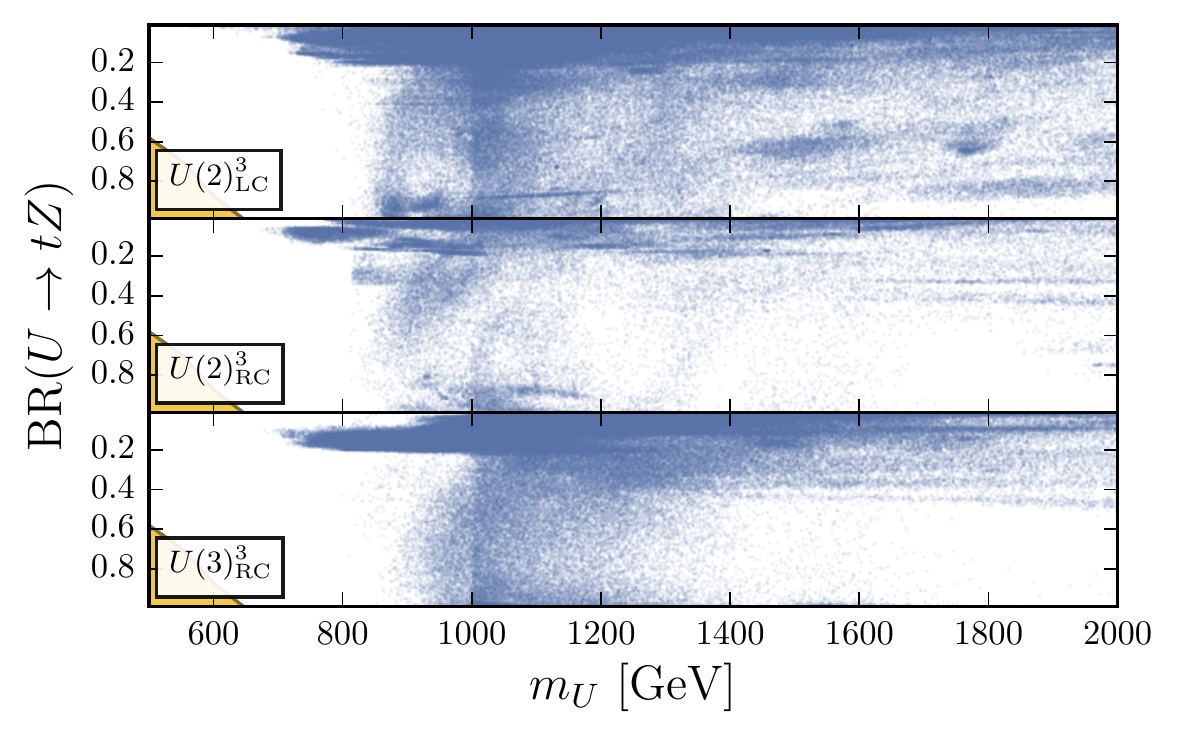}

\includegraphics[width=0.48\textwidth]{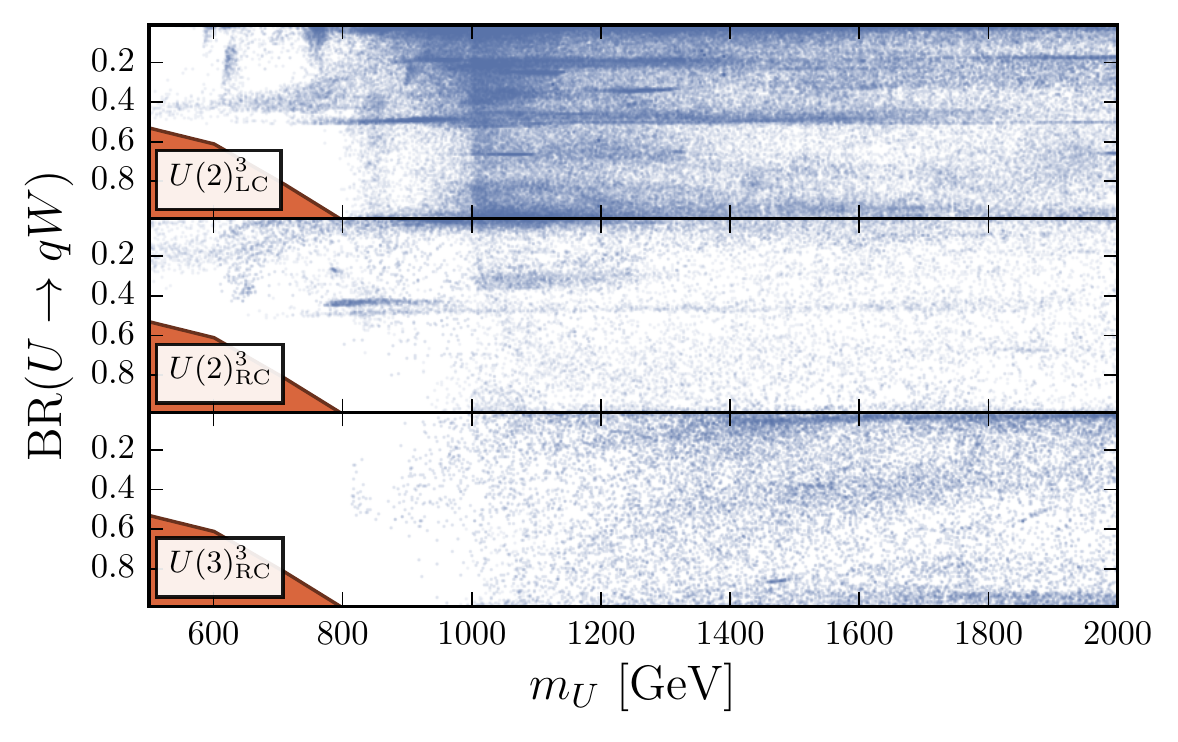}
\includegraphics[width=0.48\textwidth]{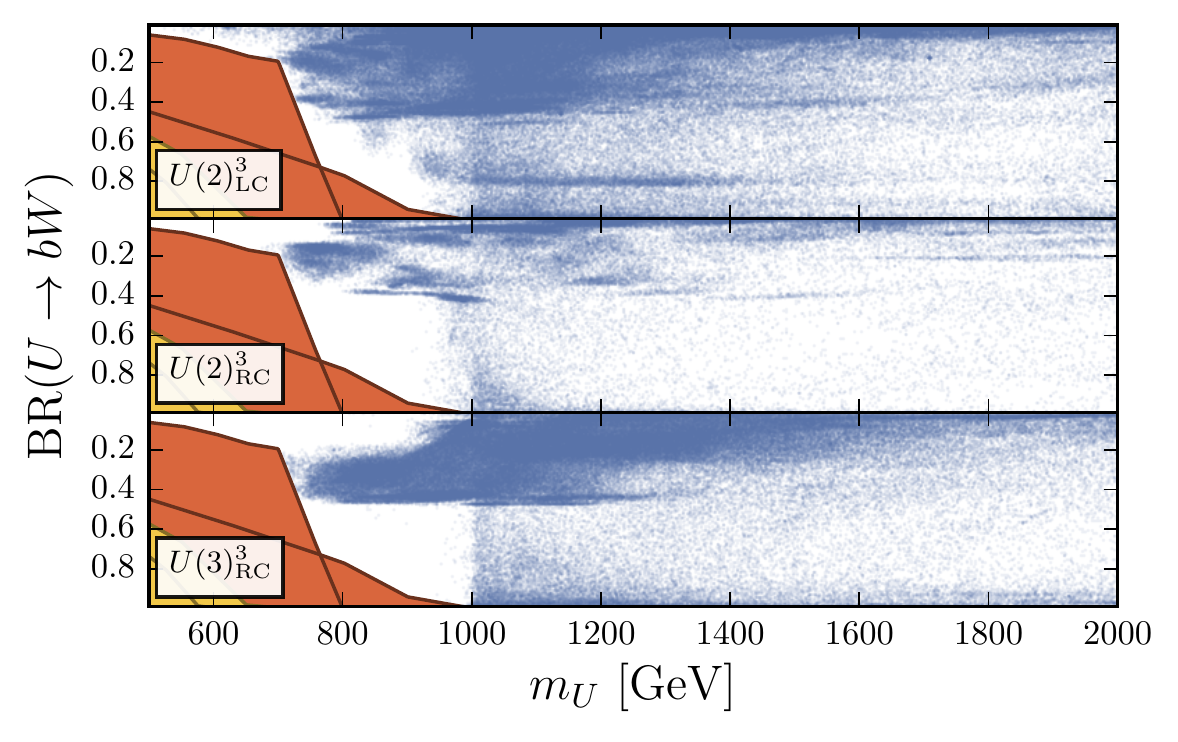}

\includegraphics[width=0.48\textwidth]{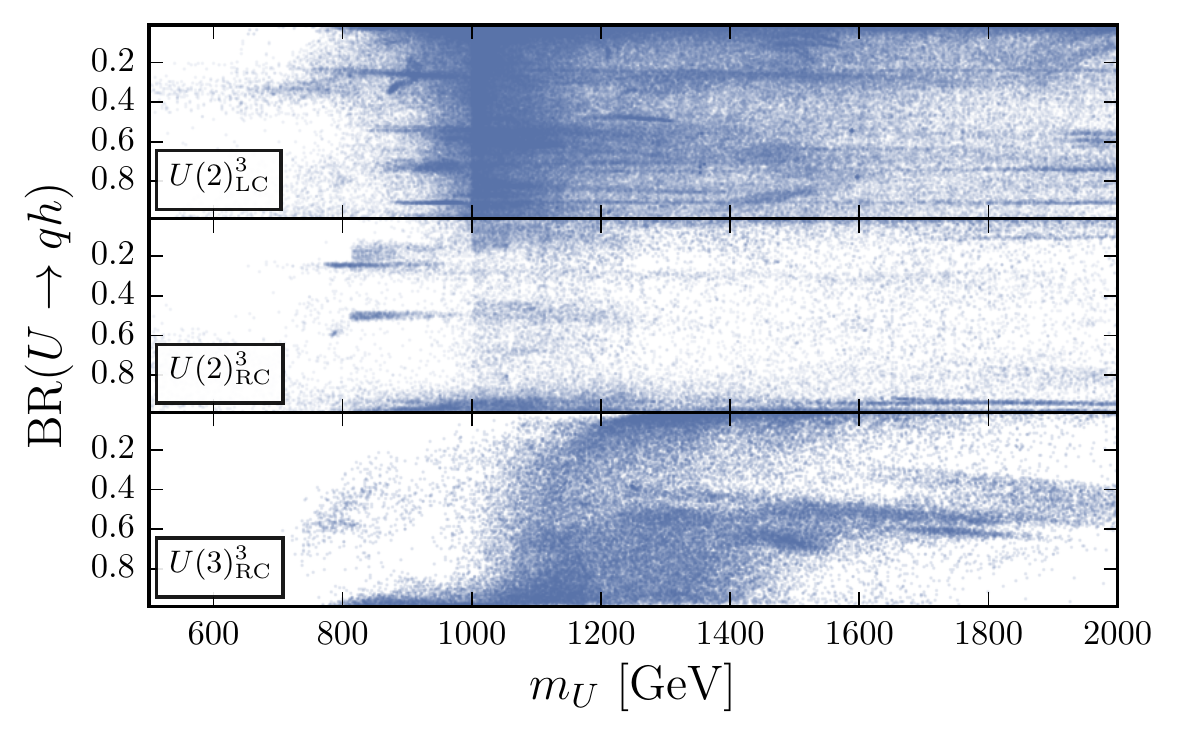}
\includegraphics[width=0.48\textwidth]{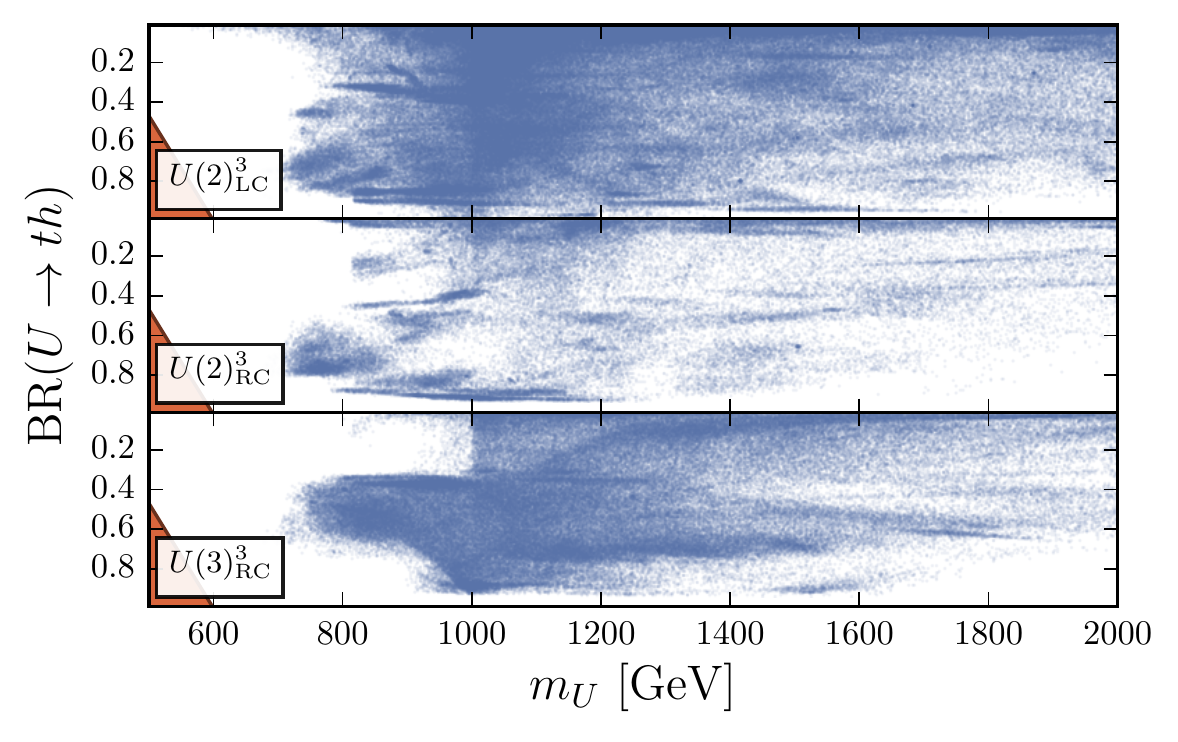}
\caption{Branching ratios of up-type quark partners to final states
involving light (left) and third generation (right) quarks as function of their 
mass for all three models.
The coloured regions are the same as in fig.~\ref{fig:fermion_br}.}
\label{fig:br-predictions_U}
\end{figure}

\begin{figure}[tbp]
\includegraphics[width=0.48\textwidth]{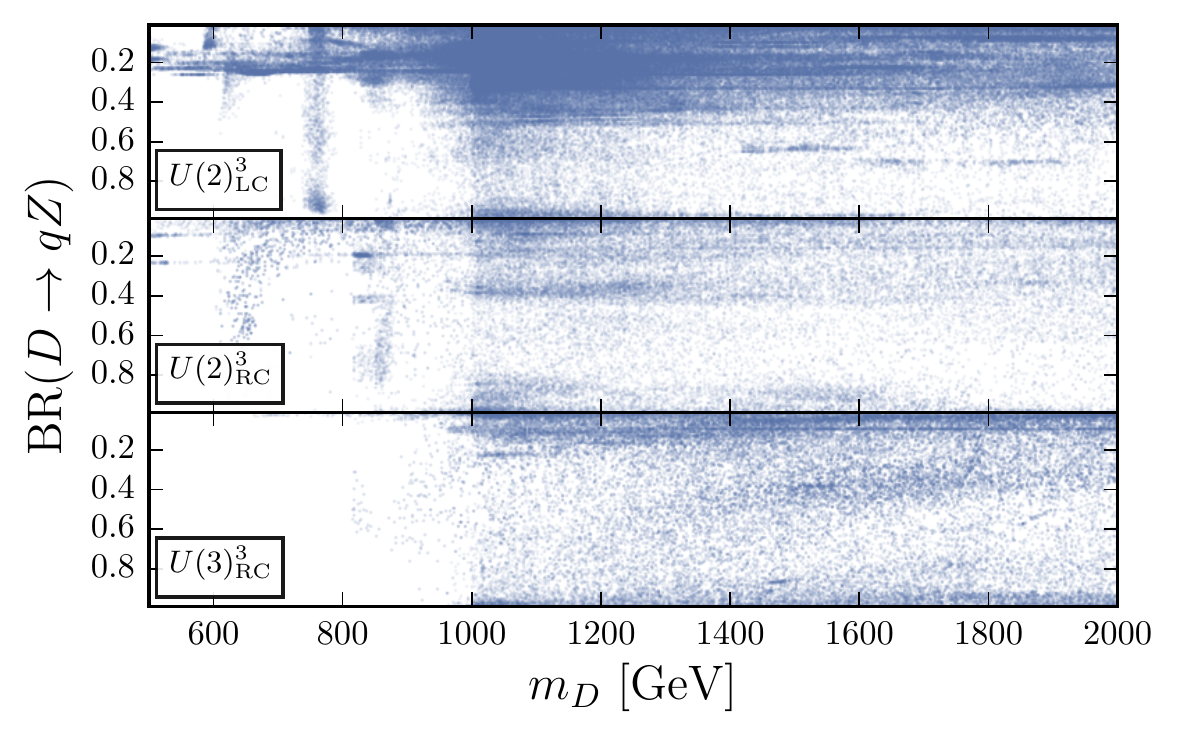}
\includegraphics[width=0.48\textwidth]{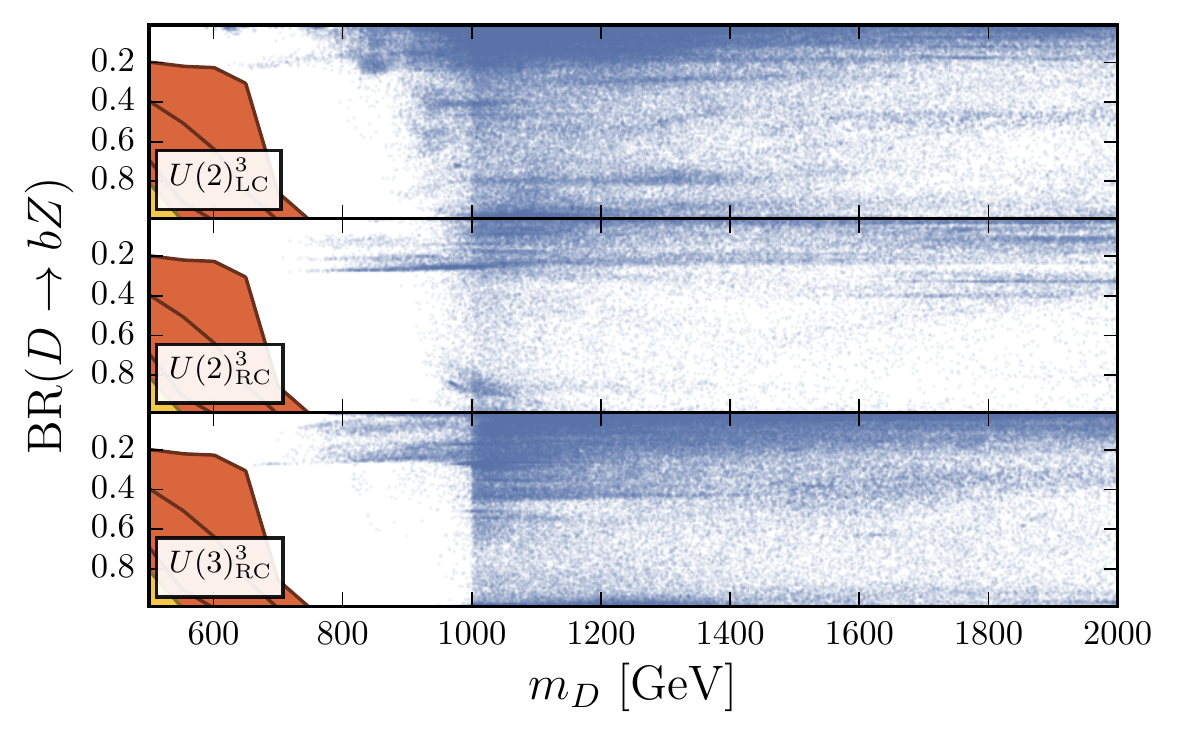}

\includegraphics[width=0.48\textwidth]{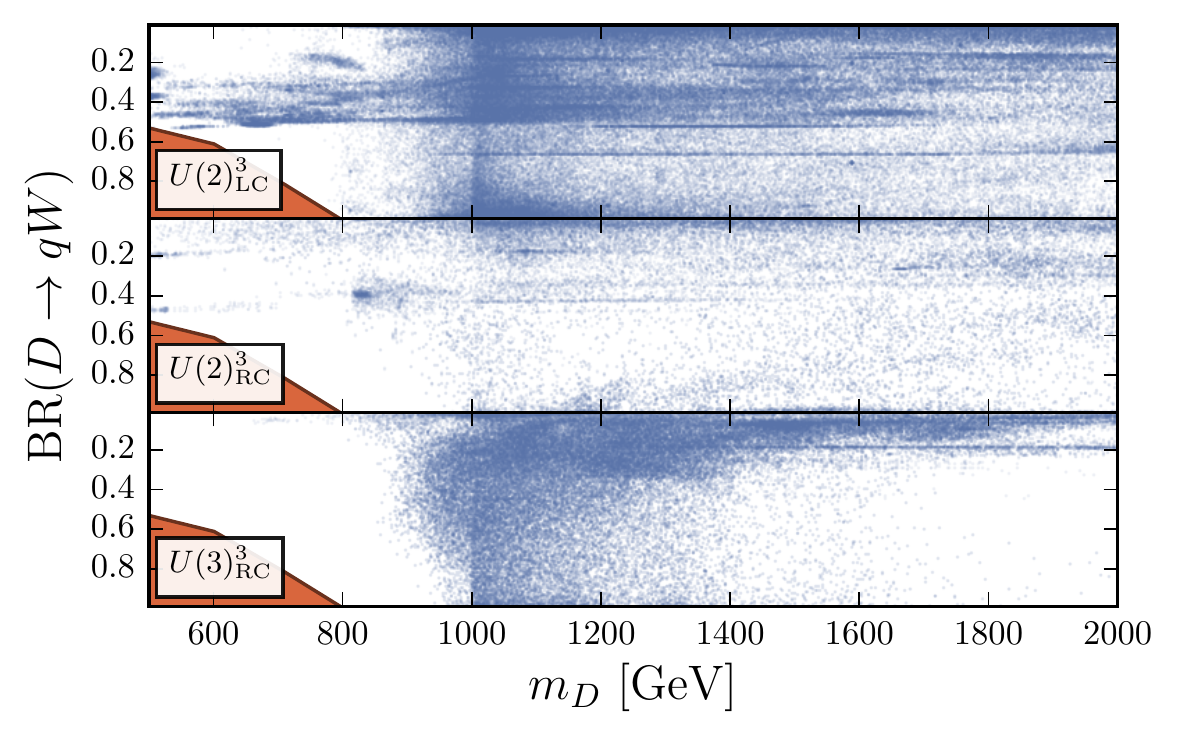}
\includegraphics[width=0.48\textwidth]{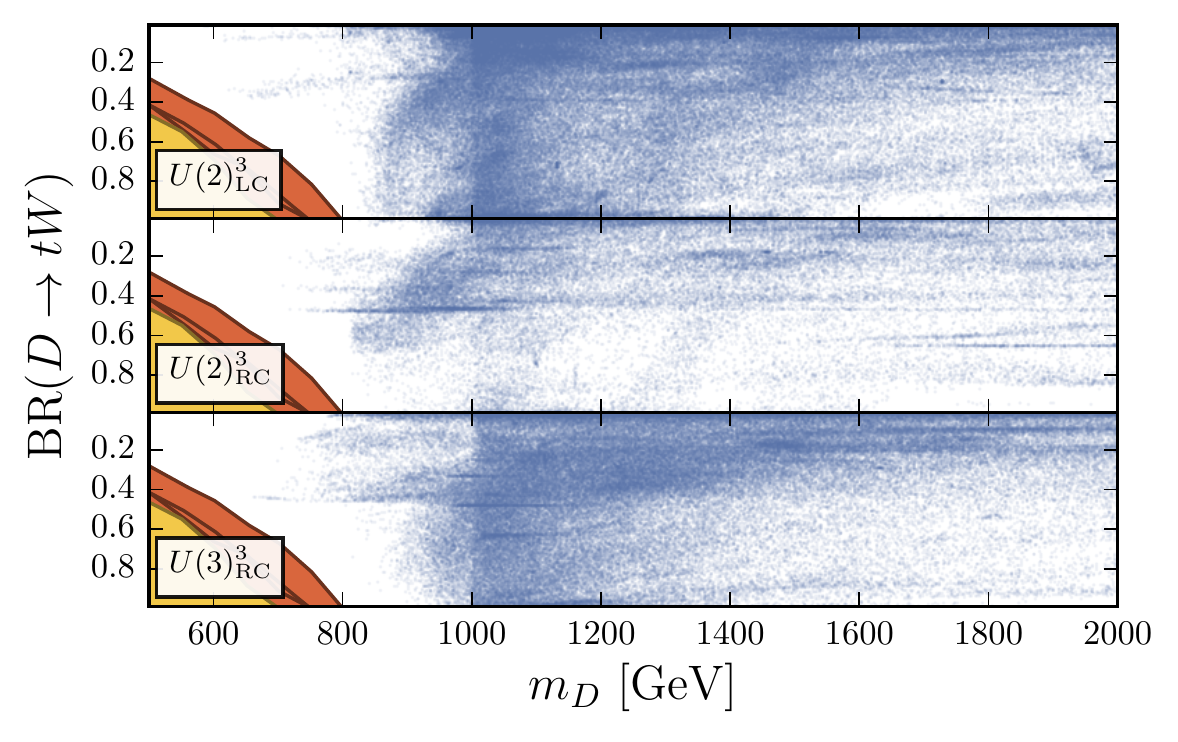}

\includegraphics[width=0.48\textwidth]{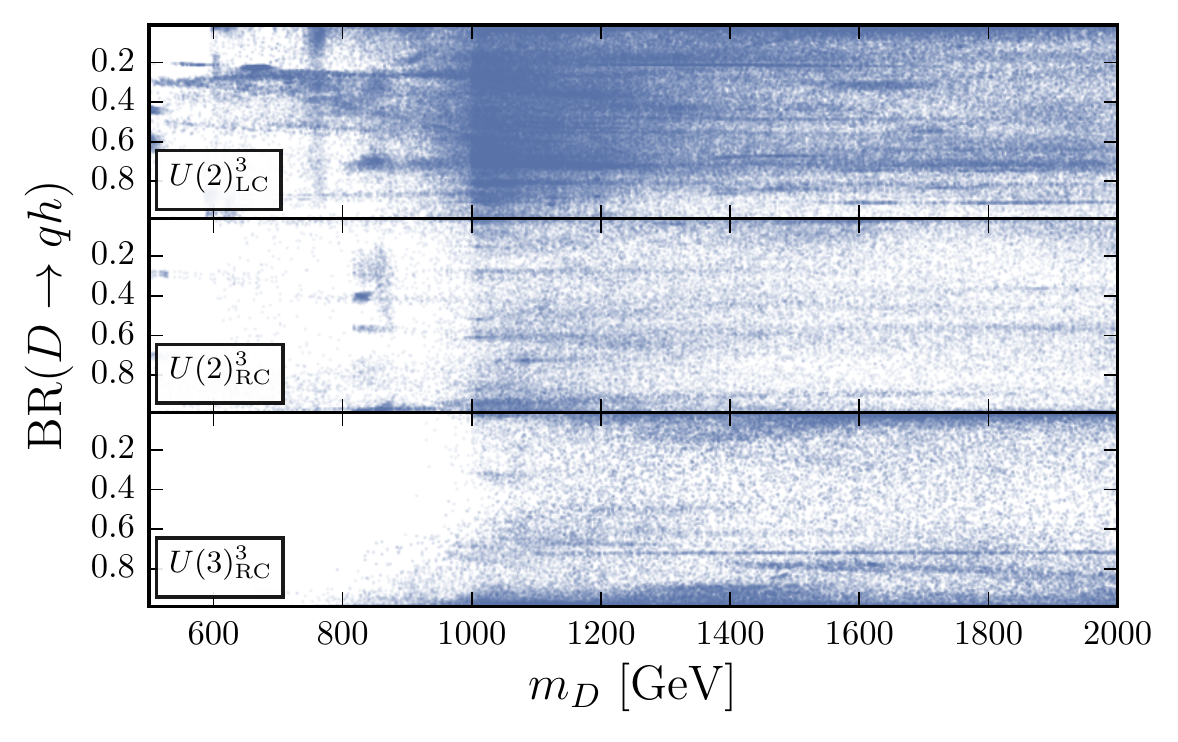}
\includegraphics[width=0.48\textwidth]{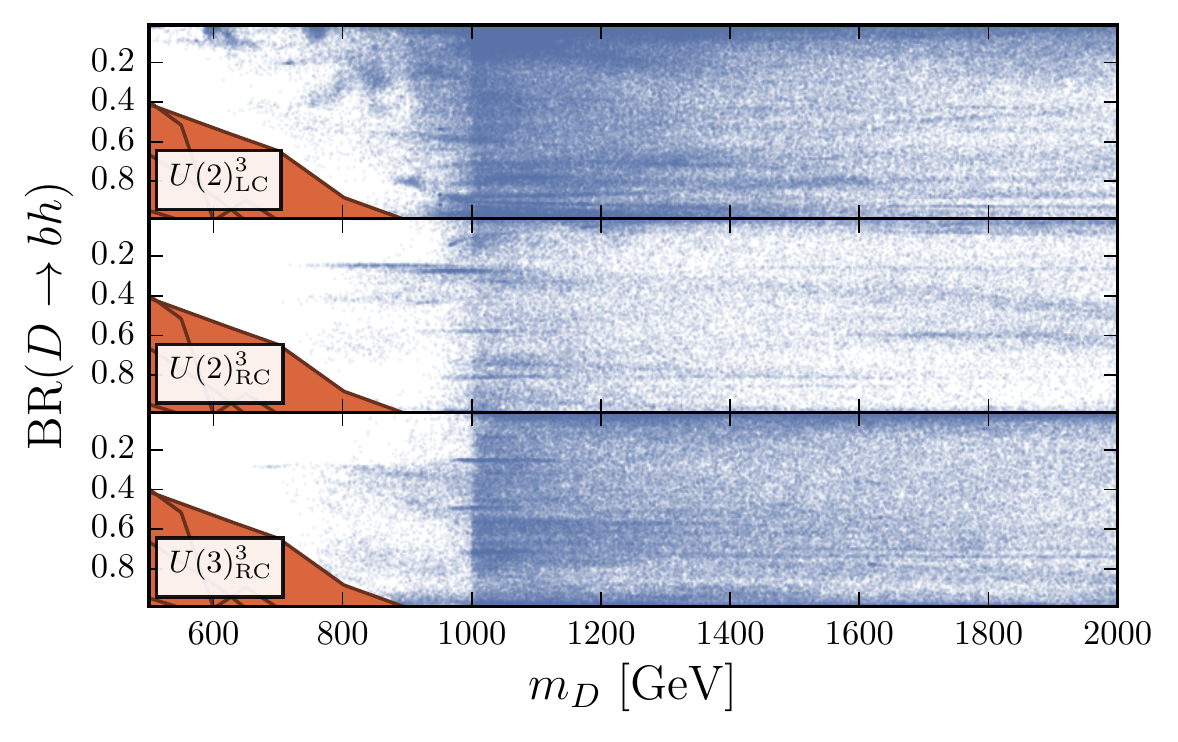}
\caption{Branching ratios of down-type quark partners to final states
involving light (left) and third generation (right) quarks as function of their 
mass for all three models.
The coloured regions are the same as in fig.~\ref{fig:fermion_br}.}
\label{fig:br-predictions_D}
\end{figure}

\paragraph{Up- and down-type quark partners}  When decaying to SM states, these 
quark partners can decay to a $W$, $Z$, or $h$ plus a SM quark.
In figures \ref{fig:br-predictions_U} and ~\ref{fig:br-predictions_D},
we show predictions for the masses and 
branching ratios in the most important channels for quark partners in the 
three viable models.

\medskip
In summary, the plots show that searches for pair-produced quark partners, both 
with exotic and with SM-like charges, are very promising, with masses and 
branching ratios just above what LHC has excluded in run~1 being viable in all 
models.

\subsubsection{LHC excesses}\label{sec:exc-res}

\begin{figure}[tbp]
\centering
\includegraphics[width=0.8\textwidth]{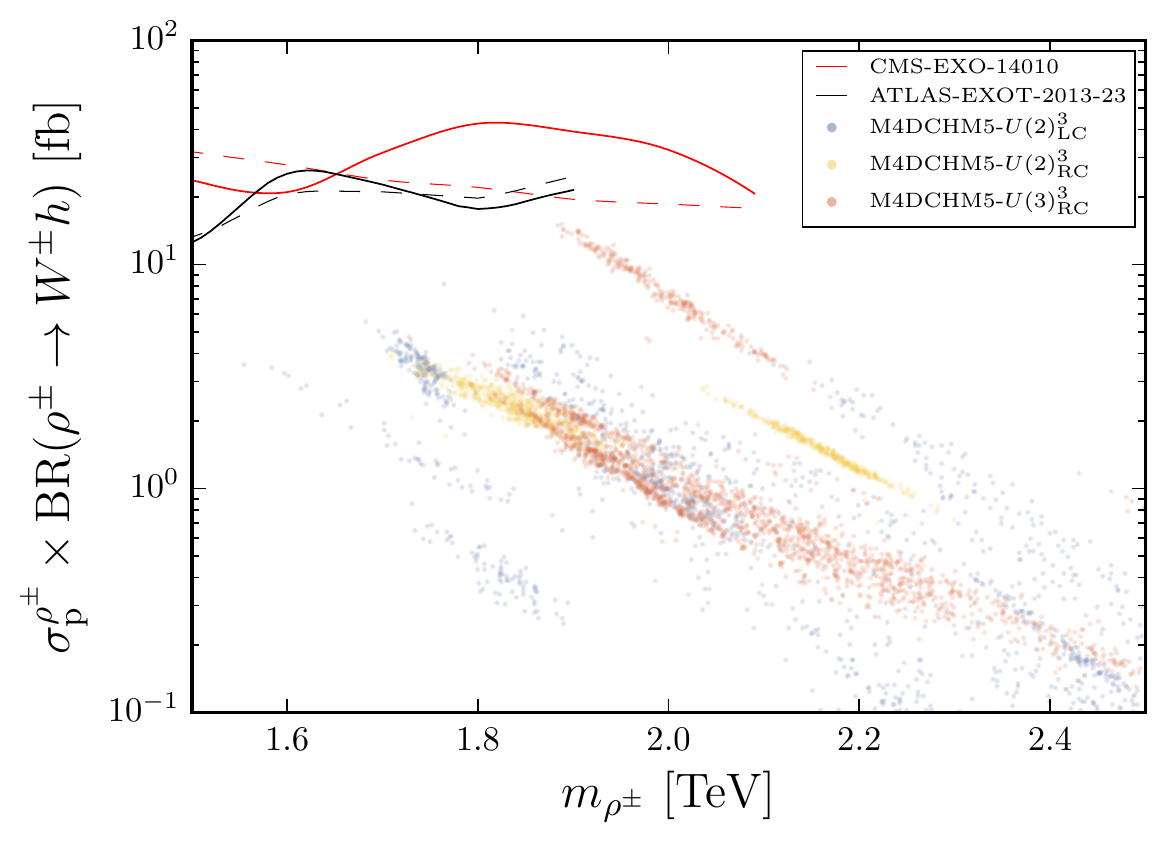}
\includegraphics[width=0.8\textwidth]{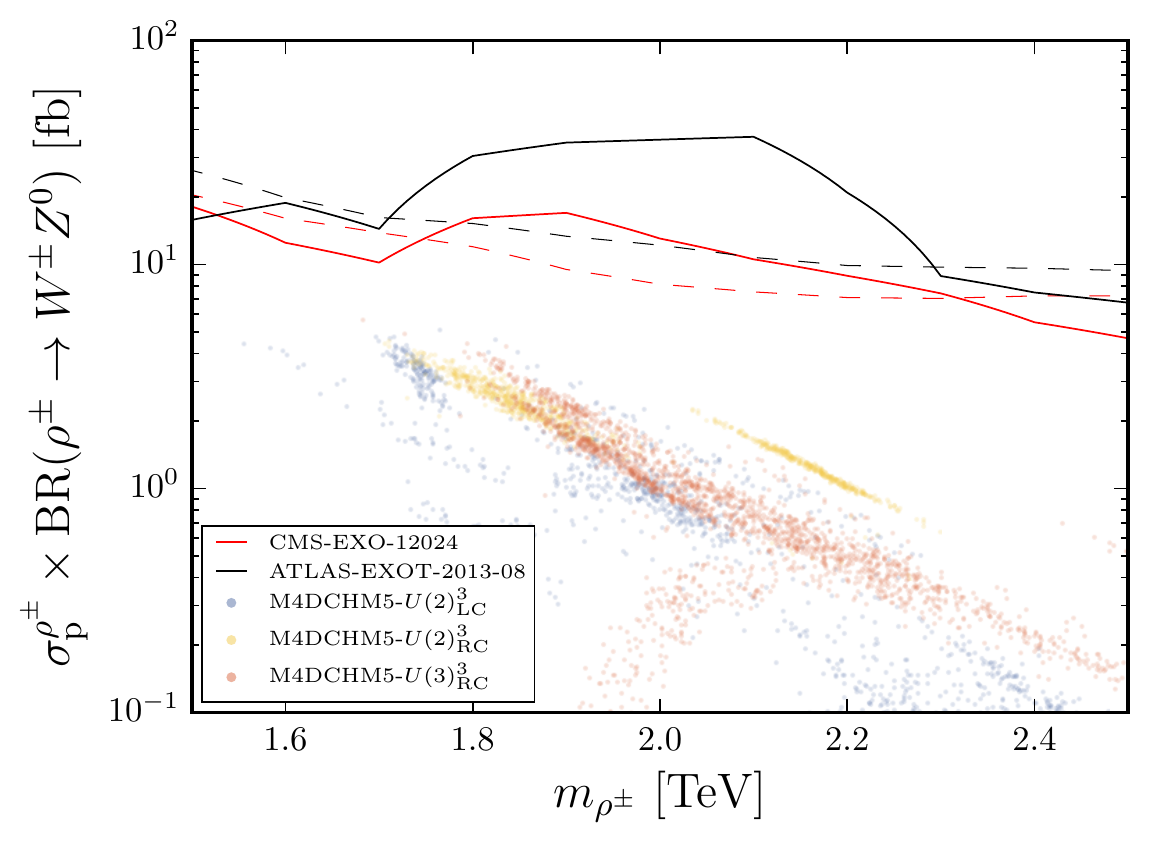}
\caption{Predictions for the production cross sections times branching ratios 
of charged electroweak vector resonances decaying to $Wh$ or $ZW$ final states 
in all three models. Only points with narrow resonances ($\Gamma/m<0.05$) are 
shown. The dashed and solid curves show the expected and observed 95\% C.L.\ 
experimental limits.}
\label{fig:excess-W}
\end{figure}

\begin{figure}[tbp]
\centering
\includegraphics[width=0.8\textwidth]{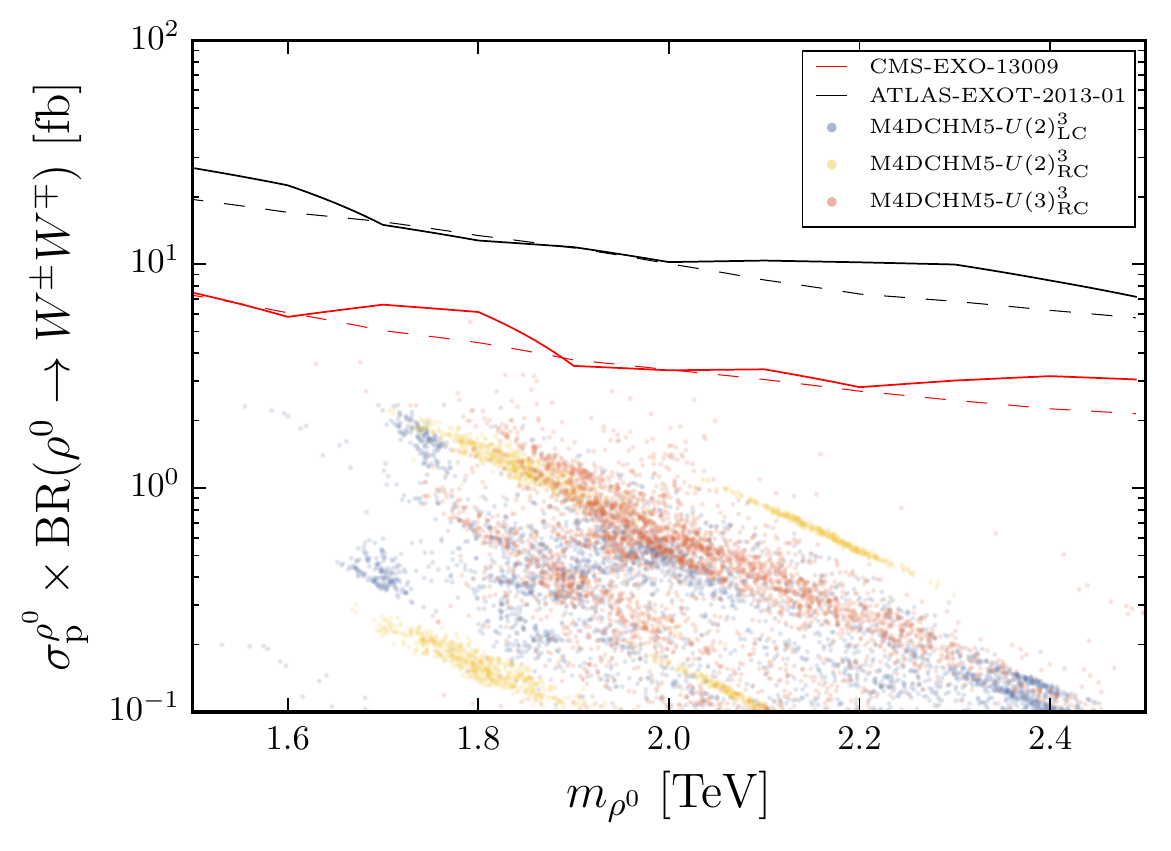}
\includegraphics[width=0.8\textwidth]{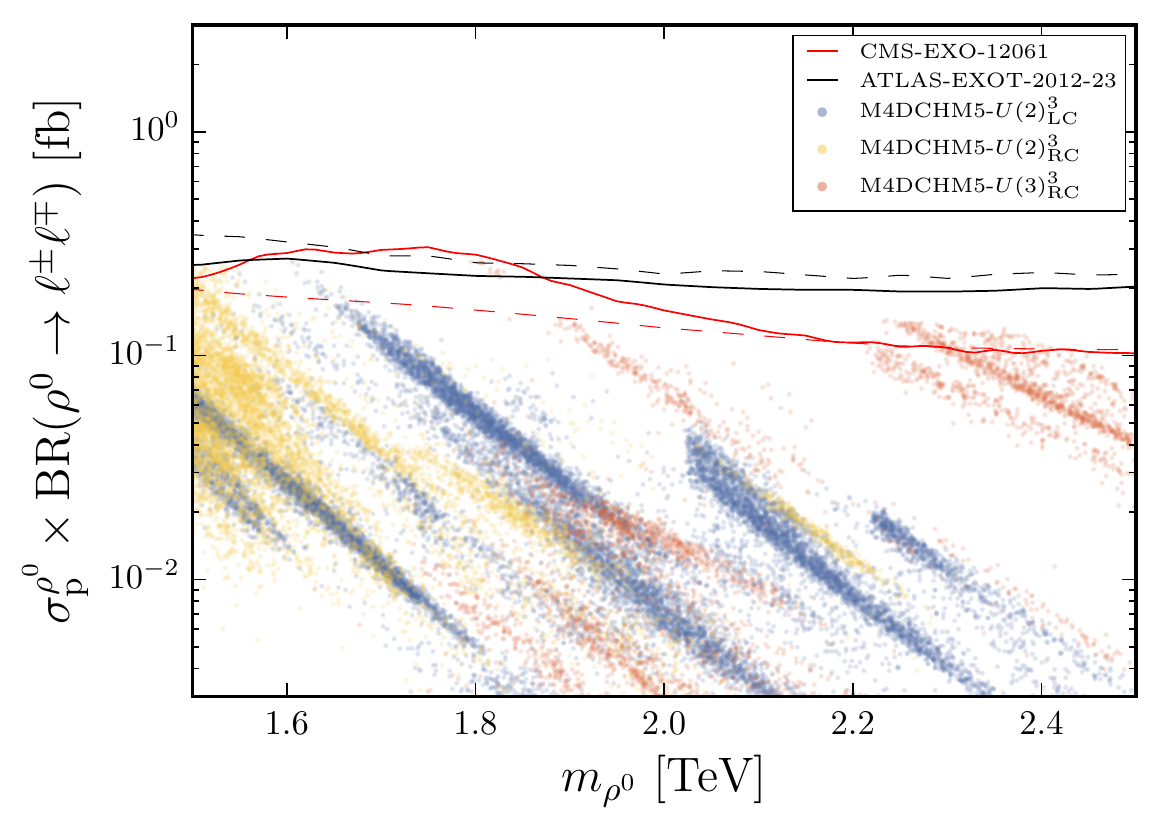}
\caption{Predictions for the production cross sections times branching ratios 
of neutral electroweak vector resonances decaying to $WW$ or dilepton final 
states 
in all three models. Only points with narrow resonances ($\Gamma/m<0.05$) are 
shown. The dashed and solid curves show the expected and observed 95\% C.L.\ 
experimental limits.}
\label{fig:excess-Z}
\end{figure}

As discussed in section~\ref{sec:excess}, several excesses with 
significances up to around $3\sigma$ have been observed by ATLAS and CMS in 
resonance searches in $Wh$, $WZ$, and $WW$ final states around a  resonance 
mass of 2~TeV. To investigate whether the models studied by us could account 
for these anomalies, we have computed the production cross sections of charged 
and neutral electroweak vector resonances times the branching ratios to 
the relevant final states. In figures \ref{fig:excess-W} and 
\ref{fig:excess-Z}, we show these predictions in the relevant mass region for 
all three viable models, compared to the expected (dashed) and observed (solid) 
limits in some of the relevant ATLAS and CMS searches (for a total list of 
searches included, see section~\ref{sec:spin1}).
In these plots, to be conservative we only show points where the decaying 
resonance has a narrow width, namely $\Gamma/m<0.05$, because, as discussed in 
section~\ref{sec:spin1}, we have not imposed any LHC constraints on broader 
resonances. We note however that there are a significant number of more points 
in the same region where the width is slightly larger than 5\%.
But even with this strong condition, we do find points in all three models 
where there are resonances with mass around 2~TeV and with cross sections of 
the order of 5~fb in the case of $\rho^\pm\to W^\pm h$ and $\rho^\pm\to W^\pm 
Z^0$, which is the right ballpark to explain the excesses (see e.g.\  
\cite{Thamm:2015csa,Carmona:2015xaa,Bian:2015ota, Lane:2015fza,Low:2015uha}).
In the case of $\rho^0\to W^\pm W^\mp$, the predicted cross section is roughly 
a factor of two smaller due to the PDF suppression, but this agrees at least 
qualitatively with the less pronounced excess in the CMS analysis, as seen in 
the upper plot of figure~\ref{fig:excess-Z}.

Interestingly, a slight excess around 2~TeV has also been observed in a CMS 
dilepton resonance search \cite{Khachatryan:2014fba}. Our predictions for this 
channel are shown in the lower plot of figure~\ref{fig:excess-Z}. Also here, we 
find a significant number of points with cross section times branching ratio of 
the order of $0.1$~fb, which could account for this excess. Also in this plot, 
we are only showing resonances with a narrow width. This is also why there are 
few points in the region of interest for the $U(3)^3_\text{RC}$ model. In this 
model, the electroweak resonances are typically broader than 5\% due to the 
stronger coupling to light quarks compared to the $U(2)^3$ 
models.%
\footnote{This does not mean that this model cannot explain the 
excesses, but a detailed analysis of the impact of broad resonances is beyond 
the scope of our study.}

\begin{figure}[tbp]
\centering
\includegraphics[width=0.5\textwidth]{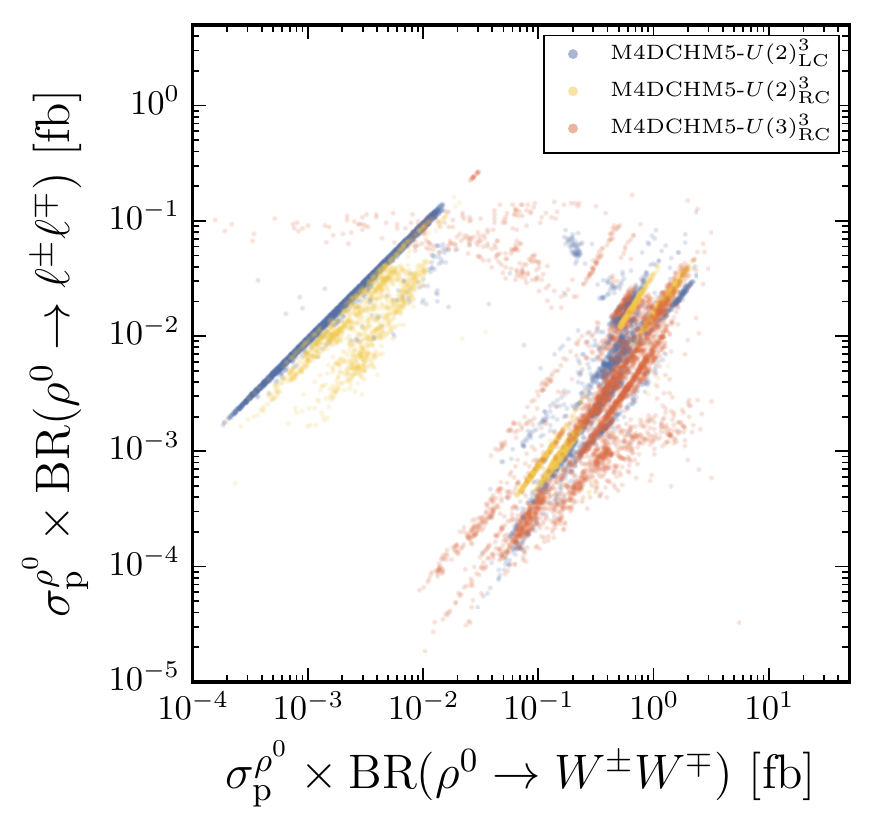}
\caption{Predictions for the production cross sections times branching ratios 
of neutral electroweak vector resonances decaying to $WW$ vs.\ dilepton final 
states 
in all three models. Only points with narrow resonances ($\Gamma/m<0.05$) are 
shown.}
\label{fig:diboson-dilepton}
\end{figure}

Given figures~\ref{fig:excess-W} and \ref{fig:excess-Z}, the question arises 
whether the points explaining the excesses in the individual plots are actually 
the same points, i.e.\ the question whether the models can explain all 
excesses simultaneously. For the final states involving bosons, this is 
obviously the case as the branching ratios are sizable only for the composite 
$SU(2)_L$ triplet, for which the branching ratios into $WW$, $WZ$, and $Wh$ 
final states are expected to be the same (see section~\ref{sec:excess}). For 
the 
diboson vs.\ dilepton final states, this is not obvious, so in 
figure~\ref{fig:diboson-dilepton}, we compare the cross sections times 
branching ratios of neutral vector resonances decaying to dileptons vs.\ $WW$ 
in all cases where the mass is between $1.7$ and $2.2$~TeV and the $\Gamma/m$ 
is at most $5\%$. We observe that the points with production cross section 
times branching ratio into $WW$ of order 1~fb typically lead to a signal in 
dileptons that is one to three orders of magnitude smaller. Comparing this to 
figure~\ref{fig:excess-Z}, we conclude that if the excesses in diboson final 
states are due to composite resonances, the excess in dileptons could be 
explained as well, but could also be absent.

\subsubsection{Prospects for vector resonance searches}
\label{sec:prospects-vector}

The discussion in the previous section has already shown that the diboson and 
dilepton final states are promising channels to look for vector resonances in 
the models studied by us. It should however be stressed that the vector 
resonances are not required to be light enough to be probed at LHC, even at
$\sqrt{s}=13$~TeV. In all three models, we have found viable points with 
moderate fine-tuning where all vector resonances are heavier than 6~TeV.
In the following, we discuss the most promising search channels for the vector 
resonances if they are light enough.

\paragraph{Gluon resonance}

$\rho_G$ can only decay to fermion pairs and usually has the largest branching 
fraction into quark partners because it couples to them through the strong 
coupling $g_G$. In that case, the most promising experimental strategy is 
to look for the quark partner pair decaying to SM particles 
\cite{Azatov:2015xqa,Araque:2015cna}.
If the decay to quark partners is kinematically disfavoured or forbidden, 
$\rho_G$ can also decay to SM quark pairs.
In the $U(2)^3$ models, we find that the decays to $t\bar t$ or $b \bar b$ 
can be up to 50\% and to light quarks up to 30\% (summing over the four light 
quark flavours).
In $U(3)^3_\text{RC}$, due to the large degree of compositeness of light 
right-handed quarks, the dominant SM decay mode are light quark pairs, with a 
branching ratio up to 40\%, while the $t\bar t$ and $b \bar b$ final states 
have branching ratios  below  10\% each.
The relative width of $\rho_G$ is around 10--50\% when 
the decays to SM states are relevant, with $U(3)^3_\text{RC}$ closer to the 
upper end of this range.\footnote{%
Note that this means that in our numerical analysis, there are effectively no 
direct bounds on $\rho_G$ due to our requirement of a narrow width in the LHC 
searches, see section~\ref{sec:spin1}.}

\paragraph{Charged resonances}

Among the three charged resonances, the lighter two are always nearly 
degenerate, with the lighter one being mostly the $\rho_R^\pm$ and the 
heavier one mostly the $\rho_L^\pm$, while the third charged resonance can be 
heavier and is mostly the axial vector resonance $\axial^\pm$.
The most  important state for collider phenomenology is the second one since it 
is the only one with an appreciable Drell-Yan production cross section.
Since its couplings to SM quarks are even weaker compared to the 
gluon resonance, it typically decays to quark partners, if kinematically 
allowed. If not, it decays to $WZ$ and $Wh$ with roughly equal branching 
ratios (cf.\ the discussion in sections~\ref{sec:excess} and \ref{sec:exc-res}).
The branching ratio into $tb$ is typically small but can reach 20\% in corners 
of the parameter space. The branching ratio to $\ell\nu$ is always below a 
percent.
The other two states could be produced via vector boson fusion that we have 
neglected in our analysis since it is expected to be very small at the LHC (see 
\cite{Mohan:2015doa} for a recent discussion).
We note that the axial vector resonance typically 
decays to $WZ$ and $Wh$ with the largest branching ratios and we find 
$\text{BR}(\rho_3^\pm\to W^\pm Z)\approx 3\,\text{BR}(\rho_3^\pm\to W^\pm h)$.

\paragraph{Neutral electroweak resonances}

Among the five neutral uncoloured resonances, the two heaviest are 
usually mostly the axial vector resonances that have a small production cross 
section in quark-antiquark collisions and (if produced via vector boson fusion) 
would decay with the largest branching ratios to $WW$ and $Zh$ with
$\text{BR}(\rho_{4,5}^0\to W^\pm W^\mp)\approx 3\,\text{BR}(\rho_{4,5}^0\to 
Z h)$.
Concerning the other three resonances, which are linear combinations of the 
$\rho_L^0$, $\rho_R^0$, and $\rho_X^0$, they again preferably decay to a 
pair of quark partners. If this is kinematically disfavoured, they can decay to 
pairs of SM quarks, leptons, or $W$ bosons, or to $Zh$. In the latter two 
cases, one typically has 
$\text{BR}(\rho_i^0\to W^\pm W^\mp)\approx \text{BR}(\rho_i^0\to Z h)$,
as expected for an $SU(2)_L$ triplet  (cf.\ the discussion in 
sections~\ref{sec:excess} and \ref{sec:exc-res}).
We find the branching ratios into electron or muon pairs to always be below 
2\%, which can however be overcome by the higher experimental sensitivity, cf.\ 
figure~\ref{fig:excess-Z} bottom.
The branching ratio to light jets can be up to 30\% in the $U(2)^3$ models 
and up to 70\% in $U(3)^3_\text{RC}$, while the one to $b\bar b$ can be up to 
40\% in all models.
In the $U(2)^3$ models,
the decay to $t\bar t$ can come close to 
100\% in parts of the 
parameter space.

\begin{figure}[tbp]
\centering
\includegraphics[width=0.8\textwidth]{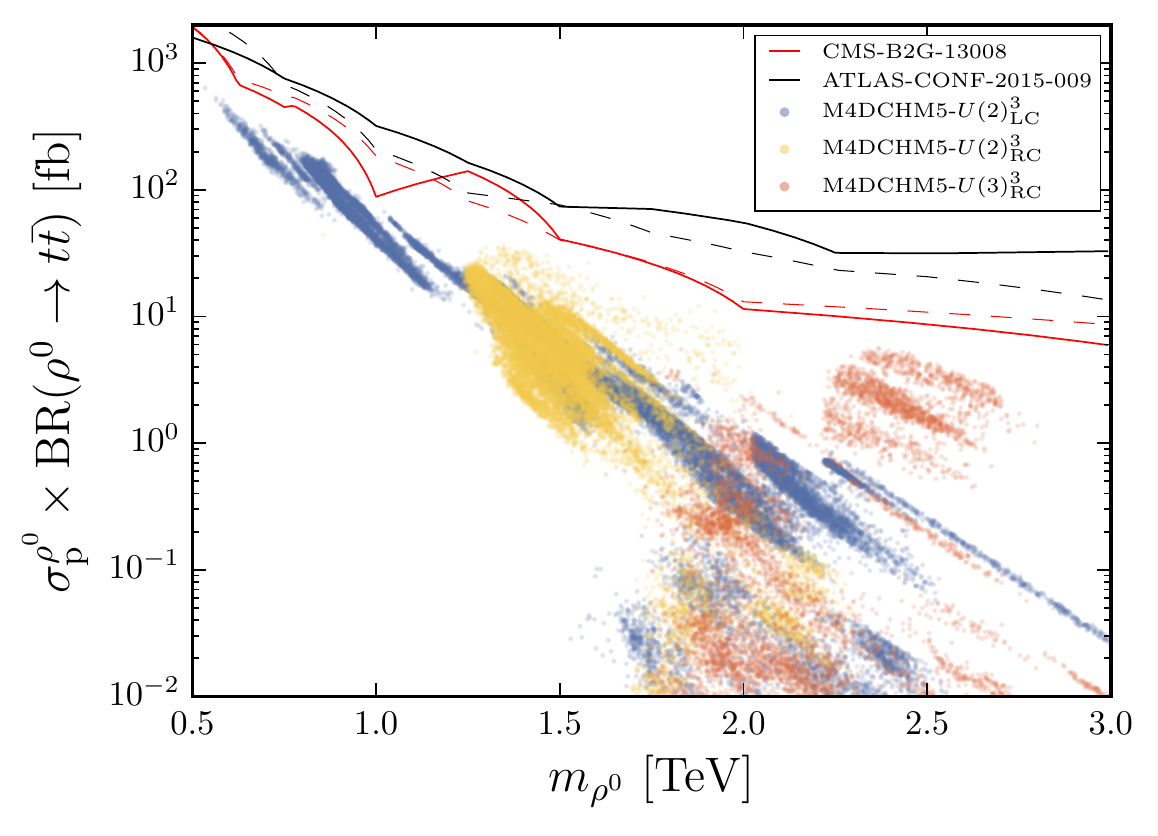}
\caption{Predictions for the production cross sections times branching ratios 
of neutral electroweak vector resonances decaying to top quarks
in all three models. Only points with narrow resonances ($\Gamma/m<0.05$) are 
shown. The dashed and solid curves show the expected and observed 95\% C.L.\ 
experimental limits.}
\label{fig:Ztt}
\end{figure}

In sections \ref{sec:u2lc-df1} and \ref{sec:unrc-df1}, we have already 
discussed that sizable NP contributions to the rare $B$ decay Wilson 
coefficient $C_9^{bs}$, as is required if one wants to solve the anomalies in 
$B$ physics discussed in section~\ref{sec:df1} in terms of new physics, requires 
a light, narrow neutral vector resonance with a large branching ratio to $t\bar 
t$.
In figure~\ref{fig:Ztt}, we show the predictions for the production cross 
section times branching ratio of the neutral electroweak resonances decaying to 
$t\bar t$ at LHC with $\sqrt{s}=8$~TeV, compared to 
existing ATLAS and CMS analyses. As in the previous plots,  we only show points 
with narrow resonances ($\Gamma/m<0.05$).
At masses below 1~TeV, the $U(2)^3_\text{LC}$ points correspond to the ones 
generating sizable NP effects in $C_9^{bs}$.
The points in right-handed compositeness only start at higher masses because 
the relative width is typically larger than 5\%.
The plot shows that cross sections not far from what LHC has probed in run 1 are 
attainable in all models. We conclude that this channel remains a promising 
probe at run 2, and discoveries are possible both for low and high masses.

\section{Summary}

In this paper, we have performed a comprehensive numerical analysis of a
four-dimen\-sional pNGB Higgs model based on the symmetry breaking coset 
$SO(5)/SO(4)$ with quark partners transforming as fundamentals of $SO(5)$. The 
model features a calculable one-loop Higgs potential and a custodial protection 
of the $Zb_L\bar b_L$ coupling. We have included constraints from electroweak 
precision tests, flavour physics, Higgs production and decay, contact 
interaction searches, as well as direct searches for quark and vector 
resonances. We have considered three different flavour symmetries, all of them 
exact in the composite sector and broken only by the composite-elementary 
mixing terms, namely $U(2)^3$ or $U(3)^3$ with left- or right-handed 
compositeness. Below, we summarize our main findings.
\begin{itemize}
\item Model-independently, we have pointed out that there are holes in existing 
experimental searches for quark partners decaying to $W$ or $Z$ plus a top or 
bottom quark, particularly for quark partners around 350~GeV which are 
not covered by Tevatron or LHC searches, see figure~\ref{fig:fermion_br}. We 
call on the experimental collaborations to close these holes by reanalyzing 
existing data. Quark partners decaying to a boson and a light quark are still 
weakly constrained.
\item In our numerical analysis, we have not found a single valid parameter 
point for the $U(3)^3_\text{LC}$ flavour structure. Although not a formal 
proof, we think this is a strong indication that this flavour structure is not 
compatible with electroweak precision tests and radiative EWSB in the model 
setup considered by us.
\item We have shown that the three other flavour structures can be made 
compatible with all relevant constraints with a fine tuning 
$\Delta_\text{BG}\lesssim100$, see figure~\ref{fig:tuning}.
\item We have demonstrated that first-row CKM unitarity is the most sensitive 
probe of light-quark compositeness in $U(2)^3_\text{LC}$, while in right-handed 
compositeness the dijet angular distribution is most sensitive to it, cf.\ 
figures \ref{fig:u2lc-ckm} and \ref{fig:unrc-jj}.
\item Higgs signal strengths are the cleanest observables to constrain the pNGB 
decay constant $f$, with small corrections due to light-quark compositeness, 
cf.\ figures \ref{fig:u2lc-h-st} and \ref{fig:unrc-h}.
\item In meson-antimeson mixing in the $U(2)^3$ models, the relations between 
$B_d$ and $B_s$ mixing that are expected from a leading-order spurion analysis 
are strongly violated for some of the valid points by terms that are formally 
of higher order in the spurion expansion.
\item In both $U(2)^3$ models, all observables in $B$, $B_s$ and $K$ mixing can 
saturate their current experimental limits, while in $U(3)^3$, this is true for 
the mass differences and $\epsilon_K$, while CP violation in the $B$ and $B_s$ 
systems is SM-like. The best means to experimentally distinguish the models 
based on $\Delta F=2$ observables alone can be read off
figures \ref{fig:u2lc-df2}, \ref{fig:u2rc-df2}, and \ref{fig:u3rc-df2}:
\begin{itemize}
\item In $U(2)^3_\text{LC}$, the relative NP effect in $\epsilon_K$ compared 
to $\Delta M_d$ is always smaller (or equal), in $U(2)^3_\text{RC}$ it is 
always larger (or equal), while in $U(3)^3_\text{RC}$ it is always equal.
\item In $U(2)^3_\text{LC}$, there can be large NP effects in $\phi_s$ which 
are typically correlated with an equal effect in $\phi_d$; in $U(2)^3_\text{RC}$ 
NP effects can be only in $\phi_s$ and not in $\phi_d$; in $U(3)^3_\text{RC}$ 
both phases are free from NP.
\end{itemize}
\item In the $U(2)^3$ models, CP violation in $D^0$-$\bar D^0$ mixing is small 
compared to the current experimental sensitivity, but could become relevant if 
the sensitivity improves by an order of magnitude.
\item The FCNC top decay $t\to cZ$ can reach a branching ratio of up to 
$10^{-5}$ in $U(2)^3_\text{LC}$ but is negligible in the other models.
\item Rare $B$ decays of the type $b\to s\ell^+\ell^-$ can not only receive
$Z$-mediated contributions, but also resonance-mediated contributions that can 
affect the Wilson coefficient $C_9^{bs}$ which is required for a NP explanation 
of various anomalies in $B$ physics data. These anomalies can be explained in 
all three models. In $U(3)^3_\text{RC}$, NP affects the Wilson coefficient 
$C_9^{bs}$ but not $C_{10}^{bs}$.
\item Explaining the $B$ physics anomalies implies the presence of a narrow 
neutral vector resonance around 1~TeV with a sizable branching ratio into 
$t\bar t$ or dijets with a production cross section just below what has been 
excluded in LHC run 1.
\item Various excesses in diboson events at a mass of roughly 2~TeV observed by 
ATLAS and CMS can be explained in all three models as well by the decay of a
2~TeV vector resonance, see figures \ref{fig:excess-W} and \ref{fig:excess-Z}. 
The solution possibly, but not necessarily, predicts a signal in dilepton 
events around the same mass as well, see figure~\ref{fig:diboson-dilepton}.
\end{itemize}

While we have limited ourselves to a single model with four different flavour 
structures in this work, there are several ways how our analysis could be 
generalized, such as studying non-minimal cosets, non-minimal couplings, 
different fermion representations, or different flavour structures, 
including more radical changes like disposing of partial compositeness for the 
first two generation quarks \cite{Matsedonskyi:2014iha,Cacciapaglia:2015dsa}. It 
would also be interesting to include a more realistic lepton sector. Finally, a 
more accurate treatment of the top quark mass, of loop corrections to the
$Z b_L \bar{b}_L$ coupling, of renormalization group effects
on FCNC operators, and of LHC constraints on singly produced fermion resonances 
would be very interesting to further scrutinize composite Higgs models in the 
future.

\section*{Acknowledgements}

We thank
Wolfgang Altmannshofer,
Frederik Beaujean,
Dario Buttazzo,
Marco Farina,
Oleksii Matsedonskyi,
Michele Redi,
Filippo Sala,
Marco Serone,
Andrea Tesi,
Andreas Weiler,
and
Andrea Wulzer
for useful discussions.
We acknowledge important code contributions by Stephan Jahn in the initial 
phase of the project.
We are grateful for the support by Jovan Mitrevski
through the Computational Center for Particle and Astrophysics (C2PAP),
where  the simulations have been carried out.
This work was supported by the DFG cluster of excellence ``Origin and Structure 
of the Universe''.

\appendix
\section{\texorpdfstring{$SO(5)$}{SO(5)} conventions} \label{sec:SO5}
For concreteness, we will present the conventions for $SO(5)$ generators and 
embeddings used by us.

The group $SO(5)$ can locally be expressed as $SO(5) \cong SU(2)_L \times 
SU(2)_R \times SO(5)/SO(4)$. Therefore, its 10 generators can be grouped into 
``left'', ``right'' and ``coset'':
\begin{eqnarray}
 \left(\T{a_L}\right)_{ij} &=& -\frac{\im}{2}\left( \frac{1}{2}\epsilon^{a_L b 
c} \left( \delta^b_i \delta^c_j - \delta^b_j \delta^c_i \right) + \left( 
\delta^{a_L}_i \delta^4_j - \delta^{a_L}_j \delta^4_i \right) \right), \\
  \left(\T{a_R}\right)_{ij} &=& -\frac{\im}{2}\left( \frac{1}{2}\epsilon^{a_R b 
c} \left( \delta^b_i \delta^c_j - \delta^b_j \delta^c_i \right) - \left( 
\delta^{a_R}_i \delta^4_j - \delta^{a_R}_j \delta^4_i \right) \right), \\
  \left( \T{\hat a} \right)_{ij} &=& -\frac{\im}{\sqrt{2}} \left( \delta^{\hat 
a}_i \delta^5_j - \delta^{\hat a}_j \delta^5_i \right),
\end{eqnarray}
Then, $SO(4)$ singlets $S$ and bidoublets $Q^{n_1, n_2}$ (with $SU(2)_L \times 
SU(3)_R$ quantum numbers $(n_1, n_2)$) can be embedded into $SO(5)$ 
fundamentals via
\begin{equation} \label{eq:SO5Embedding}
 \left( \begin{array}{c}
         Q^{1} \\ Q^{2} \\ Q^{3} \\ Q^{4} \\ Q^{5}
        \end{array}
 \right) = \frac{1}{\sqrt{2}} \left( \begin{array}{c}
                                      Q^{++} + Q^{--} \\ \im Q^{++} - \im 
Q^{--} 
\\ Q^{+-} - Q^{-+} \\ \im Q^{+-} + \im Q^{-+} \\ \sqrt{2} \, S
                                     \end{array}
 \right).
\end{equation}

\section{Mass matrices} \label{app:massmatriecs}
In this appendix we give the expressions for the mass mixing matrices that were 
obtained in the M4DCHM5.

\subsection{Boson sector}
The pNGB structure of the M4DCHM Lagrangian leads to mixings between the 
elementary and composite vector bosons of equal charge. In particular, the 
composite triplets $\rho_L^\mu$ and $\rho_R^\mu$ as well as the axial 
resonances $\axial^\mu$ will have neutral and charged components mixing with 
the elementary $\elem{W}^\mu$ and $\elem{B}^\mu$ gauge bosons. In addition, the 
neutral components will also mix with the $U(1)_X$ resonance $\rho_X^\mu$.

For the neutral and charged vector bosons one finds the 
following mass matrices given in tab. \ref{tab:BosonMassMatrices}.
\begin{table}
  \begin{sideways}
    \begin{minipage}{\textheight}
      \begin{eqnarray}
      && M^2_\mathrm{Boson, neutral} =\\
      &&\left( 
	\begin{array}{c|ccccccc}
	  &  \elem{W}^3_\mu & \elem{B}_\mu & \rho_{\mathrm{L} \, \mu} 
& \rho_{\mathrm{R} \, \mu} & \axial^3_\mu & \rho_{X\mu} & \axial^4_\mu \\
	  \hline 
	  \elem{W}^{3 \, \mu} & \frac{1}{2} \elemcoupl{g} f_1^2 & 0 
& -\frac{1}{2} \elemcoupl{g} g_\rho f_1^2 \cos^2 \left( \frac{h}{2 f} \right) & 
-\frac{1}{2} \elemcoupl{g} g_\rho f_1^2 \sin^2 \left( \frac{h}{2 f}   \right) 
& -\frac{1}{2 \sqrt{2}} \elemcoupl{g} g_\rho f_1^2 \sin \left( 
\frac{h}{f} \right)&0 &0\\
	  \elem{B}^\mu &  & \frac{1}{2} \elemcoupl{g}^{' \, 2} \left( f_1^2 
+ f_X^2 \right) & -\frac{1}{2} \elemcoupl{g}' g_\rho f_1^2 \sin^2 
\left( \frac{h}{2 f} \right) & - \frac{1}{2} \elemcoupl{g}' g_\rho f_1^2 
\cos^2 \left( \frac{h}{2 f} \right)& \frac{1}{2\sqrt{2}} \elemcoupl{g}' 
g_\rho f_1^2 \sin \left( \frac{h}{f} \right) & -\frac{1}{2} \elemcoupl{g}' 
g_X f_X^2 &0\\
	  \rho_\mathrm{L}^\mu & & & \frac{1}{2} g_\rho^2 f_1^2 & 0 & 0 & 0 & 0\\
	  \rho_\mathrm{R}^\mu & & & & \frac{1}{2} g_\rho^2 f_1^2 & 0 & 0 &0\\
	  \axial^{3 \, \mu}   & & & & & \frac{1}{2} g_\rho^2 \frac{f1^4}{ f_1^2 
- f^2} & 0 & 0 \\
	  \rho_X^\mu & & & & & & \frac{1}{2} g_X^2 f_X^2 &0 \\
	  \axial^{4 \, \mu} & & & & & & & \frac{1}{2} g_\rho^2 
\frac{f1^4}{f_1^2-f^2} 
	  \end{array} \nonumber 
	\right)
      \end{eqnarray}
      \\
      \begin{eqnarray}
      && M^2_\mathrm{Boson, charged} =\\
      && \left( 
	\begin{array}{c|cccc}
	  & \elem{W}^+_\mu & \rho^+_{\mathrm{L} \, \mu} & \rho^+_{\mathrm{R} 
\, \mu} & \axial^+_\mu \\
	  \hline
	  \elem{W}^{- \, \mu} & \frac{1}{2} \elemcoupl{g}^2 f_1^2 & 
-\frac{1}{2} \elemcoupl{g} g_\rho f_1^2\cos^2 \left( \frac{h}{2 f} 
\right)& -\frac{1}{2} \elemcoupl{g} g_\rho f_1^2\sin^2 \left( \frac{h}{2 
f} \right)   & -\frac{1}{2 \sqrt{2}} \elemcoupl{g} g_\rho f_1^2 \sin 
\left( \frac{h}{f} \right) \\
	  \rho^{- \, \mu}_\mathrm{L} & & \frac{1}{2} g_\rho^2 f_1^2 & 0 & 0\\
	  \rho^{- \, \mu}_\mathrm{R} & & & \frac{1}{2} g_\rho^2 f_1^2 & 0 \\
	  \axial^{- \, \mu} & & & & \frac{1}{2} g_\rho^2 \frac{f1^4}{f_1^2-f^2}
	\end{array}
      \right) \nonumber
      \end{eqnarray}
    \end{minipage}
  \end{sideways}
  \caption{Mass matrices for the neutral and singly charged bosons in the 
M4DCHM5.}
  \label{tab:BosonMassMatrices}
\end{table}

By the explicit mixing introduced in the Lagrangian one finds the 
following mass matrices for the gluon and their composite resonances. By 
construction this does not spoil invariance under the SM $SU(3)_c$, which
survives as a linear combination of the elementary and composite $SU(3)$ 
symmetries as can be seen from the fact that the gluon mass matrices exhibit a 
massless eigenvalue.

\begin{eqnarray}
 && M^2_\mathrm{Boson, Gluon} =\\
 && \left( \begin{array}{c|cc}
            & \elem{G}_\mu & \rho_{G \, \mu} \\
            \hline
            \elem{G}^{\mu} & \frac{1}{2} \elemcoupl{g}_3^2 f_G^2 & -\frac{1}{2} 
\elemcoupl{g}_3 g_G f_G^2 \\ 
            \rho_G^{\mu} &  & \frac{1}{2} g_{\rho _3}^2 f_G^2 \\
           \end{array}
 \right) \nonumber
\end{eqnarray}

\subsection{Fermion sector}
After EWSB the elementary quarks mix with all resonances carrying 
the same electric charge. By using the embedding (\ref{eq:SO5Embedding}) we 
express the components of the bidoublet resonances in such a way that they have 
definite quantum numbers under the $SO(4) = SU(2)_L \times SU(2)_R$ custodial 
symmetry.\footnote{For example, the field $Q_u^{+-}$ is part of a composite 
bidoublet resonance with $q_X^{(u)} = \frac{2}{3}$ and it has eigenvalues $+ 
\frac{1}{2}$ and $- \frac{1}{2}$ under $\T{3_L}$ and $\T{3_R}$, respectively.}

For the up- and down-type quarks we find the mass matrices given in tab. 
\ref{tab:QuarkMassMatrices}.

\begin{table}
  \begin{sideways}
    \begin{minipage}{\textheight}
      \begin{eqnarray}
      && M^{(\mathrm{u})}_\mathrm{fermion} = \\
      && \left( \begin{array}{c|ccccccccc}
		  & \elem{u}_R & Q^{+-}_{u R} & 
      \widetilde{Q}^{+-}_{u R} & Q^{-+}_{u R} & 
      \widetilde{Q}^{-+}_{u R}& Q^{++}_{d R} & 
      \widetilde{Q}^{++}_{d R} & S_{u R} & 
      \widetilde{S}_{u R} \\
		  \hline
		  \elem{\overline{u}}_L & 0 & -\Delta_{u_L} \cos^2 
      \left( \frac{h}{2 f} \right) & 0 & \Delta_{u_L} \sin^2 \left( 
      \frac{h}{2 f} \right) & 0 &-\Delta_{d_L} & 0 & 
      \frac{\im}{\sqrt{2}} \Delta_{u_L} \sin \left( \frac{h}{f} 
      \right) & 0 \\
		  \overline{Q}^{+-}_{u L} & 0 & m_U & 
      m_{Y_U} & 0&0&0&0&0&0 \\
		  \overline{\widetilde Q}^{+-}_{u L} & 
      -\frac{\im}{\sqrt{2}} \Delta^{\dagger}_{u_R} \, \sin \left( 
      \frac{h}{f} \right) & 0 & m_{\widetilde U}& 0&0&0&0&0&0 \\
		  \overline{Q}^{-+}_{u L}&0&0 & 0 & m_U & 
      m_{Y_U} & 0&0&0&0 \\
		  \overline{\widetilde Q}^{-+}_{u L} & 
      -\frac{\im}{\sqrt{2}} \Delta^{\dagger}_{u_R} \, \sin \left( 
      \frac{h}{f} \right) & 0 &0&0& m_{\widetilde U}& 0&0&0&0 \\
		  \overline{Q}^{++}_{d L}&0&0&0&0 & 0 &  
      m_D & m_{Y_D} & 0&0 \\
		  \overline{\widetilde Q}^{++}_{d L} & 0 & 0 
      &0&0&0&0& m_{\widetilde D}& 0&0 \\
		  \overline{S}_{u L}&0&0&0&0&0&0 & 0 & m_U 
      & m_{Y_U} + Y_U  \\
		  \overline{\widetilde S}_{u L} & 
      -\Delta^{\dagger}_{u_R} \, \cos \left( \frac{h}{f} \right) & 
      0 &0&0&0&0&0&0& m_{\widetilde U} \\
		\end{array}
      \right) \nonumber
      \end{eqnarray}		 
      \\
      \begin{eqnarray}
      && M^{(\mathrm{d})}_\mathrm{fermion} = \\
      && \left( \begin{array}{c|ccccccccc}
		  & \elem{d}_R & Q^{+-}_{d R} & 
      \widetilde{Q}^{+-}_{d R} & Q^{-+}_{d R} & 
      \widetilde{Q}^{-+}_{d R}& Q^{--}_{u R} & 
      \widetilde{Q}^{--}_{u R} & S_{d R} & 
      \widetilde{S}_{d R} \\
		  \hline
		  \elem{\overline{d}}_L & 0 & \Delta_{d_L} \sin^2 
      \left( \frac{h}{2 f} \right) & 0 & -\Delta_{d_L} \cos^2 \left( 
      \frac{h}{2 f} \right) & 0 &-\Delta_{u_L} & 0 & 
      \frac{\im}{\sqrt{2}} \Delta_{d_L} \sin \left( \frac{h}{f} 
      \right) & 0 \\
		  \overline{Q}^{+-}_{d L} & 0 & m_D & 
      m_{Y_D} & 0&0&0&0&0&0 \\
		  \overline{\widetilde Q}^{+-}_{d L} & 
      -\frac{\im}{\sqrt{2}} \Delta^{\dagger}_{d_R} \, \sin \left( 
      \frac{h}{f} \right) & 0 & m_{\widetilde D}& 0&0&0&0&0&0 \\
		  \overline{Q}^{-+}_{d L}&0&0 & 0 &  m_D & 
      m_{Y_D} & 0&0&0&0 \\
		  \overline{\widetilde Q}^{-+}_{d L} & 
      -\frac{\im}{\sqrt{2}} \Delta^{\dagger}_{d_R} \, \sin \left( 
      \frac{h}{f} \right) & 0 &0&0& m_{\widetilde D}& 0&0&0&0 \\
		  \overline{Q}^{--}_{u L}&0&0&0&0 & 0 & 
      m_U & m_{Y_U} & 0&0 \\
		  \overline{\widetilde Q}^{--}_{u L} & 0 & 0 
      &0&0&0&0& m_{\widetilde U}& 0&0 \\
		  \overline{S}_{d L}&0&0&0&0&0&0 & 0 & m_D 
      & m_{Y_D} + Y_D  \\
		  \overline{\widetilde S}_{d L} & 
      -\Delta^{\dagger}_{d_R} \, \cos \left( \frac{h}{f} \right) & 
      0 &0&0&0&0&0&0& m_{\widetilde D} \\
		\end{array}
      \right) \nonumber 
      \end{eqnarray}
    \end{minipage}
  \end{sideways}
  \caption{Mass matrices for the up- and down-type fermions in the M4DCHM5.}
  \label{tab:QuarkMassMatrices}
\end{table}

For the exotically charged fermion resonances the mass matrices are independent 
of the Higgs field. Thus, they do not give a contribution to the Higgs 
potential, which is clear since they do not mix with elementary fields.

\begin{equation}
 M^{+\frac{5}{3}}_\mathrm{fermion} = 
          \left( \begin{array}{c|cc}
            & Q^{++}_{u R} & \widetilde{Q}^{++}_{u R} \\
            \hline
            \overline{Q}^{++}_{u L}            & m_U & m_{Y_U}\\
            \overline{\widetilde Q}^{++}_{u L} &   0 & m_{\widetilde{U}}
           \end{array}
 \right), \qquad \qquad 
 M^{-\frac{4}{3}}_\mathrm{fermion} = 
      s \left( \begin{array}{c|cc}
            & Q^{--}_{d R} & \widetilde{Q}^{--}_{d R} \\
            \hline
            \overline{Q}^{--}_{d L}            & m_D & m_{Y_D}\\
            \overline{\widetilde Q}^{--}_{d L} &   0 & m_{\widetilde{D}}
           \end{array}
 \right) \nonumber
\end{equation}

Of course, the fields used above still carry flavour indices. As a consequence 
of this, all the entries tin the fermionic mass matrices actually are $3 \times 
3$ matrices in flavour space, promoting the up- and down-type mass matrices to 
$27 \times 27$ objects. The explicit form of the entries is model dependent and 
will be given in appendix \ref{sec:deltas}.

Since we took the leptons as purely elementary, their mass matrices are 
just diagonal taking the SM values.

\section{Explicit form of the composite-elementary mixings}\label{sec:deltas}

In this appendix, we give the explicit flavour structure of the 
composite-elementary mixings in the flavour symmetric models. We use bases 
where all unphysical parameters have already been rotated away and all phases 
have been made explicit.
\begin{itemize}
\item In $U(3)^3_\text{LC}$,
\begin{align}
\Delta_{u_L} &= \Delta_{Lt} ~\mathbbm{1} \,, &
\Delta_{u_R}^\dagger &= V^\dagger
\begin{pmatrix}
\Delta_{Ru} \\
& \Delta_{Rc} \\
&& \Delta_{Rt} \\
\end{pmatrix}
, \\
\Delta_{d_L} &= \Delta_{Lb} ~\mathbbm{1} \,, &
\Delta_{d_R}^\dagger &=
\begin{pmatrix}
\Delta_{Rd} \\
& \Delta_{Rs} \\
&& \Delta_{Rb} \\
\end{pmatrix}
. &
\end{align}
Here, $V$ is the CKM matrix with 3 angles and 1 phase.
\item In $U(3)^3_\text{RC}$,
\begin{align}
\Delta_{u_L} &= V^\dagger
\begin{pmatrix}
\Delta_{Lu} \\
& \Delta_{Lc} \\
&& \Delta_{Lt} \\
\end{pmatrix}
, &
\Delta_{u_R}^\dagger &= \Delta_{Rt} ~\mathbbm{1} \,, \\
\Delta_{d_L} &=
\begin{pmatrix}
\Delta_{Ld} \\
& \Delta_{Ls} \\
&& \Delta_{Lb} \\
\end{pmatrix}
, &
\Delta_{d_R}^\dagger &= \Delta_{Rb} ~\mathbbm{1} \,.
\label{eq:u3rc-deltadL}
\end{align}
\item In $U(2)^3_\text{LC}$,
\begin{align}
\Delta_{u_L} &= \begin{pmatrix}
\Delta_{Lu} \\
& \Delta_{Lu} \\
&& \Delta_{Lt} \\
\end{pmatrix} , &
\Delta_{u_R}^\dagger &=
\begin{pmatrix}
c_u \Delta_{Ru} & -s_u \Delta_{Rc} e^{i\alpha_u} \\
s_u \Delta_{Ru} e^{-i\alpha_u} & c_u \Delta_{Rc} & \epsilon_u\Delta_{Rt} 
e^{i\phi_t}\\
&& \Delta_{Rt} \\
\end{pmatrix}
, \\
\Delta_{d_L} &= \begin{pmatrix}
\Delta_{Ld} \\
& \Delta_{Ld} \\
&& \Delta_{Lb} \\
\end{pmatrix} , &
\Delta_{d_R}^\dagger &=
\begin{pmatrix}
c_d \Delta_{Rd} & -s_d \Delta_{Rs} e^{i\alpha_d} \\
s_d \Delta_{Rd} e^{-i\alpha_d} & c_d \Delta_{Rs} & 
\epsilon_d\Delta_{Rb}e^{i\phi_b}\\
&& \Delta_{Rb} \\
\end{pmatrix}
. &
\end{align}
\item In $U(2)^3_\text{RC}$,
\begin{align}
\Delta_{u_L} &=
\begin{pmatrix}
c_u \Delta_{Lu} & -s_u \Delta_{Lc} e^{i\alpha_u} \\
s_u \Delta_{Lu} e^{-i\alpha_u} & c_u \Delta_{Lc} & 
\epsilon_u\Delta_{Lt}e^{i\phi_t}\\
&& \Delta_{Lt} \\
\end{pmatrix}
, &
\Delta_{u_R}^\dagger &=
\begin{pmatrix}
\Delta_{Ru} \\
& \Delta_{Ru} \\
&& \Delta_{Rt} \\
\end{pmatrix} 
, \\
\Delta_{d_L} &=
\begin{pmatrix}
c_d \Delta_{Ld} & -s_d \Delta_{Ls} e^{i\alpha_d} \\
s_d \Delta_{Ld} e^{-i\alpha_d} & c_d \Delta_{Ls} & 
\epsilon_d\Delta_{Lb}e^{i\phi_b}\\
&& \Delta_{Lb} \\
\end{pmatrix}
, &
\Delta_{d_R}^\dagger &= \begin{pmatrix}
\Delta_{Rd} \\
& \Delta_{Rd} \\
&& \Delta_{Rb} \\
\end{pmatrix} .
\end{align}
\end{itemize}

\section{Constraints from the dijet angular distribution}\label{sec:dijetapp}

As discussed in section~\ref{sec:ci}, experimental analyses of contact 
interactions typically only quote constraints on a single operator -- or for 
individual operators, but only allowing one at a time.
To correctly treat the case with simultaneous contributions from multiple 
operators, we follow the procedure outlined in \cite{Domenech:2012ai}.
In this paper, analytical expressions are given for the dijet cross section in 
bins of the dijet mass $m_{jj}$ and the rapidity $\chi$.
The most recent ATLAS and CMS analyses use multivariate techniques rather than 
considering only a ratio of bins. In our numerical analysis, we have thus 
adopted the following procedure:

\begin{enumerate}
\item We identify the most sensitive bin in the experimental analysis;
\item We compute the numerical coefficients $\vec{\mathcal  P}$ and
$\vec{\mathcal  Q}$ defined as in \cite{Domenech:2012ai} for the 8~TeV LHC in 
the respective bin.
\item We compute the NP contribution of all operators to the cross section in 
this bin;
\item We multiply our result by an overall factor to exactly reproduce the 95\% 
C.L.\ bound on the Wilson coefficient $c_{qq}^{(1)}$ quoted in the experimental 
paper.
\end{enumerate}
In this way, our approximation of computing the cross section analytically and 
pretending that only a single bin is relevant is only used for the relative 
contributions of the individual operators, while any overall change (such as 
k-factors) cancels out since we normalize to the bound obtained for the 
$c_{qq}^{(1)}$ coefficient by the experimentalists. We have checked that the 
relative contributions are not very sensitive to changes in the bin chosen in 
the first step.

In our numerical analysis, we use the bound from the most recent ATLAS analysis
\cite{Aad:2015eha}. We assume the most sensitive bin to be the one with 
$\chi<3.32$, $m_{jj}>3.2$~TeV. We can then write the new physics 
contribution to the dijet cross section in this bin,
\begin{equation}
\sigma_{jj}^\chi = \left.
\int_1^{3.32}d\chi ~ \frac{d\sigma(pp\to jj)}{d\chi}
\right|_{m_{jj}>3.2\,\text{TeV}}^\text{NP}
\end{equation}
normalized to the 95\% C.L.\ cross section on this quantity extracted by 
reproducing the bound on $c_{qq}^{(1)}$ quoted in \cite{Aad:2015eha}, as
\begin{equation}
\frac{ \sigma_{jj}^\chi }{(\sigma_{jj}^\chi)_\text{95\% C.L.}} = 
- \frac{1}{\Lambda^2}
\vec{A}\cdot \vec{\mathcal P}'
+ \frac{1}{\Lambda^4}
\vec{B}\cdot \vec{\mathcal Q}' \,,
\end{equation}
where $\vec{A}$ and $\vec{B}$ are given in eq.~(16) of \cite{Domenech:2012ai} 
and $\vec{\mathcal P}', \vec{\mathcal Q}'$  are equal up to normalization to 
$\vec{\mathcal P}, \vec{\mathcal Q}$ defined in \cite{Domenech:2012ai}. 
Numerically, we find
\begin{align}
\vec{\mathcal P}' &= \left(
0.36 P_{uu}',0.12 P_{uu}',
0.36 P_{dd}',0.12 P_{dd}',
0.17 P_{ud}',0.74 P_{ud}'  \right) \,,
\\
\vec{\mathcal Q}' &= \left(
0.013 Q_{uu}',0.0069 Q_{uu}',
0.013 Q_{dd}',0.0069 Q_{dd}',
0.0024 Q_{ud}',0.00097 Q_{ud}' \right) \,,
\end{align}
where
\begin{align}
P_{uu}' &= (4.93 \,\text{TeV})^2 \,,
&
P_{dd}' &= (1.46 \,\text{TeV})^2 \,,
&
P_{ud}' &= (3.82 \,\text{TeV})^2 \,.
\\
Q_{uu}' &= (7.93 \,\text{TeV})^4 \,,
&
Q_{dd}' &= (4.28 \,\text{TeV})^4 \,,
&
Q_{ud}' &= (6.95 \,\text{TeV})^4 \,.
\end{align}

\bibliographystyle{JHEP}
\bibliography{chm}

\end{document}